\documentclass[twocolumn,showpacs,preprintnumbers,amsmath,amssymb,nofootinbib]{revtex4-2}
\usepackage{epsfig}

\usepackage{comment}
\usepackage[table]{xcolor}
\usepackage{graphicx}
\usepackage{dcolumn}
\usepackage{bm}

\usepackage{hyperref}
\hypersetup{colorlinks=true,linkcolor=magenta,anchorcolor=green,citecolor=cyan,filecolor=black,menucolor=black,urlcolor=brown}
\usepackage{slashed}
\newcommand{\ba}{\begin{eqnarray}}
\newcommand{\ea}{\end{eqnarray}}
\newcommand{\be}{\begin{equation}}
\newcommand{\ee}{\end{equation}}

\begin{document}

\title{3D Vortices and rotating solitons in ultralight dark matter}

\author{Philippe Brax}
\affiliation{Universit\'{e} Paris-Saclay, CNRS, CEA, Institut de physique th\'{e}orique, 91191, Gif-sur-Yvette, France}
\author{Patrick Valageas}
\affiliation{Universit\'{e} Paris-Saclay, CNRS, CEA, Institut de physique th\'{e}orique, 91191, Gif-sur-Yvette, France}

\begin{abstract}

We study the formation and the dynamics of vortex lines in rotating scalar dark matter halos,
focusing on models with quartic repulsive self-interactions. 
In the nonrelativistic regime, vortex lines and their lattices arise from the Gross-Pitaevskii 
equation of motion, as for superfluids and Bose-Einstein condensates studied in laboratory experiments.
Indeed, in such systems vorticity is supported by the singularities of the phase of the scalar field,
which leads to a discrete set of quantized vortices amid a curl-free velocity background.
In the continuum limit where the number of vortex lines becomes very large, we find that the equilibrium
solution is a rotating soliton that obeys a solid-body rotation, with an oblate density 
profile aligned with the direction of the total spin. This configuration is dynamically stable provided the 
rotational energy is smaller than the self-interaction and gravitational energies.
Using numerical simulations in the Thomas-Fermi regime, with stochastic initial conditions for a spherical
halo with a specific averaged density profile and angular momentum, we find that a rotating soliton 
always emerges dynamically, within a few dynamical times, and that a network of vortex lines aligned
with the total spin fills its oblate profile. These vertical vortex lines form a regular lattice in the equatorial plane,
in agreement with the analytical predictions of uniform vortex density and solid-body rotation.
These vortex lines might further extend between halos to form the backbone of spinning cosmic filaments.  

\end{abstract}

\date{\today}

\maketitle

\section{Introduction}
\label{sec:Introduction}

Ultralight dark matter models ($ m \ll 1$ eV) 
\cite{Matos:1999et,Peebles_2000,Goodman:2000tg,Hu:2000ke}
have been the focus of a renewed interest in recent years
\cite{Sikivie:2010bq,Harko:2011jy,Hui:2016ltb,Schive:2014dra,Shapiro:2021hjp,Harko:2011jy}.
As a typical example, axion-like-particles generically arise in string theory scenarios 
\cite{Svrcek:2006yi,Arvanitaki:2009fg,Halverson:2017deq,Bachlechner:2018gew}
and could alleviate small-scale tensions of the standard cold dark matter (CDM) model
\cite{Weinberg:2013aya,DelPopolo:2016emo,Nakama:2017ohe,Salucci:2018hqu,DiLuzio:2020wdo}.
From the point of view of structure formation, the main departure from CDM involves
wavelike effects (interferences between eigenmodes of the Schr\"odinger equation of motion
in the nonrelativistic regime) and the formation of solitons (hydrostatic equilibria or ground states
with a unique density profile)
\cite{Chavanis:2011zi,Chavanis:2011zm,Mocz:2017wlg,Veltmaat:2018dfz,Eggemeier:2019jsu,Garcia:2023abs,GalazoGarcia:2024fzq,Mirasola:2024pmw},
which lead to flat dark matter density cores at the center of virialized halos.
Additionally, the characteristic oscillations of the dark matter fields at the fast frequency $m$
can lead to distinct effects on gravitational waves
\cite{Khmelnitsky:2013lxt,Brax:2024yqh,Cai:2023ykr,Chowdhury:2023xvy,Blas:2024duy}.

Although spherically symmetric for static dark matter distributions, solitons can develop more intricate 
shapes and properties when rotation is taken into account. 
This naturally occurs in the cosmological context, where the dark matter halos that form inside the
cosmic web have a nonzero angular momentum, with the dimensionless spin parameter
\be
\lambda_J = J |E|^{1/2} /( {\cal G} M^{5/2})
\label{eq:lambda-L}
\ee
ranging from 0.01 to 0.1 \cite{Barnes:1987hu,Bullock:2000ry,Maccio-2007},
where $J$, $E$ and $M$ are the total angular momentum, energy and mass of the halo,
and ${\cal G}$ is Newton's constant.

This angular momentum is generated by the tidal field from neighbouring structures
\cite{Peebles:1969jm,Doroshkevich-1970,White-1984}.
In contrast with CDM and classical systems of collisionless particles, described by a phase-space
distribution $f(\vec x,\vec v,t)$ that can exhibit any pattern of rotation and vorticity,
scalar-field dark matter is described by a wave function $\psi(\vec x,t)$ in the nonrelativistic regime,
with a velocity field $\vec v \propto \vec\nabla S$ defined as the gradient of its phase
$S={\rm arg}(\psi)$.
As such, smooth configurations can only generate curl-free velocity fields and rather constrained
rotation patterns, such as Riemann-S ellipsoids \cite{Rindler-Daller:2011afd}.
In practice, as for superfluids and Bose-Einstein condensates (BEC) studied in the laboratory
\cite{Abo-Shaeer-2001,Fetter_2001,pitaevskii2003bose,Pethick_Smith_2008},
and as already pointed out by Feynman \cite{FEYNMAN1955},
these systems develop a nonzero vorticity by generating vortex lines, associated with singularities
of the phase $S$ (but the wave function $\psi$ remains regular and vanishes at these locations).
This allows the system to display a wide variety of rotation patterns where the vorticity carried by each
vortex tube is quantized
\cite{Silverman:2002qx,RINDLERDALLER2008,Rindler-Daller:2011afd,Kain:2010rb,Banik:2013rxa}. 

In this paper, we focus on the Thomas-Fermi limit, that is, we consider scenarios where the
de Broglie wavelength $\lambda_{\rm dB} \sim 1/(mv)$ is much smaller than the size of the system
and the scalar-field Lagrangian includes a repulsive quartic self-interaction.
This means that the soliton that forms at the center of the virialized halo is governed by the balance
between gravity and the repulsive self-interaction \cite{Chavanis:2011zi,Garcia:2023abs,GalazoGarcia:2024fzq}
rather than by the balance between gravity and the quantum pressure, as in fuzzy dark matter 
(FDM) models \cite{Hu:2000ke,Hui:2016ltb}.
In this limit, the vorticity quantum carried by each vortex line is very small.
Then, even a dimensionless angular momentum (\ref{eq:lambda-L}) below one or ten percent
leads to a large number of vortices.
This approximates a continuum limit, where the velocity field is no longer curl-free and can
mimic any rotation pattern. 
Then, in contrast with models without self-interactions or attractive self-interactions
where vortices do not naturally form \cite{Schobesberger:2021ghi,Dmitriev:2021utv}
or are unstable, in the regime studied in this paper vortices can form and sustain rotating solitons
\cite{Rindler-Daller:2011afd}.

Extending our previous study of 2D systems \cite{Brax:2025uaw},
we investigate in this paper the formation and the dynamics of solitons inside 3D virialized halos
with a nonzero initial angular momentum.
We use both numerical simulations, starting with stochastic initial conditions defined by a specific
averaged density profile and total angular momentum but without an initial soliton, and analytical
computations.
As in the 2D case, we find that for initial conditions with a nonzero total angular momentum
a rotating soliton forms in a few dynamical times.
Again, the soliton exhibits a solid-body rotation, which is generated by a regular lattice of vortices,
whereas the outer envelope of the system shows a stochastic maze of vortices.
However, whereas in the 2D case the vortices are mere isolated point vortices, 
in the 3D case they are extended vortex lines.
Thus, inside the soliton we find a regular lattice of vertical vortex lines, aligned with the total angular
momentum $\vec J$ (taken to be along the vertical axis), whereas in the outer envelope we find a tangle
of intertwining and curved vortex lines of any directions.
Moreover, the rotating soliton is now a fully three-dimensional object,
with a specific axisymmetric and oblate shape aligned with the total spin. 

This paper is organized as follows.
We first recall in Sec.~\ref{sec:GP} the Gross-Pitaevskii equation of motion satisfied by the
system in the nonrelativistic regime and the shape of the static soliton that forms at the center
of virialized halos.
Next, we describe in Sec.~\ref{sec:vortex-lines} how angular momentum and rotation can be
supported by the system through the appearance of vortex lines, that is, singularities of the
phase (but the complex scalar field $\psi$ remains regular). This is the same behavior as that
observed in the laboratory for rotating superfluids.
We consider the continuum limit in Sec.~\ref{sec:continuum}. This corresponds to the Thomas-Fermi
regime where the de Broglie wavelength is much smaller than the size of the system and of the
soliton. Then, a macroscopic rotation is supported by many vortex lines, with a core radius that
is much smaller than the inter-vortex distance, which itself is much smaller than the system size.
We derive the profile of the rotating soliton that forms in such systems and show that it
displays a solid-body rotation. We also show that these rotating solitons are stable, for rotation
rates that are not too high.
We describe our simulation setup in Sec.~\ref{sec:numerical-setup}, our choice of initial conditions
and the numerical algorithm.
We analyze our numerical results in Sec.~\ref{sec:numerical-results} and we check that we find
a good agreement with the analytical predictions.
We finally conclude in Sec.~\ref{sec:conclusion}.

We provide in App.~\ref{app:rotating-soliton} the details of the computation of the rotating soliton 
profile, based on a perturbative expansion in the rotation rate.
We detail in App.~\ref{app:initial-conditions} the building of our initial conditions, based on the
correspondence in the semi-classical limit between the scalar-field system governed by 
the Gross-Pitaevskii equation of motion and a system of classical collisionless particles described
by a phase-space distribution function.

\section{Equations of motion}
\label{sec:GP}

\subsection{Nonrelativistic equations of motion}

We consider scalar-field dark matter scenarios with quartic self-interactions,
defined by the Lagrangian
\be
\mathcal{L}_{\phi} = -\frac{1}{2}(\partial\phi)^2 - \frac{m^2}{2}\phi^2 - \frac{\lambda_4}{4} \phi^4 ,
\label{eq:lagrangian-sfdm}
\ee
where we work in natural units, $c=\hbar=1$.
We focus on the late Universe where the quadratic term dominates over both the quartic term and
the Hubble friction, so that at leading order the field oscillates at frequency $m$ within the quadratic
potential and its energy density decays as $\rho \propto a^{-3}$, as for any dark matter candidate.
We focus on the case of repulsive self-interactions, $\lambda_4 > 0$, so that the effective pressure
generated by the self-interactions can balance gravity and lead to stable hydrostatic equilibria
(solitons) as in Eq.(\ref{eq:TF-static}) below.

In the nonrelativistic regime and on subgalactic scales where we can neglect the Hubble expansion,
it is convenient to factorize the oscillations at frequency $m$ by introducing a complex scalar
field $\psi$ \citep{Hu:2000ke,Hui:2016ltb},
\be
\phi= \frac{1}{\sqrt{2m}} ( \psi e^{-imt} +  \psi^* e^{imt}) .
\label{eq:phi-psi}
\ee
Substituting into the action or the equation of motion and averaging over these fast oscillations
gives the nonrelativistic equation of motion \citep{Brax:2019fzb}
\be
i \frac{\partial\psi}{\partial t} =
- \frac{\Delta\psi}{2m} + m ( \Phi_N + \Phi_I ) \psi , \;\;\;
\Phi_I = \frac{3 \lambda_4}{4 m^3} |\psi|^2 ,
\label{eq:Schrod}
\ee
which has the form of a nonlinear Schr\"odinger equation, also known as a Gross-Pitaevskii equation.
However, the Newtonian potential $\Phi_N$ is not external but given by the self-gravity
of the scalar field.

It is useful to work with dimensionless quantities, which we define by
\be
\psi = \psi_\star \tilde\psi, \;\;\; t = T_\star \tilde t, \;\;\;
\vec r = L_\star \tilde{\vec r} , \;\;\;
\Phi = V_\star^2 \tilde\Phi,
\label{eq:dimesionless-def}
\ee
where $T_\star$, $L_\star$ and $V_\star = L_\star/T_\star$ are the characteristic time, length and
velocity scales of the system, and $T_\star = 1/\sqrt{{\cal G} m \psi_\star^2}$ where
${\cal G}$ is Newton's constant.
For instance, the characteristic length $L_\star$ and mass $M_\star = (4/\pi) m \psi_\star$ 
can be chosen as the radius and the mass of the halo at the initial time
of our numerical simulations, so that in these rescaled coordinates $\tilde R= R/L_\star$
and $\tilde M= M/M_\star$ are unity in Fig.~\ref{fig:initial} below.
This defines, in turn, $\psi_\star$, $T_\star$ and $V_\star$. Up to factors of order unity,
these are the initial amplitude of the wavefunction at the center and the initial dynamical time
and virial velocity of the halo.

Under this rescaling, the dimensionless Schr\"odinger--Poisson
equations that govern the dynamics read
\be
i \epsilon \frac{\partial\tilde\psi}{\partial\tilde t} =
- \frac{\epsilon^2}{2} \tilde\Delta \tilde\psi + (\tilde\Phi_{N} + \tilde\Phi_I ) \tilde\psi ,
\label{eq:Schrod-eps}
\ee
\be
\tilde\Delta \tilde\Phi_{N} = 4 \pi \tilde\rho , \;\;\; \tilde\Phi_I = \lambda \tilde\rho , \;\;\;
\mbox{and}  \;\;\; \tilde\rho = | \tilde\psi|^2 ,
\label{eq:Poisson-eps}
\ee
with the coupling constant $\lambda = 3\lambda_4/(4 {\cal G} m^4 L_\star^2)$.
The coefficient $\epsilon$ is given by $\epsilon = T_\star/(m L_\star^2)$; it scales as
$\epsilon \sim \lambda_{\rm dB}/L_\star$ where $\lambda_{\rm dB} = 2\pi/(m V_\star)$
is the typical de Broglie wavelength.
Thus, this parameter measures the relevance of the quantum pressure or more generally of
wave effects.
For a given halo or soliton size and mass (e.g., in the galactic context $1$ kpc and $10^{7} M_\odot$),
$\epsilon$ grows as the inverse of the scalar mass $m$. It is of the order of unity for
fuzzy dark matter models with $m \sim 10^{-21}$ eV and smaller for higher scalar masses.
Thus, in this context, our results apply to models where $m \gg 10^{-21}$ eV and 
the soliton radius $R_0 \sim 1$ kpc is due to the repulsive self-interactions, as in 
Eq.(\ref{eq:rho-TF-0}) below.

In the remainder of this article, we omit the tilde to simplify the notations.

\subsection{Hydrodynamical picture}
\label{sec:hydro}

Defining a velocity field $\vec v$ from the phase $S$ of the wave function \citep{Madelung:1927ksh},
\be
\psi = \sqrt{\rho} \, e^{i S} , \;\;\; \vec v = \epsilon \vec\nabla S ,
\label{eq:Madelung}
\ee
and substituting into the equation of motion (\ref{eq:Schrod-eps}), the real and imaginary parts
give the continuity and Euler equations,
\be
\frac{\partial\rho}{\partial t} + \nabla\cdot(\rho \vec v) = 0 ,
\label{eq:continuity}
\ee
\be
\frac{\partial\vec v}{\partial t} + (\vec v \cdot \vec\nabla) \vec v =
- \vec\nabla( \Phi_Q + \Phi_N + \Phi_I ) ,
\label{eq:Euler}
\ee
where we introduced the so-called quantum pressure defined by
\be
\Phi_Q = - \frac{\epsilon^2}{2} \frac{\Delta \sqrt{\rho}}{\sqrt{\rho}} .
\label{eq:PhiQ-def}
\ee
This provides a mapping from the Gross-Pitaevskii equation (\ref{eq:Schrod-eps})
to hydrodynamical equations of motion.
However, this mapping breaks down at locations where the density vanishes and the phase
is ill-defined. As recalled below, these singularities lead to the formation of vortices, which support the
vorticity that could not be supported otherwise by a regular curl-free velocity field defined as the gradient
of a regular phase $S$.

\subsection{Conserved quantities}
\label{sec:conserved}

The Gross-Pitaevskii equation (\ref{eq:Schrod-eps}) conserves the total mass of the system,
\be
M= \int d\vec r \, \rho ,
\label{eq:total-mass}
\ee
the momentum $\vec P = \int d\vec r \, \rho \vec v$, the angular momentum
$\vec J = \int d\vec r \, \rho \vec r \times \vec v$, with the component along the vertical axis
\be
J_z = \int d\vec r \, \rho ( x v_y - y v_x ) ,
\label{eq:total-Jz}
\ee
and the energy
\be
E[\rho,\vec v] = \int d {\vec r} \left[ \frac{\epsilon^2}{2} (\vec\nabla \sqrt\rho)^2
+ \frac{1}{2} \rho \vec v^{\,2} + \frac{1}{2} \rho \Phi_N + {\cal V}_I \right] ,
\label{eq:E-rho-v}
\ee
which also reads in terms of the wave function $\psi$ as
\be
E[\psi] = \int d {\vec r} \left[ \frac{\epsilon^2}{2} | \vec\nabla \psi |^2
+ \frac{1}{2} \rho \Phi_N + {\cal V}_I \right] .
\label{eq:E-psi}
\ee
Here the self-interaction potential $\cal V_I(\rho)$ is related to $\Phi_I(\rho)$ by
\be
\Phi_I = \frac{d {\cal V}_I}{d\rho}  \;\;\; \mbox{and} \;\;\; {\cal V}_I = \lambda \frac{\rho^2}{2} .
\label{eq:V-I-rho}
\ee

\subsection{Static soliton}
\label{sec:static-soliton}

For a given total potential $\Phi_N+\Phi_I$ that is spherically symmetric, the ground state of the 
Schr\"odinger equation (\ref{eq:Schrod-eps}) takes the form
\be
\psi_{\rm sol}(\vec r,t) = e^{-i \mu t/\epsilon} \hat\psi_{\rm sol}(r) ,
\label{eq:psi-sol}
\ee
and the Schr\"odinger equation reads as
\be
\Phi_Q + \Phi_N + \Phi_I = \mu ,
\label{eq:hydrostatic-full}
\ee
where $\Phi_Q$ is the quantum pressure defined in Eq.(\ref{eq:PhiQ-def}).
This ground state, also called a soliton, is also the hydrostatic equilibrium of the hydrodynamical 
equations (\ref{eq:continuity})-(\ref{eq:Euler}) and the minimum of the energies (\ref{eq:E-rho-v}) and
(\ref{eq:E-psi}) at fixed mass
\citep{Chavanis:2011zi,Chavanis:2011zm,Harko:2011jy,Brax:2019fzb}.

In the Thomas-Fermi regime where we can neglect the quantum pressure because $\epsilon \ll 1$
and gravity is balanced by the repulsive self-interactions, the hydrostatic equilibrium is given by
\be
\mbox{TF regime} : \;\;\; \Phi_N + \Phi_I = \mu  .
\label{eq:TF-static}
\ee
In 3D this gives the density profile \citep{Chavanis:2011zi,Chavanis:2011zm,Harko:2011jy,Brax:2019fzb}
\be
r < R_0 : \;\;\; \rho(r) = \rho_{0} j_0( \pi r/R_0) , \;\;\; R_0 = \sqrt{\lambda \pi}/2 ,
\label{eq:rho-TF-0}
\ee
where $\rho_0$ is the central density, $j_0$ is the zeroth-order spherical Bessel function,
and $\rho(r)=0$ outside of the soliton, at $r>R_0$.
The soliton radius $R_0$ does not depend on its mass, which reads
\be
M = \frac{4}{\pi} \rho_0 R_0^3 .
\ee
The gravitational potential, normalized to zero at infinity, reads
\ba
&& r < R_0 : \;\;\; \Phi_{N}(r) =  - \lambda \rho_0 [ j_0( \pi r/R_0) + 1 ] , \nonumber \\
&& r > R_0 : \;\;\; \Phi_{N}(r) =  - \lambda \rho_0 R_0/r ,
\label{eq:PhiN-TF-0}
\ea
and we have $\mu= - \lambda \rho_0$.

\section{Vortex lines}
\label{sec:vortex-lines}

Being defined as the gradient of the phase $S$, the velocity field (\ref{eq:Madelung}) introduced by
the Madelung transformation would be curl-free. As such, it would not support vorticity nor usual rotation
patterns shown by general hydrodynamical systems or classical particle systems.
However, this only holds if the phase is regular.
Therefore, as for superfluids a nonzero vorticity and rotation can be supported by the system
when it develops singularities, that is, vortices
\cite{FEYNMAN1955,Abo-Shaeer-2001,Fetter_2001,pitaevskii2003bose,Pethick_Smith_2008}.
These structures appear at the locations where the density $\rho$ and the wave function $\psi$ vanish,
so that the phase $S$ is ill-defined in the Madelung transformation (\ref{eq:Madelung}).
The two conditions ${\rm Re}(\psi) = 0$ and ${\rm Im}(\psi)=0$ define points in a 2D system and
lines in a 3D system \cite{Hui:2020hbq}.
Therefore, in the 3D case we obtain vortex lines
\cite{Silverman:2002qx,RINDLERDALLER2008,Rindler-Daller:2011afd,Kain:2010rb}, 
instead of the isolated point vortices found in our previous 2D study \cite{Brax:2025uaw}.

\subsection{Single vortex line}

In the case of a single vortex, on scales much smaller than the system size (or the soliton radius),
we can approximate the vortex line as a straight line along the vertical $z$-axis,
embedded in a homogeneous medium at density $\rho_0$.
Thus, neglecting perturbations of the vortex line (e.g., bending modes) and using cylindrical coordinates
$(r_\perp,\varphi,z)$, for configurations that are independent of $z$ we recover a 2D system
and we can use the results presented in \cite{Brax:2025uaw}.
Then, vortices of spin $\sigma$ correspond to the solutions of the Gross-Pitaevskii equation of the form
\be
\psi(\vec r,t) = e^{-i \mu t/\epsilon} \sqrt{\rho_0} f(r_\perp) e^{i \sigma \varphi} , \;\;\;
\rho(\vec r) = \rho_0 f^2(r_\perp) ,
\label{eq:single-vortex-psi}
\ee
where $\sigma \in \mathbb{Z}$ and we have the boundary condition
$f(r_\perp) \to 1$ for $r_\perp \to \infty$.
Substituting into the Gross-Pitaevskii equation (\ref{eq:Schrod-eps}) we obtain the differential
equation \cite{pitaevskii2003bose}
\be
\frac{d^2f}{d\eta^2} + \frac{1}{\eta} \frac{df}{d\eta} + \left( 1 - \frac{\sigma^2}{\eta^2} \right) f
- f^3  = 0 ,
\ee
where we introduced the rescaled radial coordinate $\eta$ and the so-called healing length $\xi$
\cite{pitaevskii2003bose,Rindler-Daller:2011afd},
\be
\eta = \frac{r_\perp}{\xi} , \;\;\;
\xi = \frac{\epsilon}{\sqrt{2\lambda\rho_0}} = \frac{\epsilon \sqrt{\pi}}{\sqrt{8\rho_0} R_0} .
\label{eq:healing-length}
\ee
The single vortex (\ref{eq:single-vortex-psi}) gives the velocity field
\be
\vec v = \frac{\epsilon \sigma}{r_\perp} \vec e_\varphi
= \epsilon \sigma \frac{\vec e_z \times \vec r_\perp}{r_\perp^2} ,
\;\;\; v_{r_\perp} = v_z = 0 , \;\; v_\varphi = \frac{\epsilon \sigma}{r_\perp}  ,
\label{eq:v-single-vortex}
\ee
while the density close to the center vanishes as
\be
r_\perp \to 0 : \;\; \rho \propto r_{\perp}^{2 | \sigma |} .
\label{eq:rho-core}
\ee
The vorticity reads
\be
\vec\omega = \vec \nabla \times \vec v = 2 \pi \epsilon \sigma \delta_D^{(2)}(\vec r_\perp) \vec e_z ,
\label{eq:vorticity-vortex}
\ee
while the circulation $\Gamma(r_\perp)$ along a circle ${\cal C}$ of radius $r_\perp$
around the vortex line is
\be
\Gamma(r_\perp) = \oint_{\cal C} \vec v \cdot \vec{d\ell} = \int_S \vec \omega \cdot \vec{dS}
= 2 \pi \epsilon \sigma .
\ee
Thus the vorticity and the circulation are quantized.
Vortices of higher spin $\sigma$ have higher energy \cite{Brax:2025uaw}, which is why in our numerical
simulations we only find vortex lines with $\sigma = \pm 1$,
in agreement with previous numerical and analytical works 
\cite{RINDLERDALLER2008,Hui:2020hbq,Nikolaieva:2021owc}.
This splitting instability of high-spin vortices into lower-spin vortices is also found in gauge theories \cite{Bogomolny:1976tp,Radu:2008pp,Nugaev:2014iva}
and Bose-Einstein condensates measured in laboratory experiments 
\cite{Skryabin2000,Kawaguchi2004,Lundh2006,Takeuchi2018,Giacomelli2020,VanAlphen2024}.
Moreover, as the vortices have the asymptotic behavior
$f(r_{\perp}) \propto r_{\perp}^{| \sigma |}$ near their center, higher-spin vortices correspond
to the cancellation of the coefficients of the Taylor expansion of the wave function near the
center up to higher order, which does not appear in random initial conditions \cite{Hui:2020hbq}.
Thus, both in our initial conditions and in the outer halo after the formation of a central soliton,
the interferences between many uncorrelated excited modes only produce vortices of spin
$\sigma = \pm 1$.

\subsection{Lattice of vortex lines}
\label{sec:lattice}

In this paper we consider systems with a macroscopic angular momentum and rotation.
Then, the associated macroscopic vorticity is supported by many vortex lines, $N_v \propto 1/\epsilon$,
as each vortex only carries a small vorticity quantum of the order of $\epsilon$ from
Eq.(\ref{eq:vorticity-vortex}). The direction of the initial angular momentum $\vec J$, which we
take aligned with the vertical $z$-axis, breaks the isotropy of the initial conditions and sets the
direction of the stable vortex lines that appear in the system.
Taking $J_z > 0$ for the initial angular momentum, we find in the numerical simulations that after
a relaxation stage where vortices of opposite signs annihilate we are left with a regular lattice of
vortices of spin $\sigma=1$ inside the soliton.
As shown in \cite{Brax:2025uaw}, a simple ansatz to describe such a system of $N_v$ vortices is
to consider wave functions of the form \cite{Fetter-1966,CRESWICK1980}
\be
\psi(\vec r,t) = \sqrt{\rho} e^{i s} \prod_{j=1}^{N_v} e^{i \sigma_j \varphi_j} ,
\label{eq:psi-ansatz-rho-s}
\ee
where we defined the angles
\be
\varphi_j(\vec r) = ( \widehat{ \vec e_x, \vec r_\perp -  } \vec r_{\perp j} ) .
\label{eq:varphi_j-def}
\ee
This is a direct generalization of Eq.(\ref{eq:single-vortex-psi}).
Whereas $\rho(\vec r,t)$ and $s(\vec r,t)$ are smooth functions that describe the regular part of the 
density and phase (e.g. a collective uniform motion), each angle $\varphi_j$ defined in 
Eq.(\ref{eq:varphi_j-def}) is singular on the vortex line, as it runs over $2\pi$ around infinitesimal
loops encircling the vortex line.
Here we take all vortex lines to be aligned with the vertical axis. Thus, they are defined by their location
$\vec r_{\perp j}$ in the $z=0$ plane and by their spin $\sigma_j$.
This is consistent with the results of our numerical simulations, which show a collection of vortices
inside the soliton that are roughly aligned with the vertical axis defined by the initial angular momentum
(up to perturbations due to finite effects and incomplete relaxation).
Then, we again recover a 2D system and as shown in \cite{Brax:2025uaw}, the vortices are simply
advected like the matter by the velocity field $\vec v$,
\be
\dot {\vec r}_i = \vec v(\vec r_i) , \;\;\; \vec v = \epsilon \vec\nabla s + \sum_{j=1}^{N_v} \vec v_j ,
\label{eq:dot-rj}
\ee
where $\vec v_j(\vec r)$ is the velocity field generated by the vortex line $j$,
as in Eq.(\ref{eq:v-single-vortex}),
\be
\vec v_j(\vec r) = \epsilon \sigma_j \vec e_z \times \frac{\vec r_\perp - \vec r_{\perp j}}
{ | \vec r_\perp - \vec r_{\perp j} |^2 } = \epsilon \sigma_j \vec e_z \times \vec\nabla \ln
| \vec r_\perp - \vec r_{\perp j} | .
\ee

Thus, as in classical hydrodynamics of ideal fluids, which obey Kelvin's circulation theorem 
\cite{Batchelor_2000}, the vortex lines follow the matter flow 
\cite{Fetter-1966,CRESWICK1980,LUND1991}.
Moreover, the density and velocity fields still obey the hydrodynamical equations
(\ref{eq:continuity})-(\ref{eq:Euler}), but the velocity field is no longer curl-free.
It includes a nonzero vorticity supported by the vortices,
\be
\vec\omega = \vec\nabla \times \vec v
= 2 \pi \epsilon \vec e_z \left( \sum_j \sigma_j \delta_D^{(2)}(\vec r_\perp - \vec r_{\perp j}) \right) .
\label{eq:vorticity}
\ee

\section{Continuum limit}
\label{sec:continuum}

In the limit $\epsilon \to 0$, each vortex line (\ref{eq:vorticity-vortex}) carries an infinitesimal vorticity,
so that a fixed macroscopic angular momentum $\vec J$ requires a number of vortices
that grows as $N_v \propto 1/\epsilon$. The width of these vortex tubes decreases as $\epsilon$
from Eq.(\ref{eq:healing-length}) whereas their nearest-neighbor distance $d_\perp$ decreases
more slowly as $\sqrt{\epsilon}$ from Eq.(\ref{eq:d-Omega-eps}) below.
Therefore, in the limit $\epsilon \to 0$ we obtain a continuum limit where the internal structure
of the vortex tubes and their discrete distribution are irrelevant and we can replace the discrete
vorticity distribution (\ref{eq:vorticity}) by a smooth and regular vorticity field.
In this limit, we no longer need to use a discrete ansatz such as
(\ref{eq:psi-ansatz-rho-s}) in terms of $N_v$ vortices. The system is again governed by the
hydrodynamical equations (\ref{eq:continuity})-(\ref{eq:Euler}) but the velocity field is no longer 
curl-free.
The number density of vortices can next be derived from the vorticity component 
$\vec\omega = \vec\nabla \times \vec v$ as in Eq.(\ref{eq:Nv-omega}) below.

\subsection{Rotating soliton profile}
\label{sec:rotating-soliton}

The hydrostatic equilibrium (\ref{eq:TF-static}), associated with the usual spherical soliton
(\ref{eq:rho-TF-0}), corresponds to a minimum of the energy
(\ref{eq:E-rho-v}) at fixed mass $M$.
The effective chemical potential $\mu$ in Eq.(\ref{eq:TF-static}) plays the role of the Lagrange 
multiplier associated with the constant mass constraint.
Because angular momentum is also conserved by the Gross-Pitaevskii equation of motion
(\ref{eq:Schrod-eps}) and we are considering rotating systems with a nonzero angular momentum,
we now look for minima of the energy $E$ at fixed mass $M$ and angular momentum $J_z$, as in the
2D case \cite{Brax:2025uaw}.
Introducing the Lagrange multipliers $\mu$ and $\Omega$, this minimization condition reads
\be
\delta^{(1)} \left( E - \mu M - \Omega J_z \right) = 0  ,
\label{eq:delta1-E-M-Jz}
\ee
where we take the first-order variation over $\delta\rho$ and $\delta\vec v$.
As we focus on the Thomas-Fermi regime where the quantum pressure is negligible,
$\epsilon \to 0$, we neglect the first term in the energy (\ref{eq:E-rho-v}).
Taking first the variations (\ref{eq:delta1-E-M-Jz}) with respect to $\delta v_x$, $\delta v_y$
and $\delta v_z$ gives
\be
v_x = - \Omega y, \;\; v_y = \Omega x, \;\; v_z = 0, \;\;
\mbox{hence} \;\; \vec v = \vec\Omega \times \vec r ,
\label{eq:v-solid-rotation}
\ee
with $\vec \Omega = \Omega \, \vec e_z$.
This is a solid-body rotation around the vertical $z$-axis, at rate $\dot\varphi = \Omega$.
Here $\varphi$ is the azimuthal angle in the spherical coordinates $(r,\theta,\varphi)$
and in the cylindrical coordinates $(r_\perp,\varphi,z)$, where $r_\perp = r \sin\theta=\sqrt{x^2+y^2}$
is the distance from the vertical axis.
Substituting this velocity field and taking the variation (\ref{eq:delta1-E-M-Jz}) with respect to
$\delta\rho$ gives
\be
\Phi_N + \Phi_I - \frac{\Omega^2 r_\perp^2}{2} = \mu .
\label{eq:mu-Omega}
\ee
This is the generalization of the equation of hydrostatic equilibrium (\ref{eq:TF-static})
to the rotating case.
From the Poisson equation and the expression of the self-interaction potential in (\ref{eq:Poisson-eps}),
the equation of equilibrium (\ref{eq:mu-Omega}) also reads
\be
\mbox{inside:} \;\;\; \frac{\lambda}{4\pi} \Delta\Phi_{N \rm in} + \Phi_{N \rm in}
= \mu + \frac{\Omega^2 r_\perp^2}{2} ,
\label{eq:PhiN-in}
\ee
which holds inside the soliton, whereas outside of the soliton where the density vanishes
(within the Thomas-Fermi approximation) we have
\be
\mbox{outside:} \;\;\; \Delta\Phi_{N \rm out} = 0.
\label{eq:PhiN-out}
\ee
This gives two second-order linear partial differential equations over the gravitational potential $\Phi_N$,
which are coupled by a matching condition at the surface of the soliton.
The difficulty is that we do not know a priori the shape of this surface.
In the 2D case \cite{Brax:2025uaw}, where by symmetry the soliton remains circular and its border
is defined by its radius $R_\Omega$, we obtain 1D linear differential equations with a trivial matching
and we can derive the exact expression of the radial profile of the rotating soliton.
In the 3D case this is no longer possible, because of the unknown shape of the soliton surface.

Therefore, as for the study of slowly rotating stars
\cite{Chandrasekhar_1933,Chandrasekhar_1962,Kovetz_1968,Chavanis:2002vt}, 
we use a perturbative computation in $\Omega^2$,
which we describe in detail in App.~\ref{app:rotating-soliton}.
We expand the gravitational potentials inside and outside of the soliton up to first order over $\Omega^2$,
as in Eqs.(\ref{eq:PhiN-in-1}) and (\ref{eq:PhiN-out-1}).
Here we take advantage of the fact that we know the general solutions of the linear differential
equations (\ref{eq:PhiN-in}) and (\ref{eq:PhiN-out}), where the homogeneous solutions can be expanded
in the eigenfunctions $j_\ell(\pi r/R_0) P_\ell(\cos\theta)$ and $r^{-\ell-1}P_\ell(\cos\theta)$, where
$j_\ell$ and $P_\ell$ are the spherical Bessel function and Legendre polynomial of order $\ell$.
We also expand the surface $R_{\Omega}(\theta)$ of the soliton in Legendre polynomials,
as in Eq.(\ref{eq:R-theta-1}), up to first order in $\Omega^2$.
Then, the surface $R_{\Omega}(\theta)$ is obtained from the inner solution $\rho_{\rm in}$
by the condition that $\rho_{\rm in}$ vanishes on this surface.
Next, the two conditions that the gravitational potential and its gradient match on the soliton surface
determine the multipole expansions of $\Phi_{N\rm in}$ and $\Phi_{N\rm out}$.
This gives
\ba
\rho_{\rm in} & = & \left( \rho_0 - \frac{\Omega^2}{2\pi} \right) j_0 \! \left( \! \frac{\pi r}{R_0} \! \right)
- \frac{5 \pi \Omega^2}{12} j_2 \! \left( \! \frac{\pi r}{R_0} \! \right) P_2(\cos\theta) \nonumber \\
&& + \frac{\Omega^2}{2\pi} ,
\label{eq:rho-in-res}
\ea
\ba
\Phi_{N\rm in} \!\! & \! = \! & \! -\lambda \left( \rho_0 \! - \! \frac{\Omega^2}{2\pi} \right)
j_0 \! \left( \! \frac{\pi r}{R_0} \! \right) \! + \! \frac{5 \pi \lambda \Omega^2}{12}
j_2 \! \left( \! \frac{\pi r}{R_0} \! \right) P_2(\cos\theta) \nonumber \\
&& + \frac{\Omega^2 r_\perp^2}{2} - \lambda \rho_0 + \frac{2-\pi^2}{4\pi} \lambda \Omega^2 ,
\label{eq:PhiN-in-res}
\ea
for the density and gravitational potential inside the soliton,
$\rho_{\rm out}=0$ and
\ba
\Phi_{N\rm out} & = & -\lambda \left( \rho_0 \! + \! \Omega^2 \frac{\pi^2-3}{6\pi} \right)
\frac{R_0}{r} + \Omega^2 \lambda  \frac{15 - \pi^2}{12 \pi} \nonumber \\
&& \times \frac{R_0^3}{r^3} P_2(\cos\theta)
\label{eq:PhiN-out-res}
\ea
outside the soliton, and
\be
R_{\Omega}(\theta) = R_0 \left( 1 + \frac{\Omega^2}{2\pi\rho_0} \right) - R_0 \frac{5\Omega^2}{4\pi\rho_0}
P_2(\cos\theta)
\label{eq:Rsol-res}
\ee
for the surface of the soliton.
Thus, at first order in $\Omega^2$ we only have monopole and quadrupole terms.
Higher multipoles would be generated at higher orders over $\Omega^2$.
The Lagrange multiplier $\mu$ is given by
\be
\mu = - \lambda \rho_0 \left( 1 + \frac{\pi^2-4}{4\pi} \frac{\Omega^2}{\rho_0} \right) .
\label{eq:mu-res}
\ee

From Eq.(\ref{eq:Rsol-res}) we obtain for the radius of the soliton along the vertical axis and in
the equatorial plane
\ba
&& \!\! R_z = R_{\Omega}(\theta=0) = R_{\Omega}(\theta=\pi)
= R_0 \left( 1 - \frac{3\Omega^2}{4\pi\rho_0} \right) ,  \hspace{0.7cm}
\label{eq:Rz-def} \\
&& \!\! R_{xy} = R_{\Omega}(\theta=\pi/2) = R_0 \left( 1 + \frac{9\Omega^2}{8\pi\rho_0} \right) .
\label{eq:Rxy-def}
\ea
As expected, the rotation flattens the soliton. At fixed central density, the size along the vertical axis
is decreased while the radius in the equatorial plane is increased.
Thus the soliton is oblate, with a deformation in the equatorial plane that is greater than that along
the vertical axis by a factor $3/2$.

The mass within a sphere of radius $r < R_z$ inside the soliton reads
\be
M(<r) = \frac{2 r^3 \Omega^2}{3} + 4 R_0^2 r^2 \left( \rho_0 - \frac{\Omega^2}{2\pi} \right)
j_1\left(\frac{\pi r}{R_0} \right) ,
\ee
while the angular momentum reads
\ba
&& \hspace{-0.3cm} J_z(\! < \! r) = \frac{2 r^3 \Omega}{45 \pi^2} \biggl [ 6 \pi^2 r^2 \Omega^2
\! + \! 5 \pi r R_0 [ 12 \pi \rho_0 \! - \! (6 \! + \! \pi^2) \Omega^2 ]  \hspace{0.3cm} \nonumber \\
&& \hspace{-0.3cm} \times j_1(\pi r \! / \! R_0) \! - \! 5 R_0^2   [ 24 \pi \rho_0 \! - \! (12 \! + \! 5 \pi^2)
\Omega^2 ] j_2(\pi r \! / \! R_0) \biggl ] . \hspace{0.3cm}
\label{eq:Jz-r-Om}
\ea

\subsection{Dynamical stability}
\label{sec:stability}

The equilibrium solution (\ref{eq:mu-Omega}) is dynamically stable if it is a true minimum of the
energy (as opposed to a maximum or a saddle-point).
The second variation of the energy (\ref{eq:E-rho-v}), with respect to the density and velocity fields,
reads
\ba
&& \hspace{-0.5cm}  \delta^{(2)}E \! = \! \int \! d\vec r \left[ v_\varphi \delta\rho \delta v_\varphi \! + \!
 \frac{1}{2} \rho (\delta \vec v)^{\,2} \! + \! \frac{1}{2} \delta \rho \delta \Phi_N
 \! + \! \frac{\lambda}{2} (\delta\rho)^2 \right]   \nonumber \\
&& = \Omega \delta J_z^{(2)} + \frac{1}{2} \int d\vec r \left[ \rho (\delta \vec v)^{\,2}
+ \delta \rho \delta \Phi_N +\lambda (\delta\rho)^2 \right] .
\label{eq:d2E-1}
\ea
The first term in the second line vanishes as we consider variations at fixed angular momentum.
Then, the necessary and sufficient condition for dynamical stability is
\be
\int d\vec r \left[ \delta \rho \delta \Phi_N +\lambda (\delta\rho)^2 \right] > 0 ,
\ee
for all nonzero $\delta\rho(\vec r)$.
This means that the eigenvalues $\nu$ of the operator
$K \cdot \delta\rho = ( 4\pi \Delta^{-1} + \lambda ) \delta\rho$ are strictly positive,
that is, the eigenvalues $\beta$ of the eigenvalue problem
\be
\Delta \delta\rho + \beta \delta\rho = 0 , \;\;\; \beta = 4\pi/(\lambda-\nu) ,
\ee
are greater than $4\pi/\lambda$.
The eigenvectors are
\be
\delta\rho_{n\ell m}(\vec r) = j_\ell(\sqrt{\beta_{n,\ell}} r ) Y_\ell^m(\theta,\varphi) .
\ee
Let us first consider the modes $\ell \geq 1$, which automatically conserve the mass,
$\delta M=0$. We consider perturbations that vanish outside the soliton.
Therefore, we can take $\delta\rho(\vec r)=0$ for $r>R_{xy}$, as the soliton is oblate, that is,
we can impose the boundary condition $\delta\rho=0$ on the sphere $r=R_{xy}$.
Then, the eigenvalues are $\beta_{n,\ell} = (x_{n,\ell}/R_{xy})^2$, where $x_{n,\ell}$ is the
$n^{\rm th}$ zero of $j_\ell(x)$.
The stability condition, $\beta > 4\pi/\lambda$, means that all these modes are dynamically stable
if we have
\be
R_{xy} < x_{1,1} \sqrt{\frac{\lambda}{4\pi}} , \;\; \mbox{i.e.} \; R_{xy} < \frac{x_{1,1}}{\pi} R_0 .
\label{eq:Rxy-stable}
\ee
Let us now consider the modes $\ell=0$. As we obtain
$\delta M(<r) \propto r^2 j_1(\sqrt{\beta_{n,0}}r)$ they no longer automatically conserve the mass.
Therefore, we now impose the boundary condition $\delta M(<R_{xy})=0$, that is,
$j_1(\sqrt{\beta_{n,0}} R_{xy})=0$.
This gives the eigenvalues $\beta_{n,0} = (x_{n,1}/R_{xy})^2$ and we recover the stability condition
(\ref{eq:Rxy-stable}).
From Eqs.(\ref{eq:rho-TF-0}) and (\ref{eq:Rxy-def}), we find at leading order over $\Omega^2$
the dynamical stability criterion
\be
\mbox{stable for} \;\; \Omega^2 \lesssim \frac{8 (x_{1,1}-\pi)}{9} \rho_0 , \;\; {\rm i.e.} \;
| \Omega | \lesssim 1.1 \sqrt{\rho_0} .
\label{eq:Omega-stable}
\ee
This is not strictly a necessary condition. Indeed, even though the mode $\delta\rho_{1,1,1}(\vec r)$,
which saturates the condition (\ref{eq:Rxy-stable}), automatically verifies the conservation of mass
and angular momentum, and as such is allowed by both mass and angular momentum constraints,
it corresponds to a perturbation that is nonzero in a small region outside the soliton, in the domain
$R_\Omega(\theta) < r < R_{xy}$.
To derive a strictly necessary condition we should look for unstable modes with a support that is fully
inside the soliton boundary $R_\Omega(\theta)$.
However, we can expect the necessary condition to be of the same order.

In the 2D case, all rotating solitons (i.e. with a well-defined solution of the equation of equilibrium)
are dynamically stable \cite{Brax:2025uaw}.
The upper bound on the rotation rate is thus determined by the condition to have an equilibrium profile
of finite mass, i.e. the density must vanish at some finite radius.
This again gives a scaling $\Omega_{\rm max} \sim \sqrt{\rho_0}$.
In our 3D case, at leading order in $\Omega^2$ the condition of a well-defined density profile
reads $R_z>0$. From Eq.(\ref{eq:Rxy-def}) we obtain
\be
R_z > 0 : \;\; \Omega^2 \lesssim \frac{4\pi}{3}\rho_0 , \;\; {\rm i.e.} \;| \Omega | \lesssim 2 \sqrt{\rho_0} .
\label{eq:Rz-0}
\ee
Therefore, the maximum allowed and stable rotation rate would be between the two conditions
(\ref{eq:Omega-stable})-(\ref{eq:Rz-0}).
Because these values are only obtained from a leading-order analysis, we can only estimate the
order of magnitude $\Omega_{\rm max} \sim \sqrt{\rho_0}$, within a factor two or so.

The bound $\Omega_{\rm max} \lesssim \sqrt{\rho_0}$ corresponds to the condition that
in the equation of equilibrium Eq.(\ref{eq:mu-Omega}), or in the energy (\ref{eq:E-rho-v}),
the rotational energy, $\Omega^2 r_\perp^2/2 \sim \Omega^2 R_0^2 \sim \lambda\Omega^2$,
is at most of the order of the self-interaction energy $\Phi_I \sim \lambda \rho_0$
(and whence of the order of the gravitational energy).
The rotating solitons (\ref{eq:mu-Omega}) considered in this paper are quite different
from the rotating boson stars studied in \cite{Dmitriev:2021utv}.
The latter correspond to a single vortex with a possibly large spin $\sigma$ and have a toroidal
shape, with a vanishing density at the center due to the centrifugal barrier as in
Eq.(\ref{eq:single-vortex-psi}).
Instead, in our case the rotating soliton or boson star is made of a lattice of many vortices
of unit spin and it shows an oblate spheroidal shape with a maximum density at the center.

\subsection{Uniform density of vortex lines}
\label{sec:uniform-lattice}

Within the equatorial plane $z=0$ and inside the soliton, the solid-body rotation
(\ref{eq:v-solid-rotation}) gives for the circulation
$\Gamma(r_\perp)$ along the circle of radius $r_\perp$
\be
\Gamma(r_\perp) = \oint_{\cal C} \vec v \cdot \vec{d\ell} = 2\pi r_\perp^2 \Omega .
\ee
As each vortex line (\ref{eq:vorticity-vortex}) carries a vorticity quantum
$2\pi\epsilon\sigma$, the circulation also reads
\be
\Gamma(r_\perp) = \int_S \vec \omega \cdot \vec{dS} = 2 \pi \epsilon N_v(<r_\perp) ,
\label{eq:Nv-omega}
\ee
where $N_v(<r_\perp)$ is the total number of vortex lines within radius $r_\perp$
that cross the equatorial plane, weighted by their spin,
$N_v = \sum_\sigma \sigma N_{v,\sigma}$.
This gives
\be
N_v(<r_\perp) = \frac{r_\perp^2\Omega}{\epsilon} , \;\;\; n_v(\vec r_\perp) = \frac{\Omega}{\pi \epsilon} ,
\label{eq:vortex-density}
\ee
hence a constant number density $n_v(\vec r_\perp)$ of vortex lines running through the equatorial plane.
Thus, as pointed out by Feynman \cite{FEYNMAN1955} for rotating superfluid $^4{\rm He}$,
a uniform lattice of vortex lines develops to mimic a solid-body rotation \cite{Fetter-2009}.
We will check that this agrees with our numerical simulations, as in Figs.~\ref{fig:2D-rho-mu1-0p01}
and \ref{fig:3D-vortices-mu1-0p01} below.

In our case, this uniform density of vortex lines is due to the solid-body rotation (\ref{eq:v-solid-rotation}),
which arises as a minimum of the energy at fixed mass and angular momentum.
This, in turn, ensures its dynamical stability. This equilibrium is thus the rotating counterpart,
in systems with a non-vanishing angular momentum, 
of the usual static soliton profile (\ref{eq:rho-TF-0}) associated with halos without global rotation.

Since $\nabla \cdot \vec v=0$, the velocity field (\ref{eq:v-solid-rotation}) inside the soliton
is entirely due to the vortex lines. The smooth phase component $s$ in Eq.(\ref{eq:psi-ansatz-rho-s})
is uniform, such as $s=-\mu t/\epsilon$ with a constant $\mu$, and $\vec\nabla s = 0$.

The number density (\ref{eq:vortex-density}) gives for the typical distance $d$
in the equatorial plane between vortex lines
\be
d_\perp = \sqrt{\pi\epsilon/\Omega} .
\label{eq:d-Omega-eps}
\ee
It does not depend on the soliton mass and decreases as $\sqrt{\epsilon}$ in the semi-classical limit
$\epsilon\to 0$, that is, much more slowly than the healing length introduced in
Eq.(\ref{eq:healing-length}), which decreases as $\epsilon$. Therefore, the assumption of
well-separated vortex lines and the neglect of their internal structure, as in
Sec.~\ref{sec:lattice}, is justified in the Thomas-Fermi regime that we consider
in this paper.

\section{Simulations setup}
\label{sec:numerical-setup}

\subsection{Expansion over eigenfunctions}
\label{sec:eigenfunctions}

As in \cite{Garcia:2023abs,GalazoGarcia:2024fzq}, because we are interested in the
formation of solitons and their generic features we start our simulations with stochastic
initial conditions, which correspond to a collisionless virialized halo in the
semiclassical limit without soliton.
As in \cite{Garcia:2023abs}, we choose for the target initial density profile
\be
0 \leq r \leq R : \;\;\; \rho(r) = \rho_{0} j_0(\pi r/R)  ,
\label{eq:rho-class}
\ee
where $R$ is the halo radius, which is greater than the expected soliton radius.
Although this profile takes the same form as the hydrostatic soliton (\ref{eq:rho-TF-0}),
it is not related to the soliton and merely a simple model for a flat-core halo.
The total mass reads
\be
M = (4/\pi) \rho_0 R^3 ,
\label{eq:M-halo}
\ee
and the gravitational potential reads
\be
r \leq R: \;\; \Phi_N(r) = \Phi_{N0} j_0(\pi r/R)  , \;\;\; \Phi_{N0} = - (4/\pi) \rho_0 R^2 ,
\label{eq:PhiN-class}
\ee
which we normalize to zero at the radius $R$ of the halo.
Then, we take for the initial wave function \citep{Lin:2018whl,Yavetz:2021pbc,Garcia:2023abs}
a sum with random coefficients $a_{n\ell m}$ over the eigenmodes
$\hat \psi_{n\ell m}(\vec r)$ of the time-independent Schr\"odinger equation defined by this target
gravitational potential $\Phi_{N}$,
\be
\psi(\vec r) = \sum_{n\ell m} a_{n\ell m} \hat\psi_{n \ell m}(\vec r) , \;\;
\hat\psi_{n \ell m}(\vec r) = {\cal R}_{n \ell}(r) Y_{\ell}^m(\theta,\varphi) ,
\label{eq:psi-halo-a_nlm}
\ee
where we use spherical coordinates and the radial parts satisfy the radial Schr\"odinger equation
\be
\left[ - \frac{\epsilon^2}{2} \frac{1}{r^2} \frac{d}{dr} \left( r^2 \frac{d}{dr} \right)
+ \frac{\epsilon^2 \ell (\ell+1)}{2 r^2} + \Phi_N \right] {\cal R}_{n \ell}
= E_{n \ell} {\cal R}_{n \ell}
\label{eq:Rnl-def}
\ee
and form an orthonormal basis
\be
\int_0^{\infty} dr r^2 \, {\cal R}_{n_1 \ell} {\cal R}_{n_2 \ell} = \delta_{n_1,n_2} .
\label{eq:norm-R}
\ee
If the coefficients $a_{n\ell m}$ are such that the associated density profile is close to the target
(\ref{eq:rho-class}), the wave function (\ref{eq:psi-halo-a_nlm}) would be an equilibrium configuration
of the Schr\"odinger equation (\ref{eq:Schrod-eps}) without the self-interaction $\Phi_I$,
each mode evolving with a factor $e^{-i E_{n\ell}t/\epsilon}$.
This would be the scalar-field analog of the virial equilibrium for a system of collisionless particles.
In practice, the high density at the center of the halo generates significant self-interactions that will
lead to the formation of a central soliton, inside of such a virialized envelope made of a
superposition of many excited modes.

We take the coefficients $a_{n\ell m}$ to have deterministic amplitudes but random phases $\Theta_{n\ell m}$
that are uncorrelated with a uniform distribution over $0 \leq \Theta < 2\pi$,
\ba
&& a_{n\ell m} = | a_{n\ell m} | e^{i\Theta_{n\ell m}} , \nonumber \\
&& \langle a_{n_1\ell_1 m_1}^{\star} a_{n_2\ell_2 m_2} \rangle
= | a_{n_1\ell_1 m_1} |^2 \delta_{n_1,n_2} \delta_{\ell_1,\ell_2} \delta_{m_1,m_2} , \hspace{1cm}
\label{eq:a-nlm-Theta}
\ea
where the statistical average $\langle\dots\rangle$ is taken over the random phases
$\Theta_{n\ell m}$.
The mean density is then
\be
\langle \rho \rangle = \langle |\psi|^2 \rangle = \sum_{n\ell m} | a_{n\ell m} |^2
{\cal R}_{n \ell}^2 | Y_{\ell}^m |^2 .
\label{eq:rho-psi-squared}
\ee
The current associated with the wave function $\psi$ reads
\be
\vec j = \rho \vec v = \frac{i\epsilon}{2} ( \psi \vec\nabla \psi^\star - \psi^\star \vec\nabla \psi ) .
\ee
Substituting the expansion (\ref{eq:psi-halo-a_nlm}) we obtain
\be
\langle j_r \rangle = \langle \rho v_r \rangle = 0 , \;\;\;
\langle j_\theta \rangle = \langle \rho v_\theta \rangle = 0 ,
\ee
and
\be
\langle j_\varphi \rangle = \langle \rho v_\varphi \rangle = \sum_{n\ell m} \frac{\epsilon m}{r \sin\theta}
| a_{n\ell m} |^2 {\cal R}_{n\ell}^2 | Y_{\ell}^m |^2 .
\label{eq:j-phi-psi-squared}
\ee
Whereas the density (\ref{eq:rho-psi-squared}) depends on the even component over $m$
of $| a_{n\ell m} |^2$, the azimuthal current (\ref{eq:j-phi-psi-squared}) depends on the odd component.

As described in App.~\ref{app:initial-conditions}, using the WKB approximation and in the continuum limit
we can relate the initial condition (\ref{eq:psi-halo-a_nlm}) to the phase-space distribution $f(E,L_z)$
of a classical system of collisionless particles.
Choosing for the amplitude of the expansion coefficients $a_{n\ell m}$ the values
\be
| a_{n\ell m} |^2 = (2\pi\epsilon)^3 f(E_{n\ell},L_z) , \;\;\; L_z = \epsilon m ,
\label{eq:a_nellm-f_E-Lz}
\ee
we recover the density profile $\rho(r)$ (\ref{eq:rho-f0})
and the total angular momentum $J_z$ (\ref{eq:Jz-fm})
of the classical system defined by the phase-space distribution $f(E,L_z)$, which we take of the form
\be
f(E,L_z) = f_0(E) + f_-(E,L_z) ,
\label{eq:f-f0-fm}
\ee
with
\be
f_-(E,-L_z) = - f_-(E,L_z) ,
\label{eq:fm-Lz-def}
\ee
that is, the symmetric part over $L_z$ only depends on $E$.
This means that the density profile is spherically symmetric,
\be
\langle \rho(r) \rangle = 4\pi \int_{\Phi_N(r)}^{\infty} dE \sqrt{2 (E-\Phi_N(r))} f_0(E) ,
\label{eq:rho-f0-av}
\ee
while the total angular momentum along the vertical axis, defined as
\be
J_z = \int d\vec r \; r \sin\theta \rho v_\varphi ,
\label{eq:Jz-def}
\ee
reads
\ba
\langle J_z \rangle & = & 16 \pi^2 \int_0^R dr \, r \int_{\Phi_N}^{\infty} dE
\int_0^{r \sqrt{2 (E-\Phi_N)}} dL_z \, L_z  \nonumber \\
&& \times {\rm Arccos}\left( \frac{L_z}{r \sqrt{2 (E-\Phi_N)}} \right) f_-(E,L_z)  .
\label{eq:Jz-fm-av}
\ea

\subsection{Isotropic initial conditions}
\label{sec:isotropic}

For isotropic initial conditions the classical phase-space distribution only depends
on the energy $E$,
\be
f(E,L_z) = f_0(E) .
\ee
Then, the total angular momentum vanishes, $J_z=0$, while the density is given
by Eq.(\ref{eq:rho-f0}), which can be inverted by Eddington's formula \cite{Binney2008}
\be
f_0(E) = \frac{1}{2\sqrt{2} \pi^2} \frac{d}{dE} \int_E^0 \frac{d\Phi_N}{\sqrt{\Phi_N-E}}
\frac{d\rho}{d\Phi_N} .
\label{eq:Eddington}
\ee
For the density and potential profiles (\ref{eq:rho-class}) and (\ref{eq:PhiN-class}) this gives
\be
- \frac{4 \rho_0 R^2}{\pi} < E < 0 : \;\;\;  f_0(E) = \frac{1}{8\pi R^2 \sqrt{-2E }} .
\label{eq:f0-isotropic}
\ee
Here we normalize the gravitational potential as $\Phi_N(R)=0$, as in
Eq.(\ref{eq:PhiN-class}), so that the halo is made of particles with negative energy.
This provides the coefficients $a_{n\ell m}$ that define our initial conditions
through Eqs.(\ref{eq:a_nellm-f_E-Lz}) and (\ref{eq:a-nlm-Theta}).
As seen in Eq.(\ref{eq:rho-f0-av}), this gives a halo that is statistically
spherically symmetric, with the average density profile (\ref{eq:rho-class}).

\subsection{Anisotropic initial conditions}
\label{sec:anisotropic}

For anisotropic initial conditions of the form (\ref{eq:f-f0-fm}), with a nonzero angular momentum
$J_z$, we take
\be
f(E,L_z) = f_0(E) [ 1 + \alpha \, {\rm sign}(L_z) ] , \;\;\; -1 \leq \alpha \leq 1 ,
\label{eq:f-ELz-alpha}
\ee
where $f_0(E)$ is the isotropic distribution (\ref{eq:f0-isotropic}), that is, the odd component
in the decomposition (\ref{eq:f-f0-fm}) reads
\be
f_-(E,L_z) = \alpha f_0(E) {\rm sign}(L_z) .
\ee
This recovers the same spherically symmetric target density (\ref{eq:rho-class}),
but with a different proportion of particles with positive and negative vertical angular momentum.
In other words, we keep the same classical orbits as in the isotropic case, whence the same
density profile, but modify the direction of rotation of some of the particles, or in the wave function
framework (\ref{eq:psi-halo-a_nlm}) the relative weights of eigenmodes with opposite values of
the azimuthal orbital number $m$.
If $\alpha =\pm 1$ we only keep the particles that have a positive/negative vertical angular momentum.
From Eq.(\ref{eq:Jz-fm}) we obtain for the total angular momentum of the initial halo
\be
J_{z,\rm init} \simeq 0.174 \alpha M^{3/2} R^{1/2} .
\label{eq:Jz-init}
\ee

\begin{figure}
\centering
\includegraphics[height=4cm,width=0.3\textwidth]{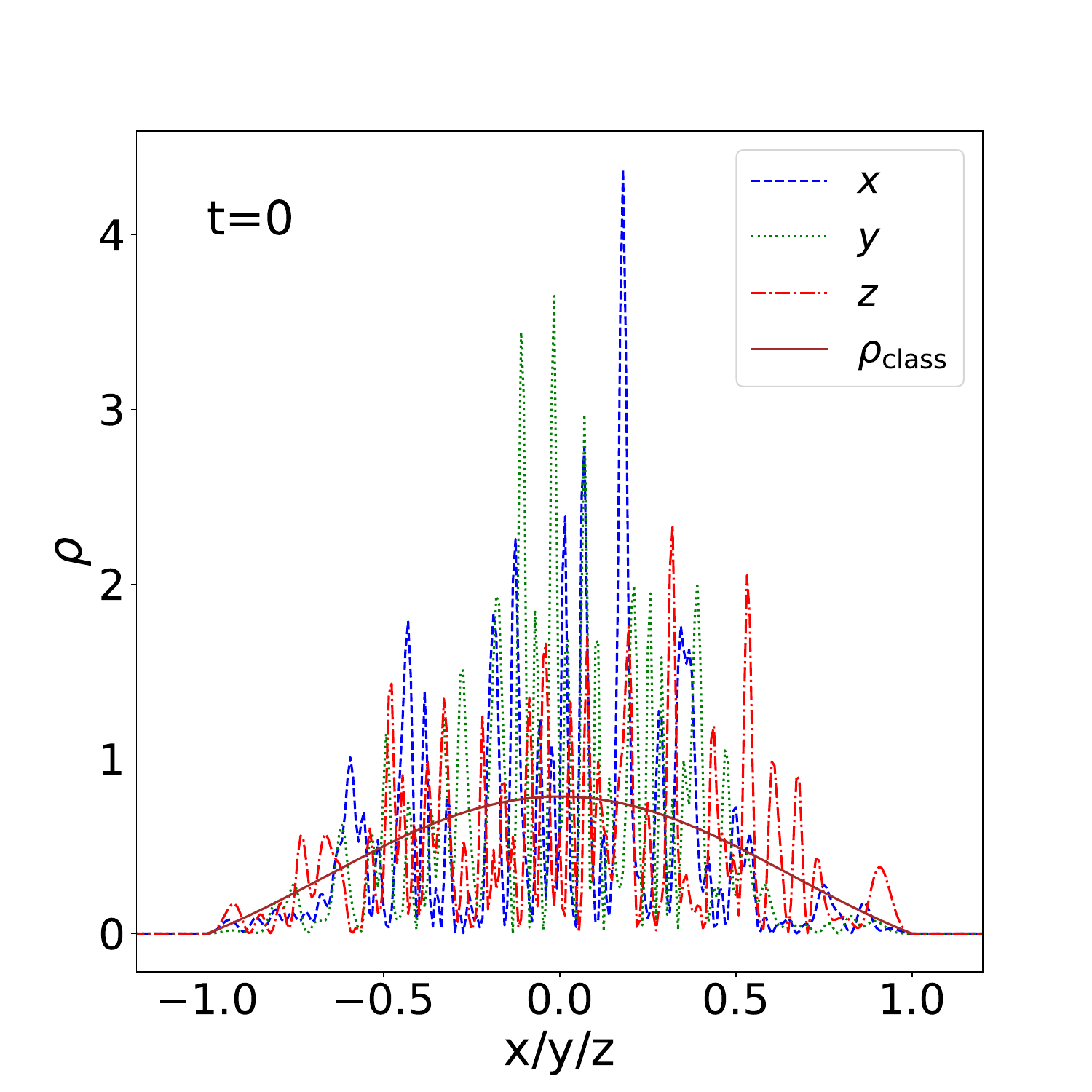}\\
\includegraphics[height=3.6cm,width=0.235\textwidth]{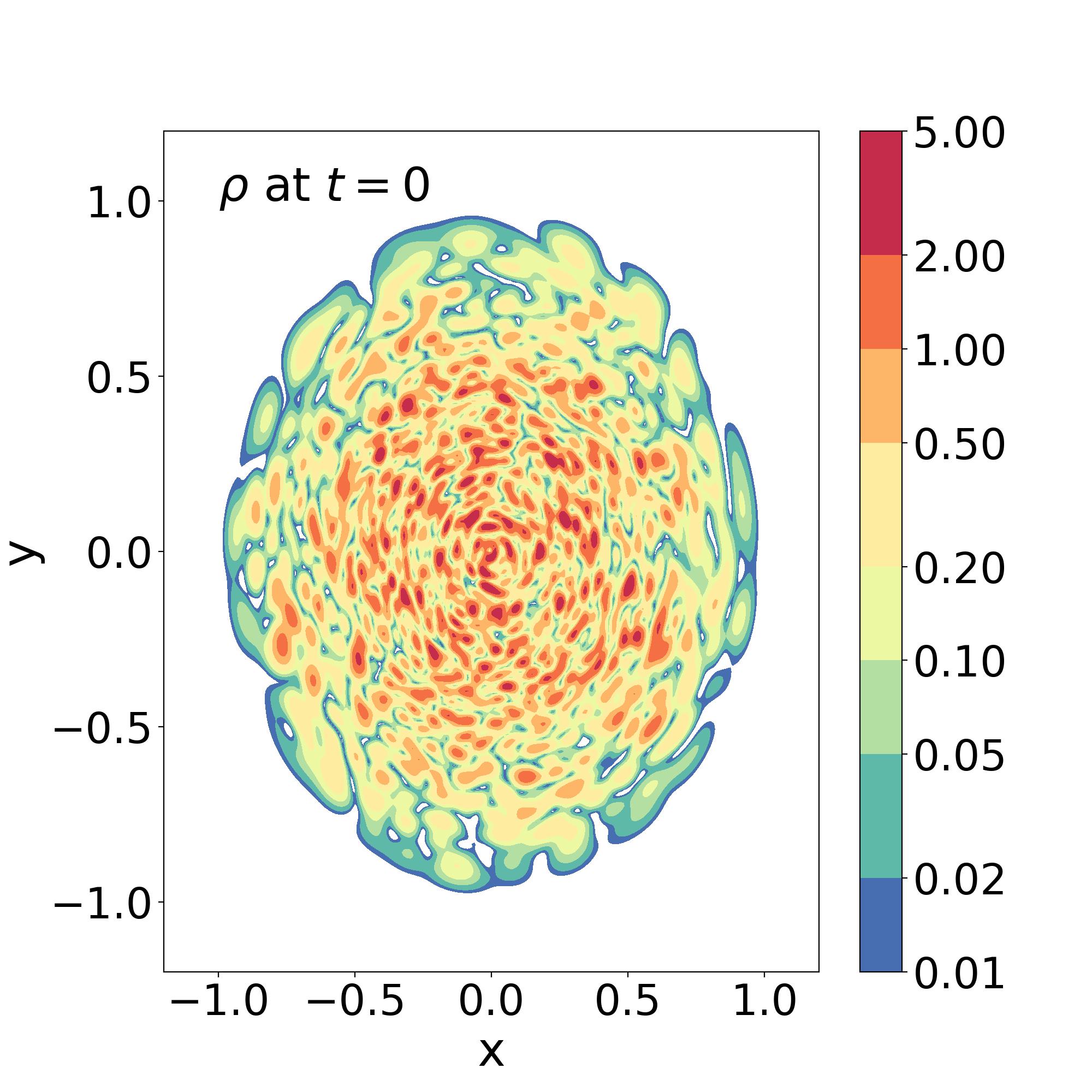}
\includegraphics[height=3.6cm,width=0.235\textwidth]{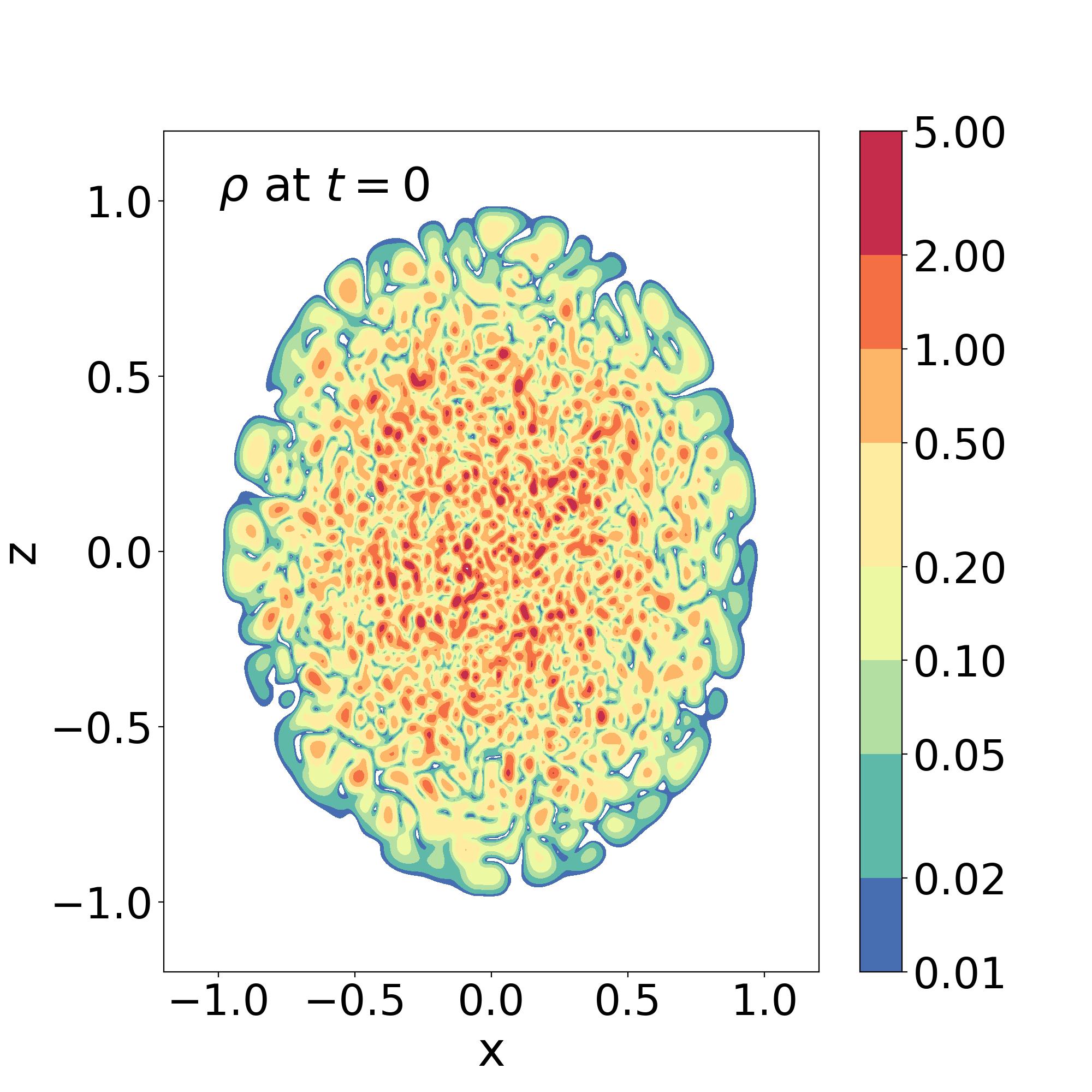}\\
\includegraphics[height=3.6cm,width=0.235\textwidth]{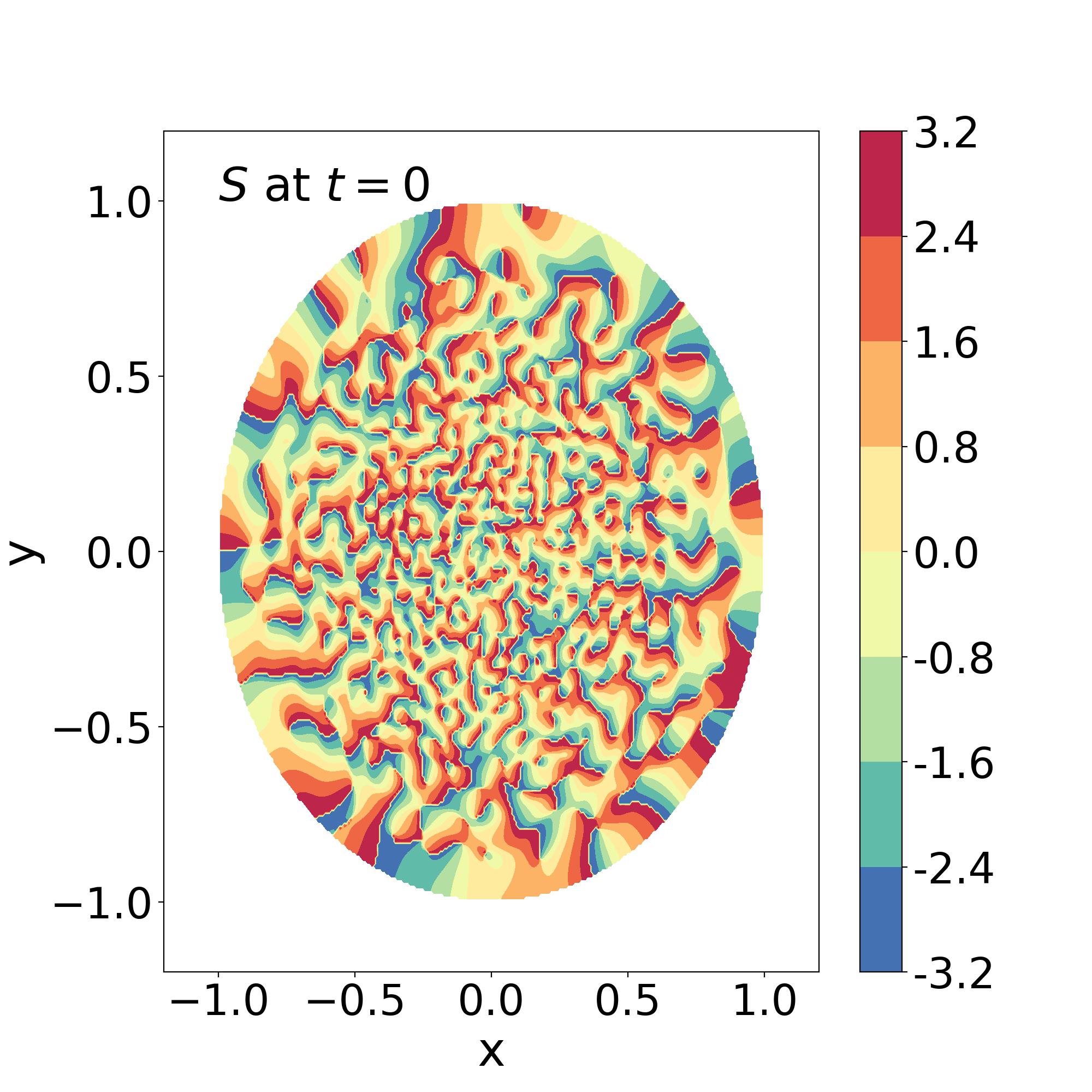}
\includegraphics[height=3.6cm,width=0.235\textwidth]{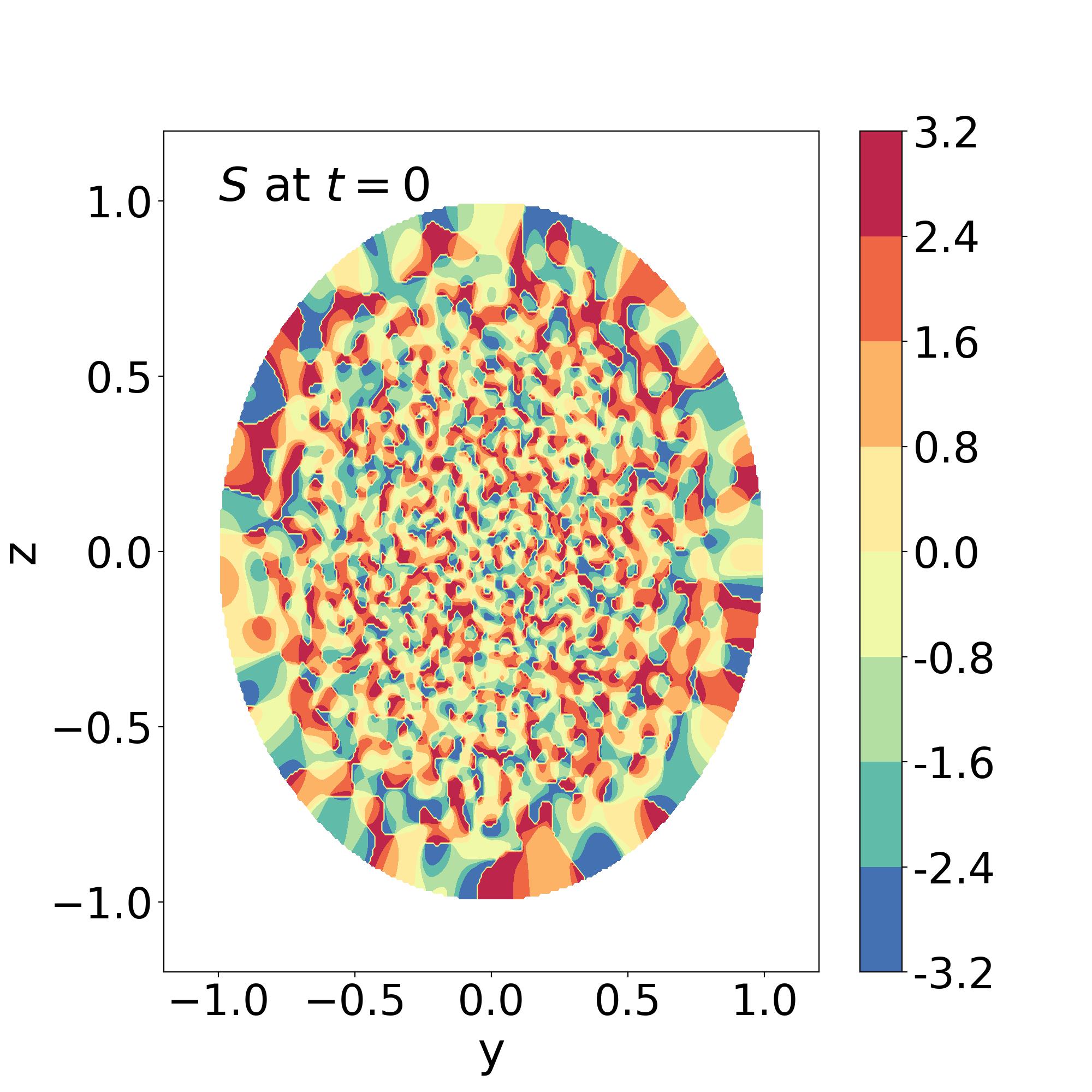}
\caption{
Initial condition of our simulation for the case $[\epsilon=0.01, \alpha=1]$.
{\it Upper panel:} density profiles along the $x$, $y$ and $z$ axes (dashed, dotted and dash-dotted
lines with large fluctuations) and classical density profile $\rho_{\rm class}$ (smooth brown line)
of Eq.(\ref{eq:rho-class}).
{\it Middle row:} 2D maps of the density $\rho$ in the planes $(x,y)$ and $(x,z)$ that run
through the center of the system.
{\it Bottom row:} 2D maps of the phase $S$, defined over $]-\pi,\pi]$, in the planes $(x,y)$ and $(x,z)$
}
\label{fig:initial}
\end{figure}

\subsection{Simulation parameters}
\label{sec:parameters}

As in our 2D paper \cite{Brax:2025uaw} we focus on the semi-classical regime and
we take as our reference the case studied in Sec.~\ref{sec:eps-0p01-alpha-1} below
with the values $\epsilon=0.01$ and $\alpha=1$.
We also take $R=1$ and $M=1$ for the initial target halo radius and mass
in Eqs.(\ref{eq:rho-class})-(\ref{eq:M-halo}).
We shall also consider the cases $\epsilon=0.005$ in Sec.~\ref{sec:eps-0p005},
to study the dependence on $\epsilon$, and $\alpha=0.5$ and $0$ in Sec.~\ref{sec:alpha-0p5-0},
to study the dependence on $\alpha$.

As explained in Sec.~\ref{sec:eigenfunctions}, with these values of $R$ and $M$, 
each choice of parameters $(\epsilon, \alpha)$ defines a classical phase-space distribution
(\ref{eq:f-ELz-alpha}). This determines the amplitude of the coefficients $a_{n\ell m}$
as in Eq.(\ref{eq:a_nellm-f_E-Lz}). For a set of random phases $\Theta_{n\ell m}$
in Eq.(\ref{eq:a-nlm-Theta}) this provides our initial condition given by the expansion 
(\ref{eq:psi-halo-a_nlm}). In this expression, we only use the WKB approximation to derive the
coefficients $a_{n\ell m}$. For the eigenfunctions $\hat\psi_{n \ell m}(\vec r)$
we numerically solve the Schr\"odinger equation (\ref{eq:Rnl-def}).
This provides an eigenfunction basis that is smooth and well-behaved and does not suffer
from the inaccuracies of the WKB approximation near the turning points.

We show in Fig.~\ref{fig:initial} our initial condition for the case $\epsilon=0.01$
and $\alpha=1$.
As in \cite{Garcia:2023abs,GalazoGarcia:2024fzq}, where we studied 3D isotropic systems,
the initial density field shows strong fluctuations of order unity around the target
classical density (\ref{eq:rho-class}), in agreement with the theoretical variance
$\langle ( \rho - \langle \rho \rangle )^2 \rangle = \langle \rho \rangle^2$.
The de Broglie wavelength $\lambda_{\rm dB} \sim \epsilon$ sets the spatial width of the fluctuations.

We obtain the phase $S$ (which we define in the interval $]-\pi,\pi]$) from the wave function $\psi$
as in Eq.(\ref{eq:Madelung}).
It again shows strong fluctuations, on the same scale as the density.
The interferences between the many modes in the sum (\ref{eq:psi-halo-a_nlm}) give rise to many
points inside the halo where the amplitude $|\psi|$ vanishes and the phase $S$ is ill-defined.
They typically correspond to vortex lines of spin $\sigma= \pm 1$ \cite{Hui:2020hbq},
where the phase is singular as it rotates by a multiple of $2\pi$ in a small circle around
the vortex line.

The 2D density maps in the planes $(x,y)$ and $(x,z)$ are rather similar, as the averaged initial
density $\langle \rho(\vec r) \rangle$ in Eq.(\ref{eq:rho-f0-av}) is spherically symmetric.
However, we can see traces of the initial rotation around the vertical axis.
The high-density spots (red regions in the figure) appear slightly less disordered in the
$(x,y)$ plane than in the $(x,z)$ plane, as they seem to be dominantly oriented and elongated
along a circular pattern around the center.
The initial rotation can be more clearly seen in the phase maps. They look quite
similar at first glance, with strong fluctuations on the de Broglie scale. However,
running counterclockwise along the border of the halo, on the circle $r=1$,
whereas in the $(x,z)$ plane we encounter about as many jumps of the phase from $\pi$ to
$-\pi$ (red to blue) as jumps from $-\pi$ to $\pi$ (blue to red),
in the $(x,y)$ plane we only encounter jumps from $\pi$ to $-\pi$ (red to blue).
This is due to our initial condition (\ref{eq:f-ELz-alpha}) with $\alpha=1$, where we only
keep modes with positive angular momentum $L_z>0$. This gives a positive total phase shift $\Delta S$
over the circle $r_\perp = 1$ in the equatorial plane, as each jump from $\pi$ to $-\pi$
corresponds to an increment of the winding number by $1$ (e.g., starting from a phase $S$
in the interval $]-\pi,\pi]$, a jump from $\pi$ to $-\pi$ actually corresponds to the entry
into the higher-phase sheet defined over $]\pi,3\pi]$).

As in the 2D case \cite{Brax:2025uaw}, the fast fluctuations of the phase (associated with large
gradients for the local velocity) imply that the hydrodynamical picture is meaningless.
Instead, the system mimics a gas of collisionless particles, as seen from the construction
presented in Sec.~\ref{sec:eigenfunctions}.
After the formation of a central soliton, this will still apply to the outer halo where the quartic 
self-interactions remain negligible and the dynamics must be described by the Vlasov equation 
rather than by hydrodynamical equations
\cite{Widrow:1993qq,Mocz:2018ium,GalazoGarcia:2022nqv,Garcia:2023abs,Liu:2024pjg}.
In the central region, the self-interactions will lead to a soliton that can instead by described
by hydrodynamics and the equilibrium solution (\ref{eq:v-solid-rotation})-(\ref{eq:mu-Omega}).

For the self-interaction coupling constant $\lambda$ in Eq.(\ref{eq:Poisson-eps})
we take the value associated with a static soliton radius $R_0=0.5$ in Eq.(\ref{eq:rho-TF-0}),
$\lambda=4 R_0^2/\pi$.
Thus, within a few dynamical times the halo typically collapses to form a soliton, with a radius
that is one-half of the initial halo, embedded within a remaining virialized envelope made of many
excited states as in (\ref{eq:psi-halo-a_nlm}).

\begin{figure*}
\centering
\includegraphics[height=5cm,width=0.3\textwidth]{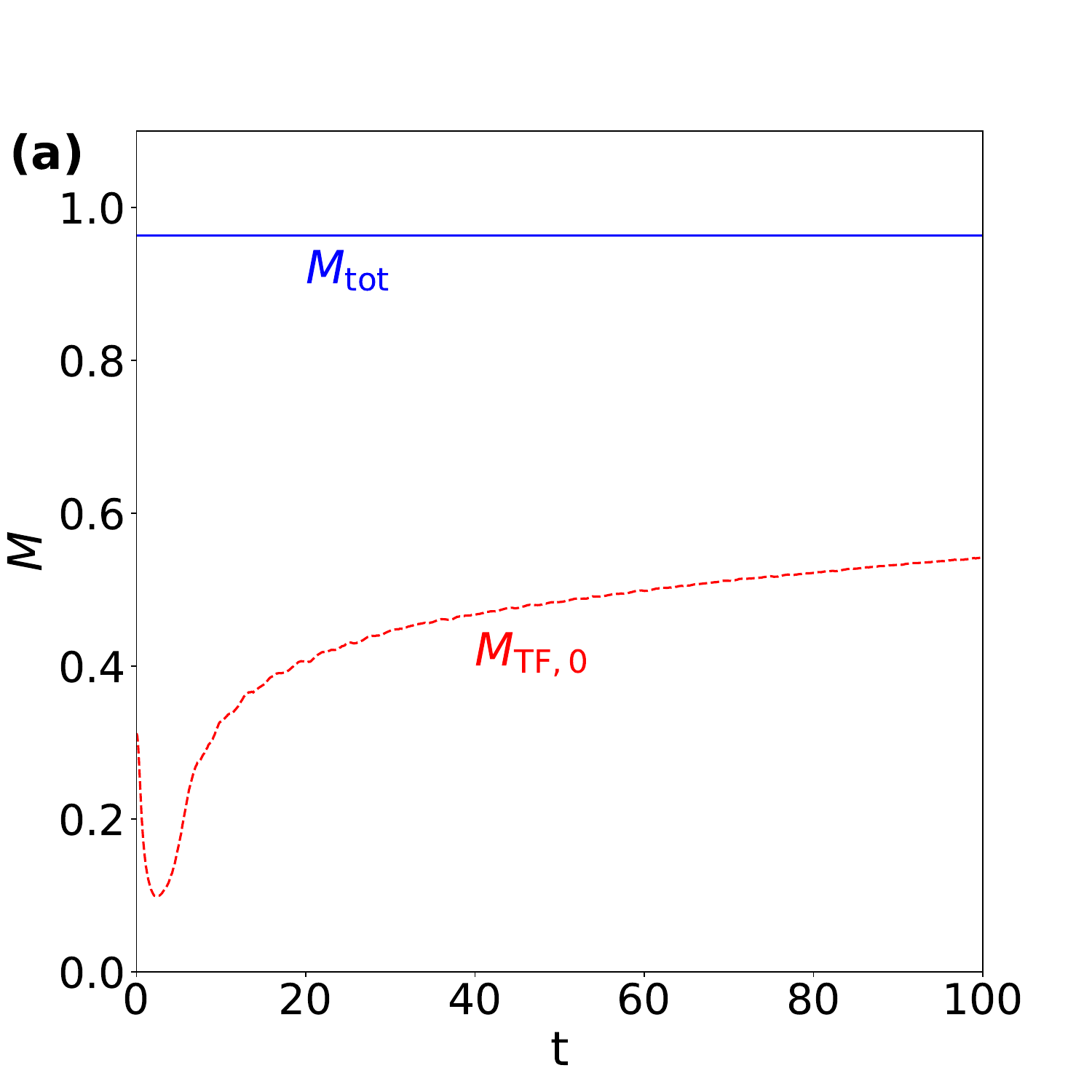}
\includegraphics[height=5cm,width=0.34\textwidth]{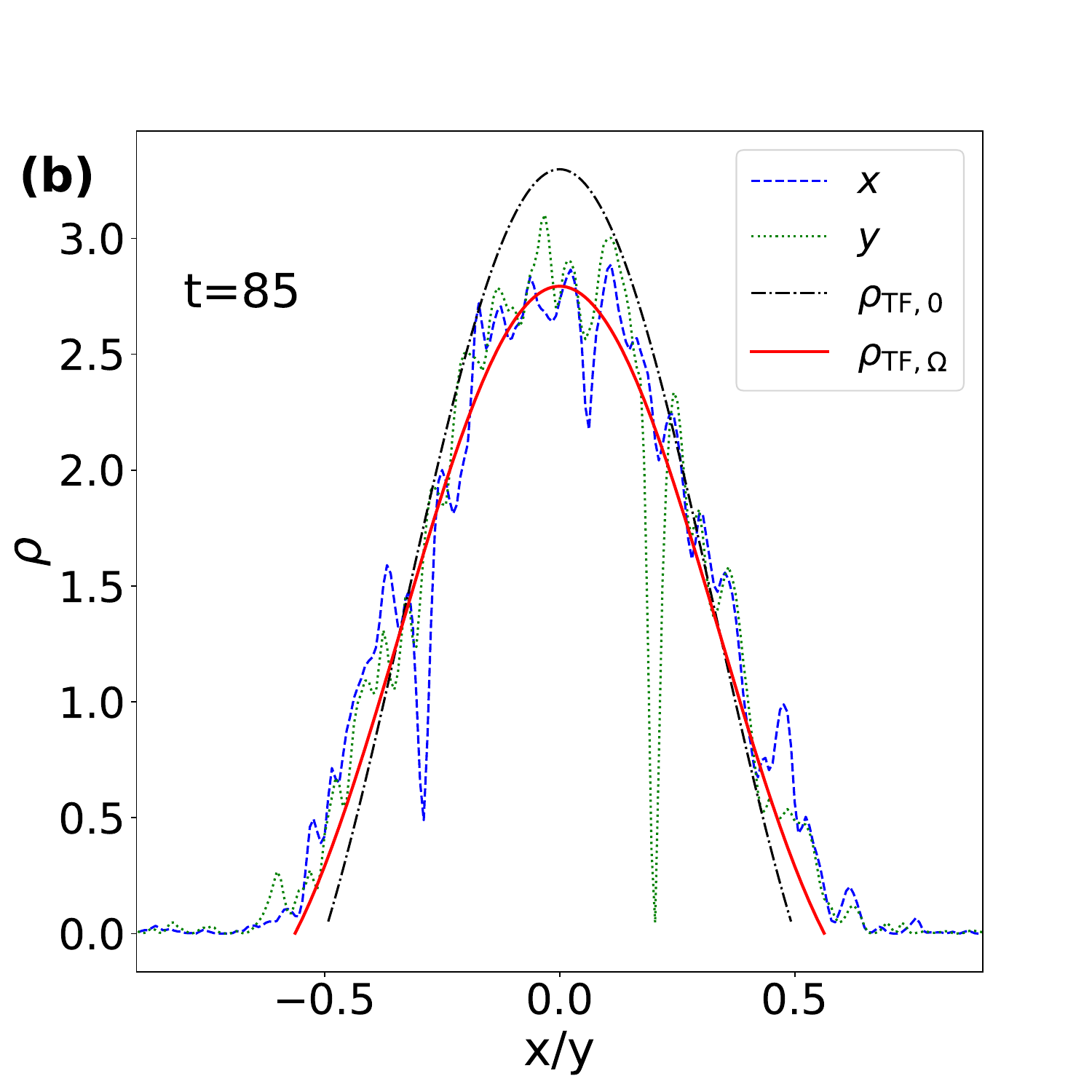}
\includegraphics[height=5cm,width=0.34\textwidth]{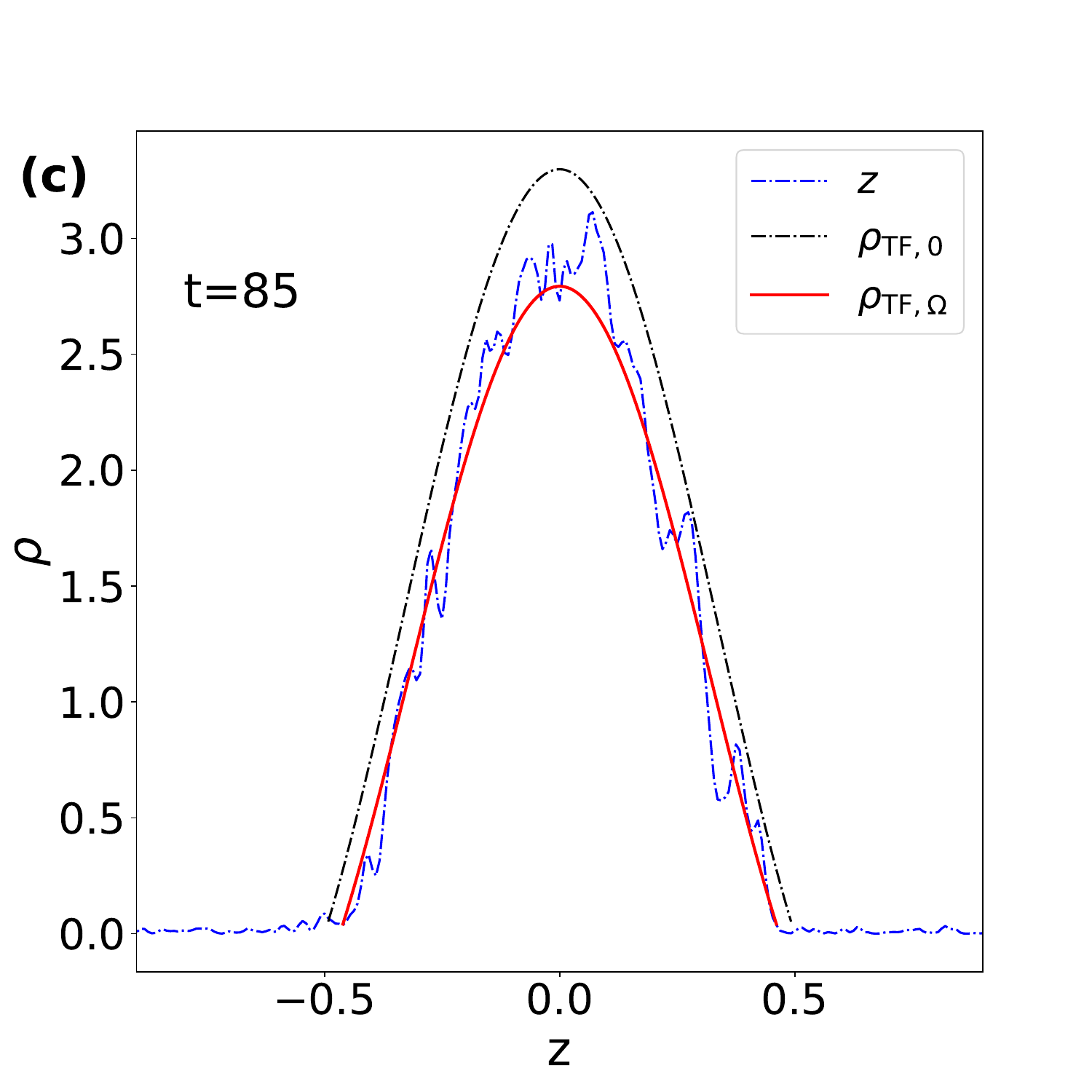}\\
\includegraphics[height=5cm,width=0.3\textwidth]{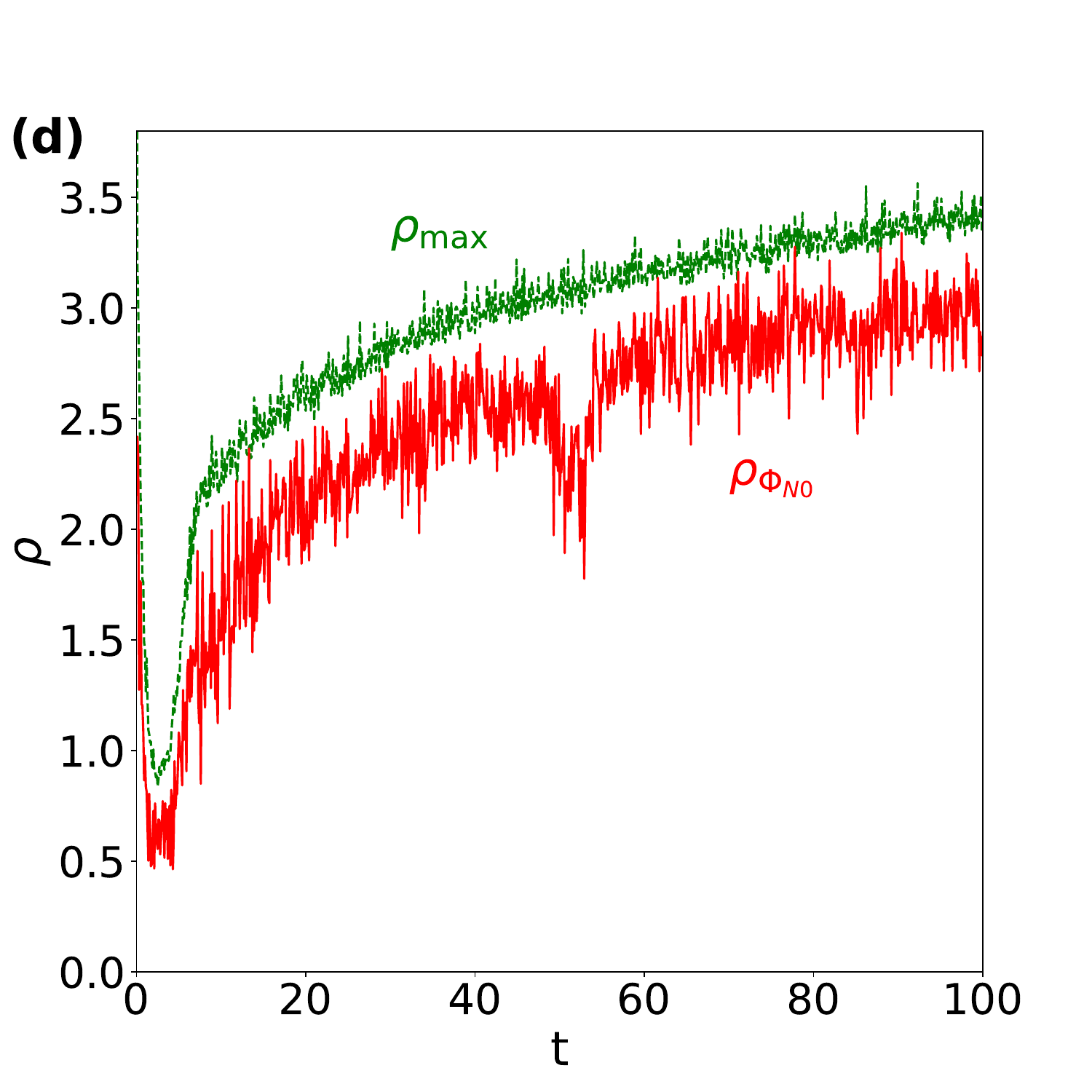}
\includegraphics[height=5cm,width=0.34\textwidth]{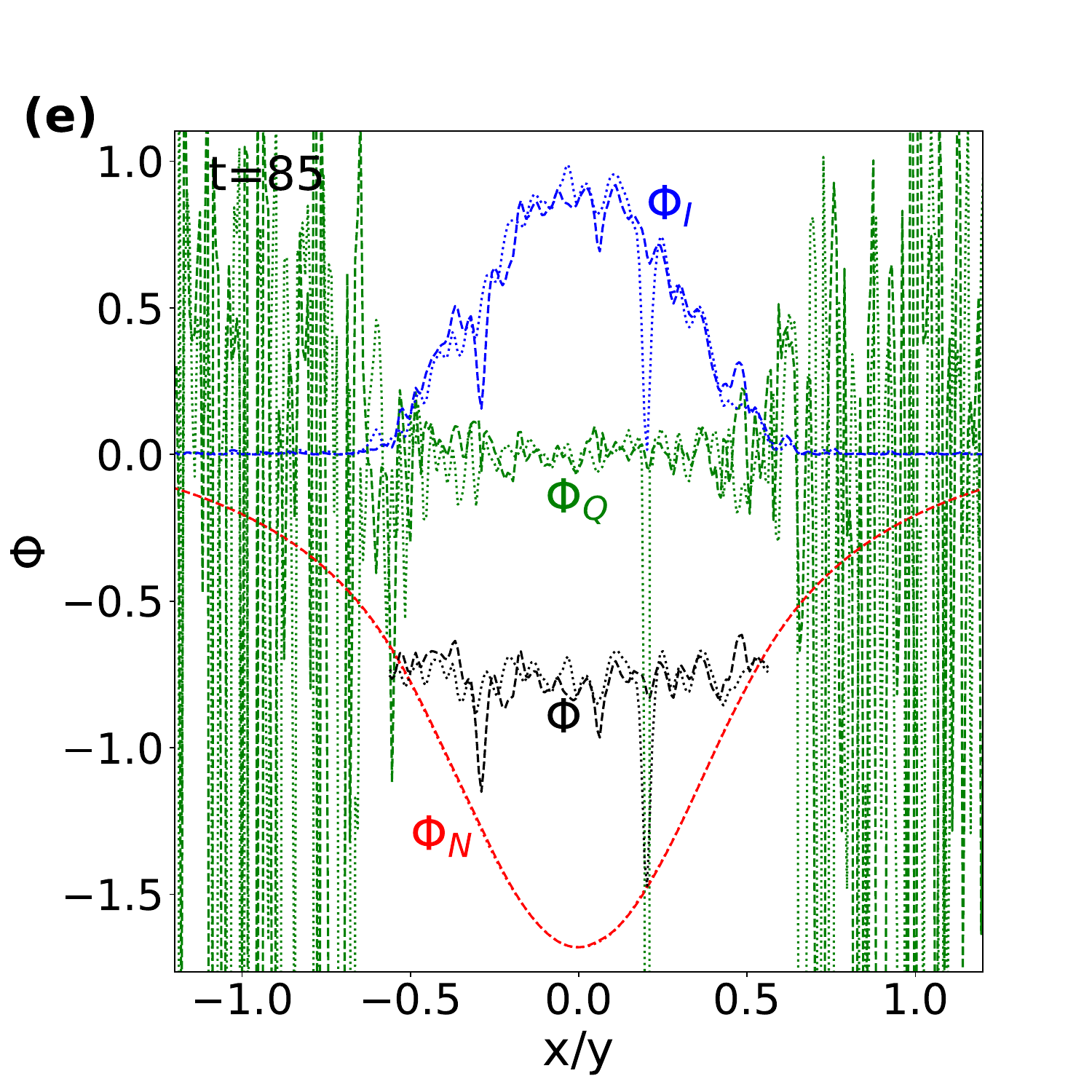}
\includegraphics[height=5cm,width=0.34\textwidth]{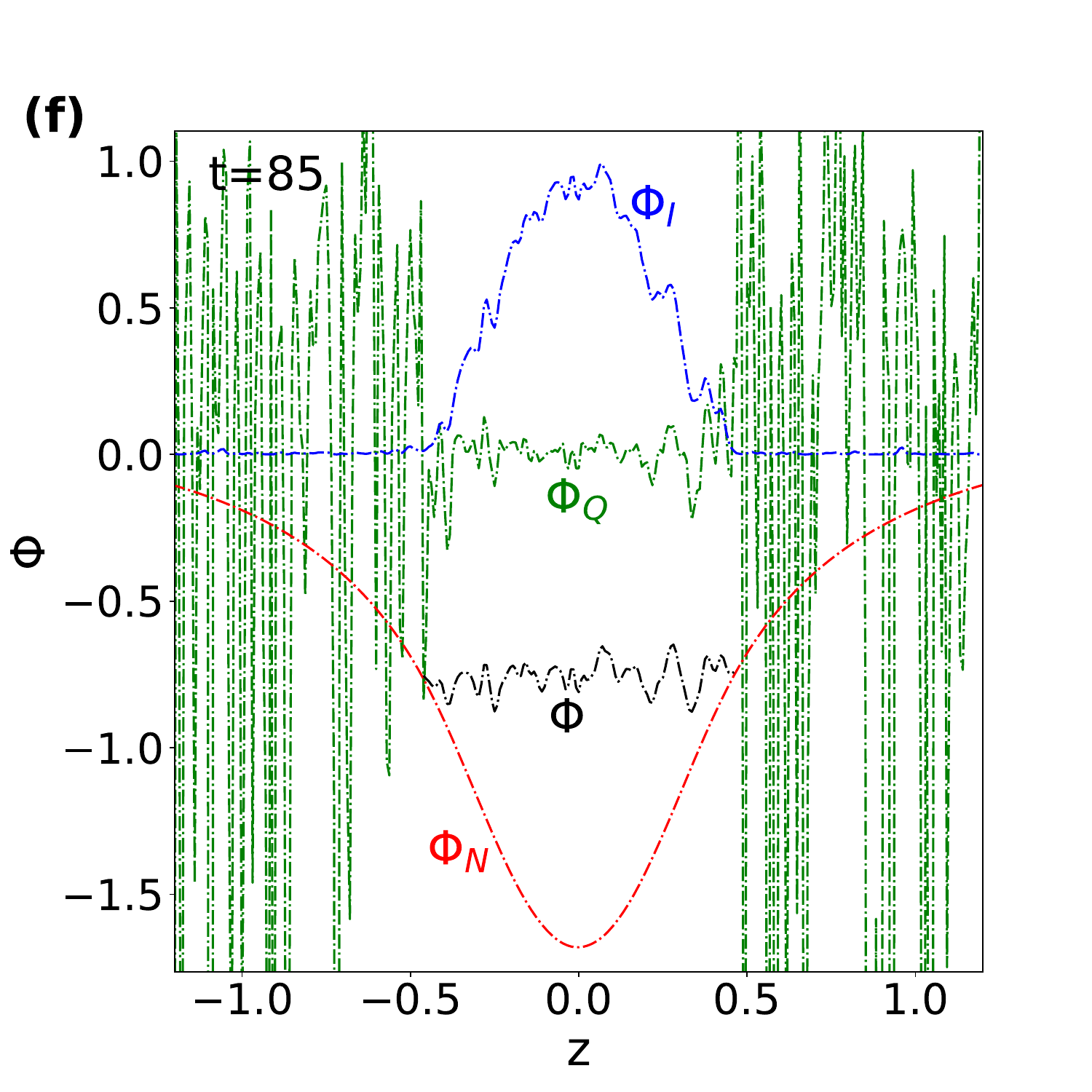}
\caption{
$[\epsilon=0.01, \alpha=1]$.
{\bf (a)} growth of the soliton mass with time, approximated by the central mass $M_{\rm TF,0}$
within the characteristic radius $R_0$ (red dashed line). The total mass of the
system (blue solid line) remains constant.
{\bf (b)} density profiles along the $x$ (blue dashed line) and $y$ (green dotted line) axes,
at time $t=85$.
We also show the profile (\ref{eq:rho-TF-0}) of a static soliton (black dash-dotted line $\rho_{\rm TF,0}$),
and the profile (\ref{eq:rho-in-res}) of a rotating soliton (red solid line $\rho_{\rm TF,\Omega}$).
{\bf (c)} density profile along the $z$ axis (blue dash-dotted line),
profile (\ref{eq:rho-TF-0}) of a static soliton
and profile (\ref{eq:rho-in-res}) of a rotating soliton.
{\bf (d)} growth of the maximum density $\rho_{\rm max}$ and of the central density $\rho_{\Phi_{N0}}$.
{\bf (e)} potentials $\Phi_Q$ (green dashed/dotted line),
$\Phi_I$ (blue dashed/dotted line), $\Phi_N$ (red dashed/dotted line)
and the sum $\Phi = \Phi_N+\Phi_I-r_\perp^2\Omega^2/2$ (black dashed/dotted line within the soliton
radius), along the $x/y$ axis.
{\bf (f)} potentials $\Phi_Q$, $\Phi_I$, $\Phi_N$ and the sum $\Phi = \Phi_N+\Phi_I-r_\perp^2\Omega^2/2$
(within the soliton radius) along the $z$ axis.
}
\label{fig:evol-mu1-0p01}
\end{figure*}

\subsection{Numerical algorithm}
\label{sec:numerical-algorithm}

As in \cite{Garcia:2023abs,Brax:2025uaw}, we solve the Gross-Pitaevskii equation
(\ref{eq:Schrod-eps}) with a pseudo-spectral method, using a split-step algorithm
\cite{Pathria-1990,Zhang-2008,Edwards-2018}.
The wave function after a time step $\Delta t$ is obtained as
\ba
&&\psi(\vec{x},t+ \Delta t) = \exp\left[-\frac{i \Delta t}{2\epsilon}
\Phi(\vec{x},t+ \Delta t)\right] \times \nonumber \\
&& \mathcal{F}^{-1} \exp\left[-\frac{i \epsilon\Delta t}{2} k^2\right]\mathcal{F}
\exp\left[-\frac{i \Delta t }{2\epsilon}\Phi(\vec{x},t)\right]\psi(\vec{x},t) \qquad .
\label{eq:step}
\ea
where the sequence of the operations is from right to left.
Here $\mathcal{F}$ and $\mathcal{F}^{-1}$ are the discrete Fourier
transform and its inverse, $k$ is the wavenumber in Fourier space and $\Phi=\Phi_N+\Phi_I$.
We also solve the Poisson equation (\ref{eq:Poisson-eps}) by Fourier transforms
and we apply periodic boundary conditions to the simulation box.

In the figures, we define the center of the system as the location of the minimum of the gravitational potential.
This removes the effects associated with the fluctuations of the position of the soliton and with
the random perturbations of the density around the averaged soliton profile.
The 1D profiles, as in Fig.~\ref{fig:evol-mu1-0p01} below, correspond to the $x$, $y$ and $z$
axes that run through this center.
The 2D planes $(x,y)$ and $(x,z)$, as in Fig.~\ref{fig:2D-rho-mu1-0p01} below, also run through
this center.
We do not show the plane $(y,z)$ as by symmetry it displays the same statistical properties
as the plane $(x,z)$ (as we checked in the numerical simulations).
The angular momentum $J_z(<r)$ within radius $r$ is also computed with respect to this center.

\section{Numerical results}
\label{sec:numerical-results}

\subsection{Results for $\epsilon=0.01$ and $\alpha=1$}
\label{sec:eps-0p01-alpha-1}

\subsubsection{Soliton mass and density profiles}

For the reference case $\epsilon=0.01$ and $\alpha=1$, we show in Fig.~\ref{fig:evol-mu1-0p01} 
how the system evolves from the initial condition displayed in Fig.~\ref{fig:initial}.
The conservation of the total mass in panel (a) provides a numerical check of the simulation
(we also checked that momentum and energy are conserved during the simulations).
Within a few dynamical times a soliton forms at the center of the system, as in the 3D isotropic
simulations presented in \cite{Garcia:2023abs}.
We show in panel (a) the mass $M_{\rm TF,0}$ within radius $R_0$, which remains a good approximation
of the soliton mass as the slow rotation does not distort the soliton too much.
We can see that after a violent relaxation the system settles in a few dynamical times to
a quasi-stationary state, with a central soliton within an excited virialized halo.
The soliton contains about $40\%$ of the mass of the system after its formation.
Afterwards it shows a secular growth until the end of our simulation, with a growth rate that decreases
with time.

As seen in panel (d), the central density $\rho_{\Phi_{N0}}$ (defined as the location where the
gravitational potential is minimum) and the maximum density $\rho_{\max}$ follow the same behavior,
but with significant fast fluctuations.
This is due to the incomplete relaxation and to the excited states in the outer halo that also
extend over the central region and give rise to fluctuations on top of the smooth soliton profile.
The central density $\rho_{\Phi_{N0}}$ shows a significant decrease at $t\simeq 50$ when
a vortex line happens to be close to the center, as the density decays as $\rho \propto r_\perp^2$
close to the vortex line from Eq.(\ref{eq:rho-core}).
Indeed, even though the vortices roughly move along circular orbits, in agreement with the
predictions (\ref{eq:dot-rj}) and (\ref{eq:v-solid-rotation}), and as checked in 
Fig.~\ref{fig:3D-vortex-motion-mu1-0p01} below, their trajectories
are perturbed by finite size effects and the random fluctuations generated by the outer halo modes.
The maximum density $\rho_{\max}$ is not affected by the location of the vortices inside the soliton
and it shows a more regular growth.

\begin{figure*}
\centering
\includegraphics[height=4.8cm,width=0.24\textwidth]{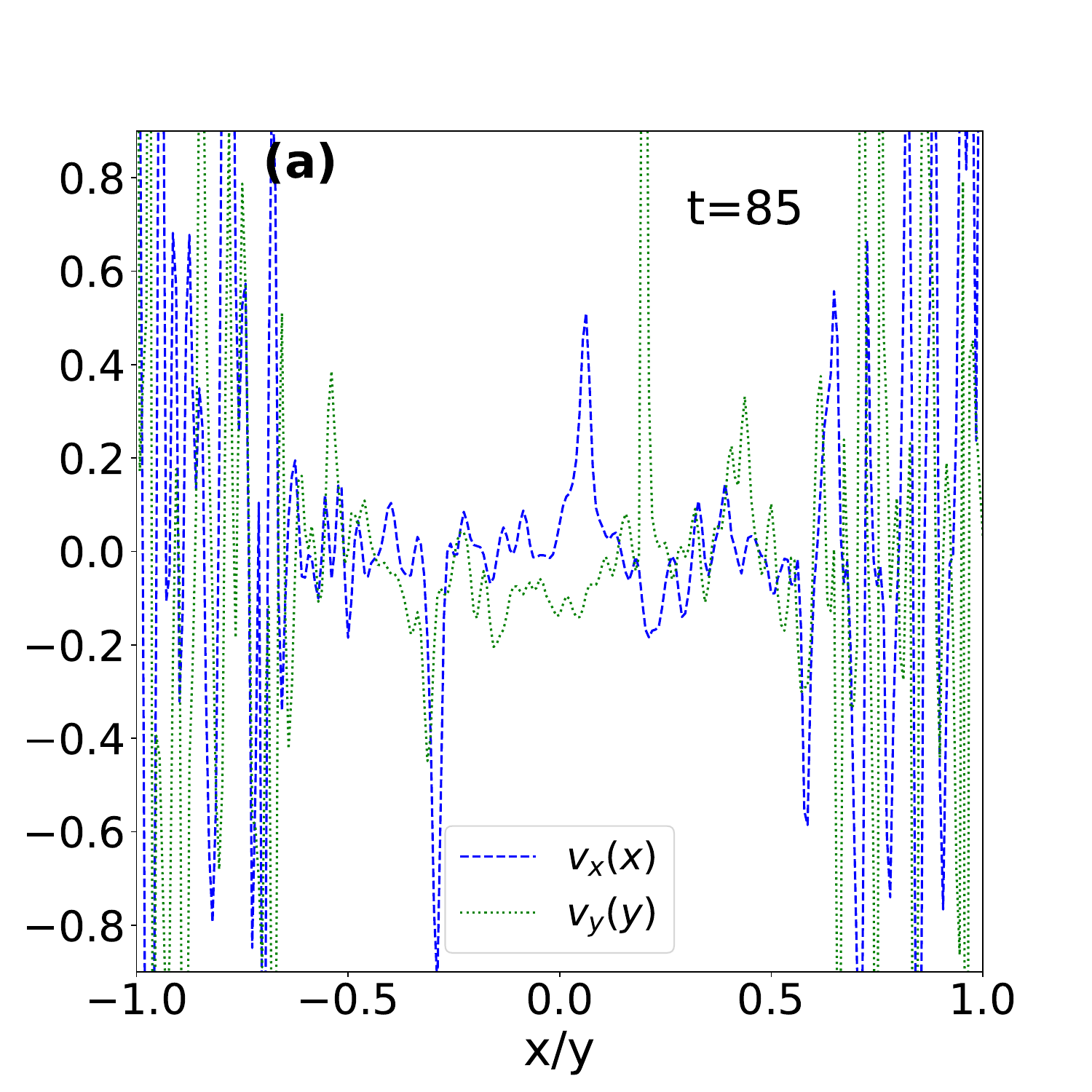}
\includegraphics[height=4.8cm,width=0.24\textwidth]{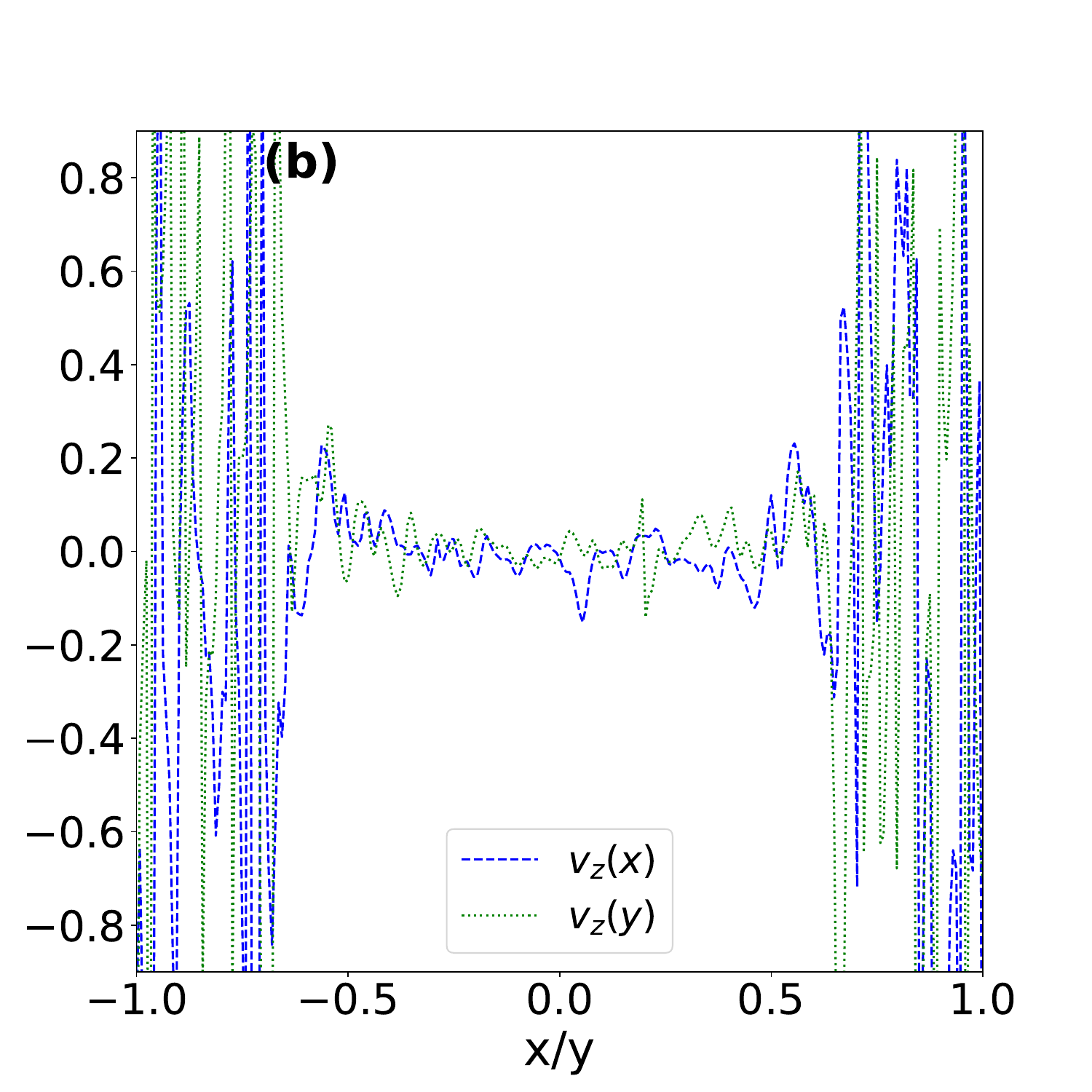}
\includegraphics[height=4.8cm,width=0.25\textwidth]{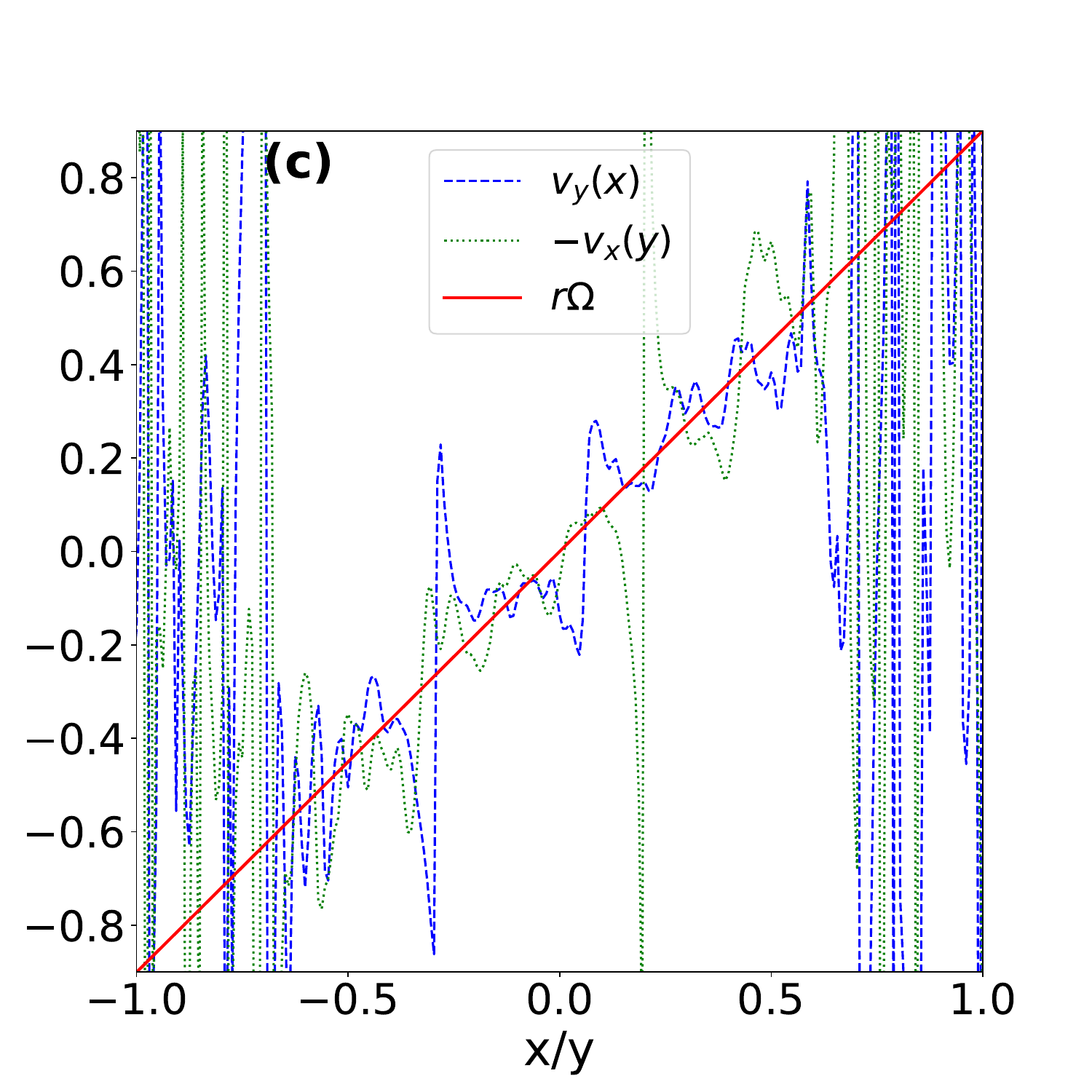}
\includegraphics[height=4.8cm,width=0.25\textwidth]{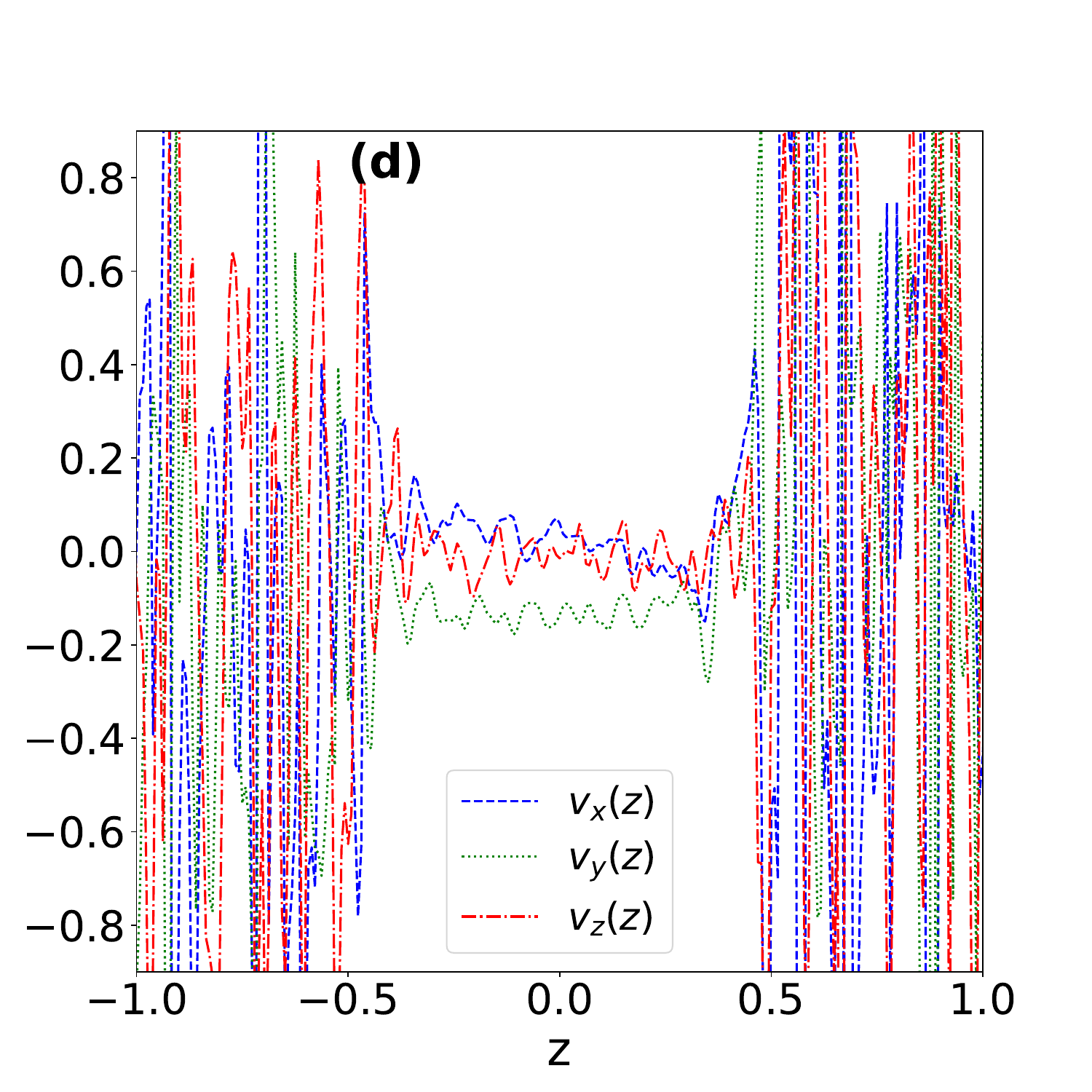}
\caption{
$[\epsilon=0.01, \alpha=1]$.
{\bf (a)} parallel velocities $v_x/v_y$ along the $x/y$ axis, at $t=85$.
{\bf (b)} vertical velocity $v_z$ along the $x/y$ axis.
{\bf (c)} perpendicular velocities in the plane $(x,y)$ along the $x/y$ axis.
The red solid line is the best fit $\Omega r_\perp$ in the central region, which provides our measurement
of the rotation rate $\Omega$.
{\bf (d)} velocity components $v_x$, $v_y$ and $v_z$ along the $z$ axis.
}
\label{fig:v-mu1-0p01}
\end{figure*}

The density profile is shown in panel (b), along the $x$ and $y$ axes, and in panel (c), along the $z$ axis.
We also show the static soliton profile (\ref{eq:rho-TF-0}), which is identical along the three axis,
and the rotating soliton profile (\ref{eq:rho-in-res}), which is broader along the $x/y$ axis
($\theta=\pi/2$) and narrower along the $z$ axis ($\theta=0$).
We normalize the static profile (\ref{eq:rho-TF-0}) by the central mass $M_{\rm TF,0}$.
We normalize the rotating density profile by the central density $\rho_0$, which we estimate from
the simulation by taking the mean within a radius $r<0.03$ from the center.
We estimate the rotation rate $\Omega$ from the velocity profile shown in Fig.~\ref{fig:v-mu1-0p01},
as explained in Sec.~\ref{sec:velocity} below.
We can see that the soliton density profile agrees reasonably well with the first-order prediction
(\ref{eq:rho-in-res}), with a flattening as compared with the static profile and an oblate shape.
Even though the rotation is not so fast, we can clearly see that the size is increased in the equatorial
plane and decreased along the vertical rotation axis.
The two low-density spikes at $x \simeq -0.3$ and $y=0.2$ in panel (b) correspond to two vortex lines
located at $\vec r_\perp \simeq (-0.3,0)$ and $(0,0.2)$, which happen to be close to the
$x$ and $y$ axes in the equatorial plane.
The vortex line at $\vec r_\perp \simeq (0,0.2)$ almost intersects with the $y$ axis as
we can see that the density almost vanishes along the $y$ axis.
These features are absent in the density profile along the $z$ axis.
Indeed, since the vortex lines are almost vertical, as can be seen in
Fig.~\ref{fig:3D-vortices-mu1-0p01} below, if the point $\vec r = (0,0,0)$ in the equatorial
plane is far from a vortex line this remains so as we move vertically along the $z$ axis,
$\vec r = (0,0,z)$.

\begin{figure*}
\centering
\includegraphics[height=4.8cm,width=0.35\textwidth]{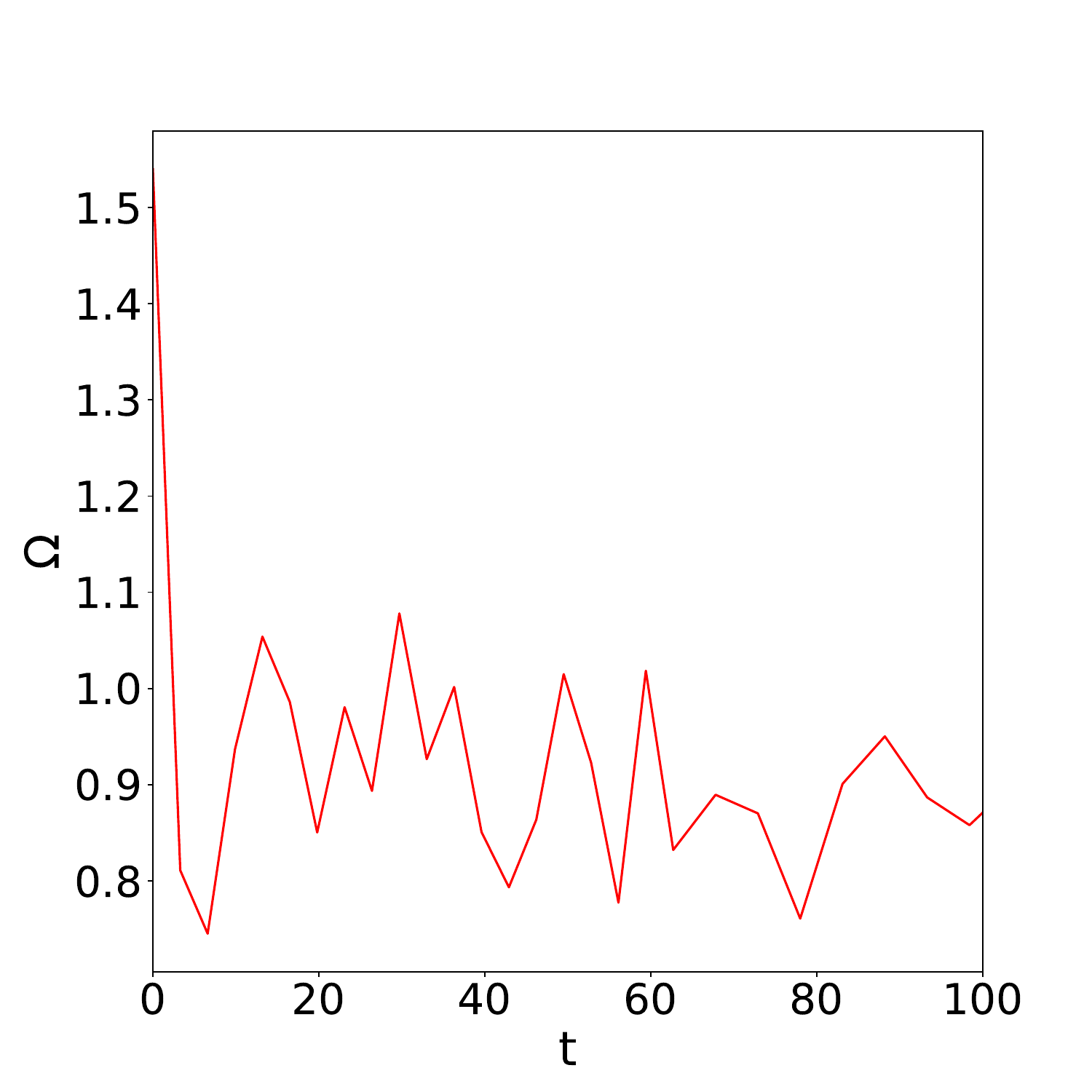}
\includegraphics[height=4.8cm,width=0.31\textwidth]{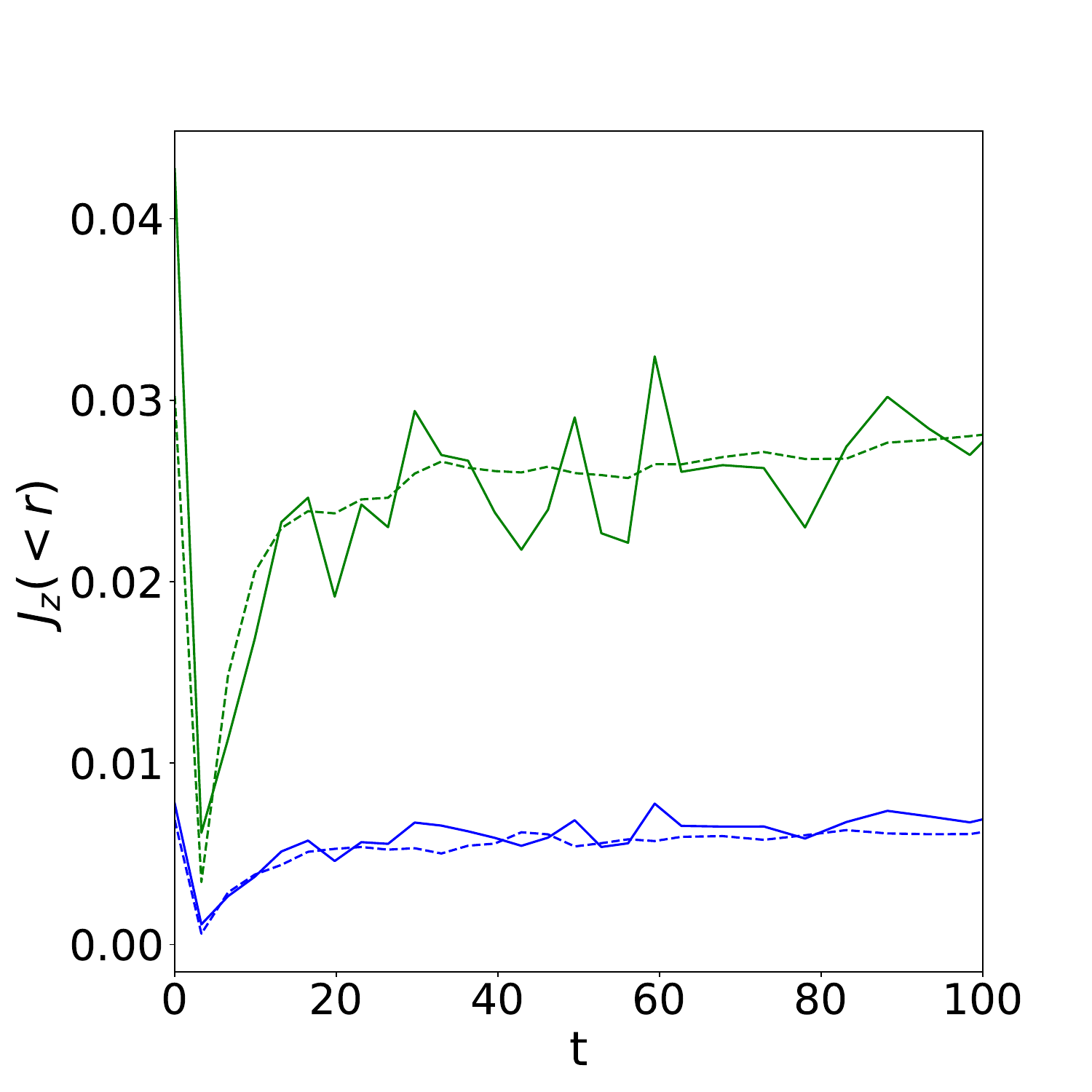}
\includegraphics[height=4.8cm,width=0.31\textwidth]{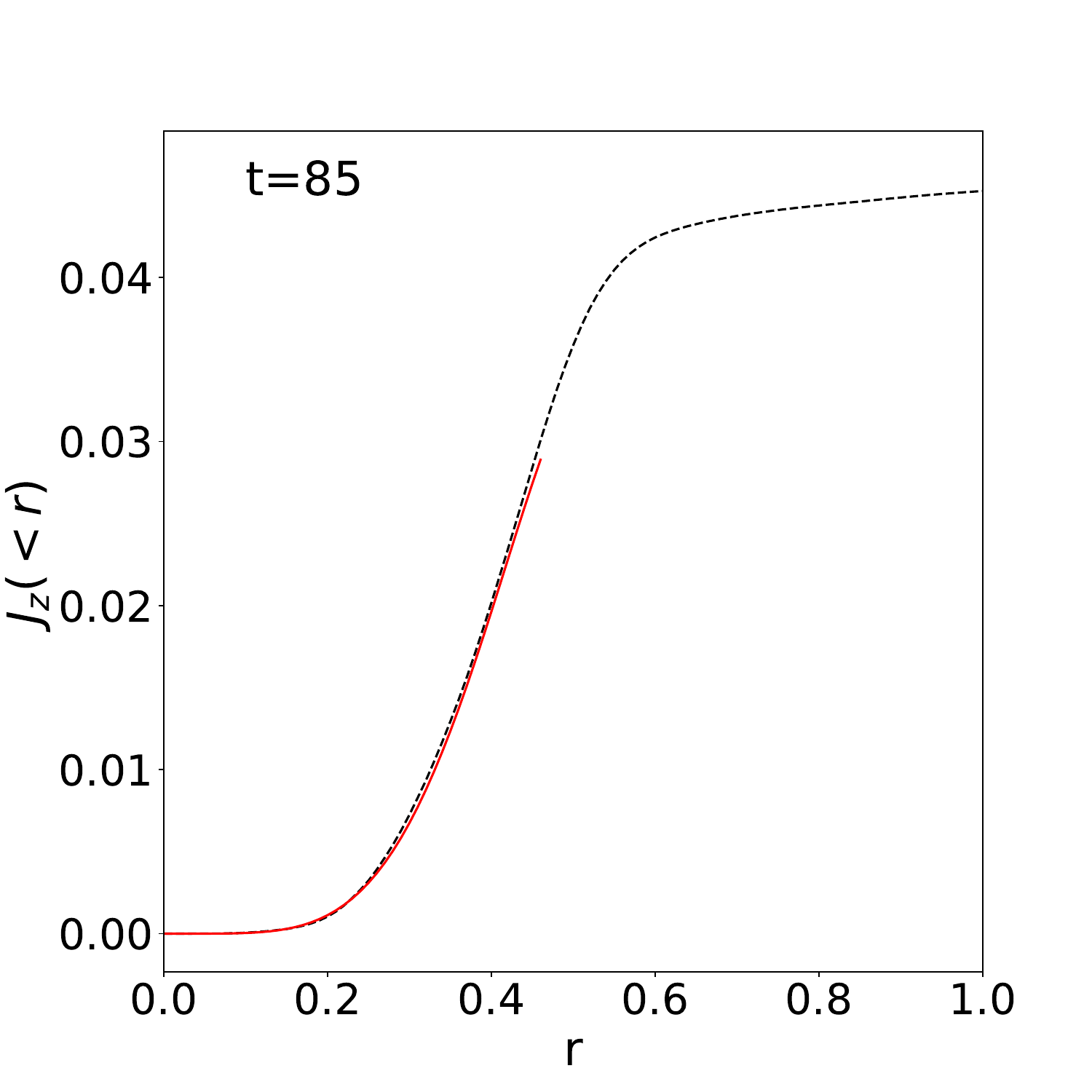}
\caption{
$[\epsilon=0.01, \alpha=1]$.
{\it Left panel:} rotation rate $\Omega(t)$ of the soliton.
{\it Middle panel:} angular momentum inside the soliton, within radii $r<0.3$ and $r<0.45$,
measured from the simulations (dashed lines). The solid lines are the theoretical prediction
(\ref{eq:Jz-r-Om}), which fluctuates with the measured rotation rate $\Omega$.
{\it Right panel:} radial distribution of the angular momentum.
}
\label{fig:Om-mu1-0p01}
\end{figure*}

We show in panels (e) and (f) the potentials along the $x$, $y$ and $z$ axes.
We also display the combination $\Phi=\Phi_N+\Phi_I-r_\perp^2\Omega^2/2$, associated
with the solid-body rotation equilibrium (\ref{eq:mu-Omega}).
As in the 2D case analysed in \cite{Brax:2025uaw}, inside the soliton the quantum pressure is negligible
as compared with the gravitational and self-interaction potentials. Moreover, in agreement with
the equation of rotating equilibrium (\ref{eq:mu-Omega}) the sum $\Phi$ is roughly constant.
As for the density profiles, there remain significant perturbations to $\Phi_I$ and $\Phi$
inside the soliton, because of the contributions to the local density from the excited modes
that extend over the outer halo.
As usual, the gravitational potential is much smoother as it is an integral over the density field.
As for the density profile, we can see that the size of the rotating soliton is larger in the equatorial
plane than along the vertical rotation axis (i.e., an oblate shape).
The self-interaction potential, $\Phi_I \propto \rho^2$, and the quantum pressure,
$\Phi_Q \propto  \Delta\sqrt{\rho}/\sqrt{\rho}$, are very sensitive to the local value
of the density. Therefore, they show distinct spikes in panel (e), at the same locations
as the density profiles obtained in panel (b), at $\vec r_\perp \simeq (-0.3,0)$ and $(0,0.2)$.
Again, such features are absent in panel (f), along the vertical axis.

Outside the soliton, the self-interaction potential becomes negligible while the quantum pressure
becomes large and shows wild fluctuations. As in the initial condition (\ref{eq:psi-halo-a_nlm}),
this regime is dominated by a sum over many uncorrelated eigenmodes, which mimics
a virialized halo supported by its velocity dispersion in the semi-classical picture.

\subsubsection{Velocity profiles}
\label{sec:velocity}

We show in Fig.~\ref{fig:v-mu1-0p01} the velocity profiles along the $x$, $y$ and $z$ axes.
Like the quantum pressure displayed in Fig.~\ref{fig:evol-mu1-0p01},
the velocity field exhibits large fluctuations in the outer halo and is much more regular inside
the soliton. Again, this is because the outer halo is made of uncorrelated excited modes,
with density and velocity fluctuations on the de Broglie scale $\lambda_{\rm dB}$, whereas
the soliton is a coherent ground state.
As for the density profiles displayed in Fig.~\ref{fig:evol-mu1-0p01}, we can clearly see the
flattening of the rotating soliton, with a greater radius in the equatorial plane than along the vertical
axis.

In agreement with the solid-body rotation (\ref{eq:v-solid-rotation}), we recover in the third
panel the linear dependence $v_{\perp} = \Omega r_{\perp}$ along the $x$ and $y$ axes
inside the equatorial plane.
We fit the velocity in the central region ($r_\perp<R_0$) by a straight line and its slope
provides our estimate of $\Omega$ from the simulation.
This is the value that we used to predict the rotating soliton density profiles shown above in
Fig.~\ref{fig:evol-mu1-0p01}.
The other two components of the velocity measured along the $x$ and $y$ axes (shown in the
first two panels) fluctuate around zero, as well as the three velocity components along the $z$ axis.
These behaviors confirm the solid-body rotation (\ref{eq:v-solid-rotation}) around the vertical axis,
with small fluctuations due to the incomplete relaxation and the impact of the outer modes.

In agreement with the $1/r_\perp$ divergence of $v_\varphi$ close to the vortex lines in
Eq.(\ref{eq:v-single-vortex}), the velocity measured in the equatorial plane in panels (a) and (c)
shows large spikes near the vortex lines, at the same locations $\vec r_\perp \simeq (-0.3,0)$ 
and $(0,0.2)$ as those found in the density profiles in panel (b) of Fig.~\ref{fig:evol-mu1-0p01}.
As for the density these features do not appear along the $z$ axis, where we remain
far from the vertical vortex lines. They are also absent in the velocity component
$v_z$ shown in panel (b), because only the velocity $\vec v_\perp$ in the plane transverse to the
vortex line becomes large, whereas $v_z$ remains unaffected and equal to zero if the vortex
line is exactly vertical.

\subsubsection{Angular momentum}

We show in Fig.~\ref{fig:Om-mu1-0p01} the evolution with time of the soliton rotation rate
and of the angular momentum inside the soliton, as well as its radial distribution at time
$t=85$.
As for the central mass and density displayed in Fig.~\ref{fig:evol-mu1-0p01}, the rotation rate
and the angular momentum in the central region quickly settle to quasi-stationary values
along with the formation of the rotating soliton.
Moreover, the radial distribution of the angular momentum closely follows the prediction
(\ref{eq:Jz-r-Om}), associated with the solid-body rotation equilibrium (\ref{eq:mu-Omega}).

In the middle panel, the solid lines show the theoretical prediction (\ref{eq:Jz-r-Om}), which
depends on the measured rotation rate $\Omega$ displayed in the left panel, whereas the
dashed lines are the angular momentum within radius $r$ directly measured by integrating the
density and velocity profiles in the simulation.
While these integrated values are rather stable, the former theoretical predictions show significant
fluctuations that follow those of $\Omega(t)$ shown in the left panel. 
As can be seen in panel (c) in Fig.~\ref{fig:v-mu1-0p01}, this is because
the measurement of $\Omega$, from the slope of the velocity field in the central region, 
is sensitive to the perturbations due to passing-by of excited wave packets from the outer halo and
of vortices inside the soliton.
The direct measurement of the angular momentum is much smoother, as usual for integrated quantities
as compared with derivative quantities.

We can see in the right panel that about $70\%$ of the angular momentum of the system is contained
inside the soliton.
As seen in Fig.~\ref{fig:evol-mu1-0p01}, at formation the soliton has a mass $M_{\rm sol} \simeq 0.4$.
From Eq.(\ref{eq:Jz-fm-av}), the angular momentum $J_z(<M_{\rm sol})$ contained in the core
of the initial halo associated with this mass is $J_{z,\rm init}(<M_{\rm sol}) \simeq 0.075$.
However, we find in Fig.~\ref{fig:Om-mu1-0p01} that the angular momentum of the soliton is
$J_{z,\rm sol} \simeq 0.03$. Therefore, there is a significant redistribution of angular momentum
to larger radii, beyond the rotating soliton. This is consistent with the radial profile
shown in Fig.~\ref{fig:Om-mu1-0p01}, where $J_z(<r)$ keeps rising beyond the soliton radius.
For $\rho_0 \simeq 1.5$ at the formation of the soliton, the stability and existence upper bounds
(\ref{eq:Omega-stable}) and (\ref{eq:Rz-0}) give $\Omega \lesssim 1.3$ and $\Omega \lesssim 2.4$.
The rotation rate obtained in the left panel fluctuates between $0.8 \lesssim \Omega \lesssim 1.1$.
As expected, this is inside the stability range (\ref{eq:Omega-stable}).
On the other hand, the initial angular momentum for the same mass,
$J_{z,\rm init}(<M_{\rm sol}) \simeq 0.075$, which is slightly more than twice the actual value
$J_{z,\rm sol} \sim 0.03$, is above the bound (\ref{eq:Omega-stable}) and close to the
existence bound (\ref{eq:Rz-0}). Therefore, the reduced value of the final soliton angular momentum,
as compared with the initial halo, can be partly explained by these upper bounds.
However, because the final value is significantly below these bounds, part of the reduction
is beyond these stability or existence criterions and must be due to the details of the dynamics.
It is likely that the fast formation of the soliton, akin to a violent relaxation, generates a significant
exchange of angular momentum and energy and that mass shells do not exactly keep their initial
ordering.

\begin{figure*}
\centering
\includegraphics[height=4.8cm,width=0.33\textwidth]{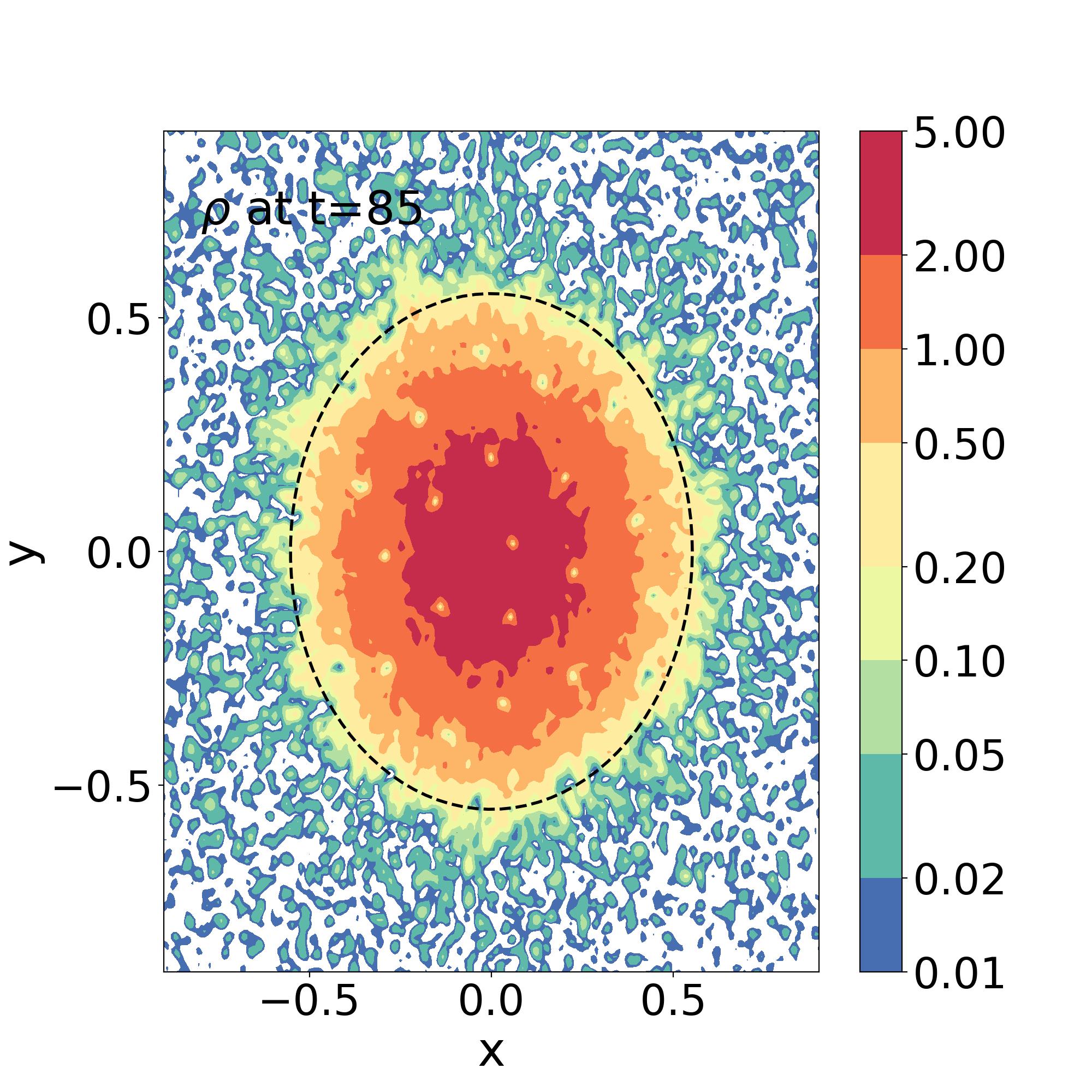}
\includegraphics[height=4.8cm,width=0.33\textwidth]{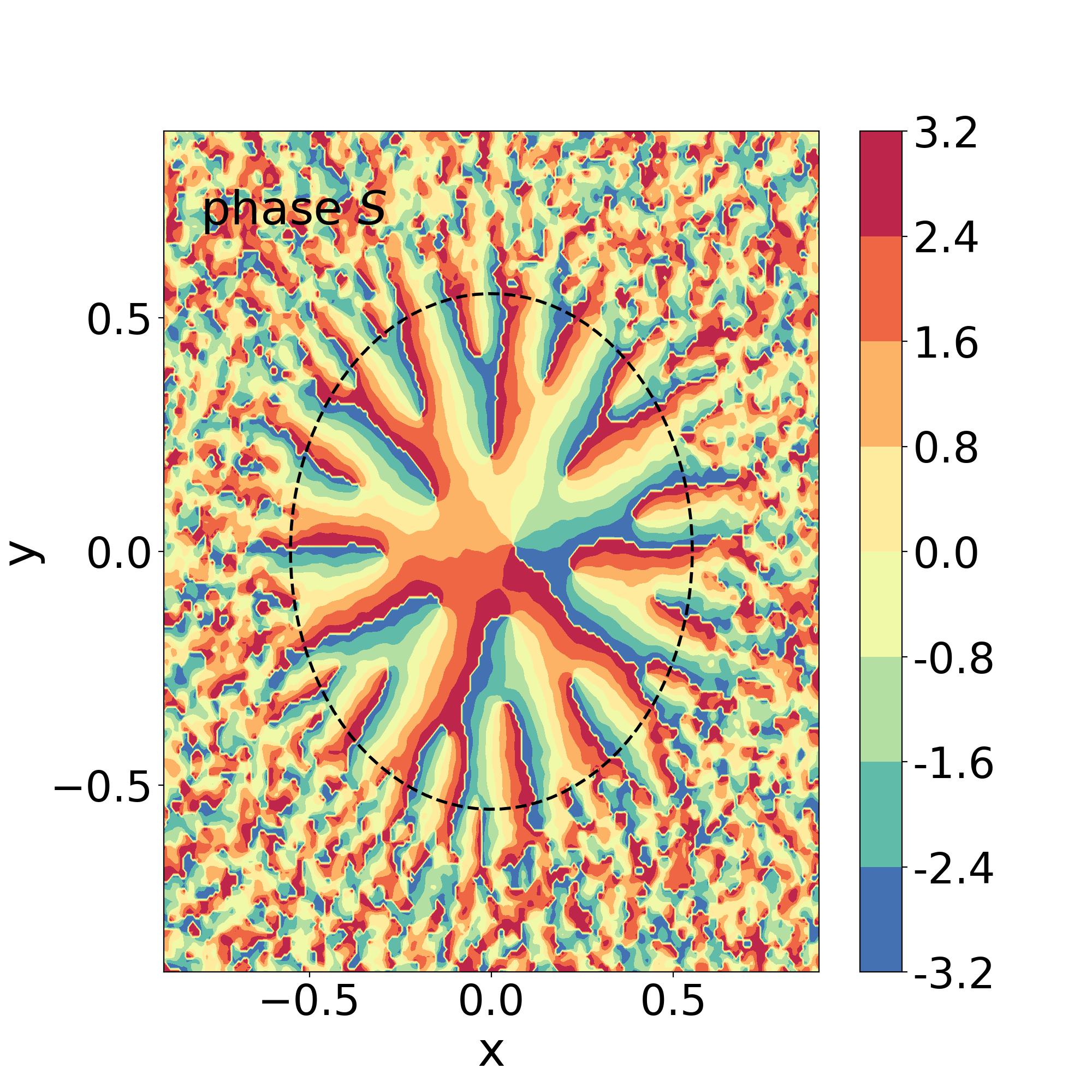}
\includegraphics[height=4.8cm,width=0.28\textwidth]{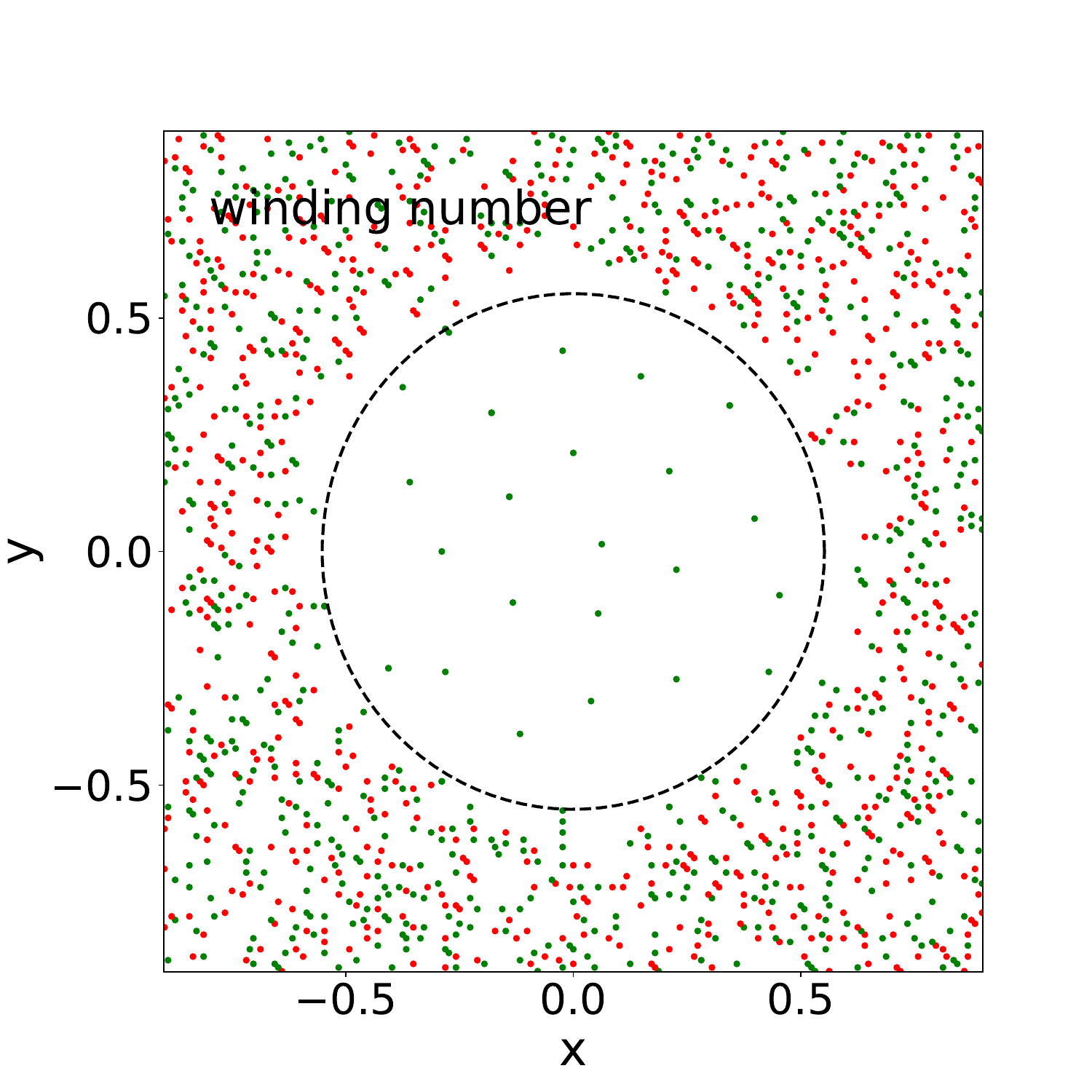}\\
\includegraphics[height=4.8cm,width=0.33\textwidth]{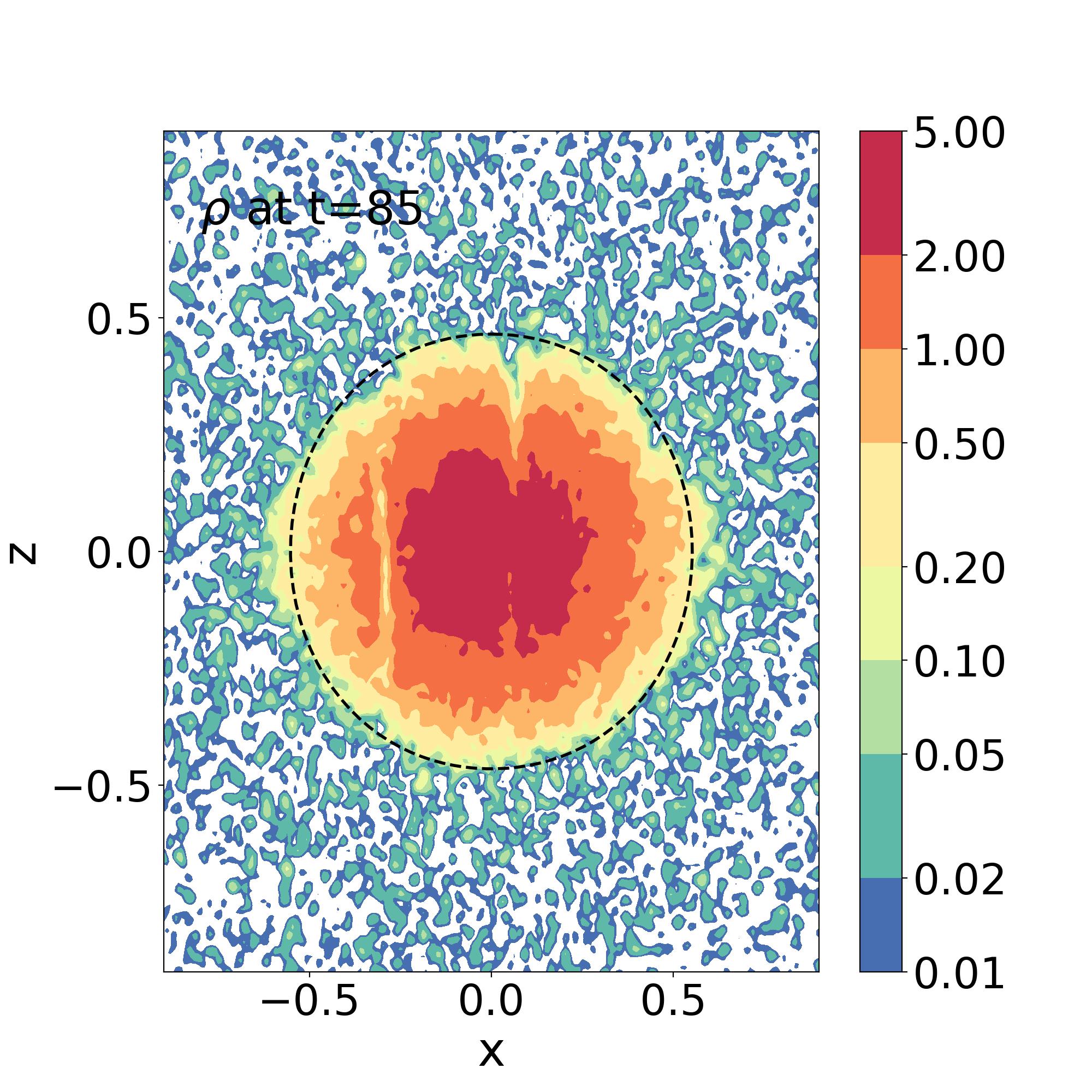}
\includegraphics[height=4.8cm,width=0.33\textwidth]{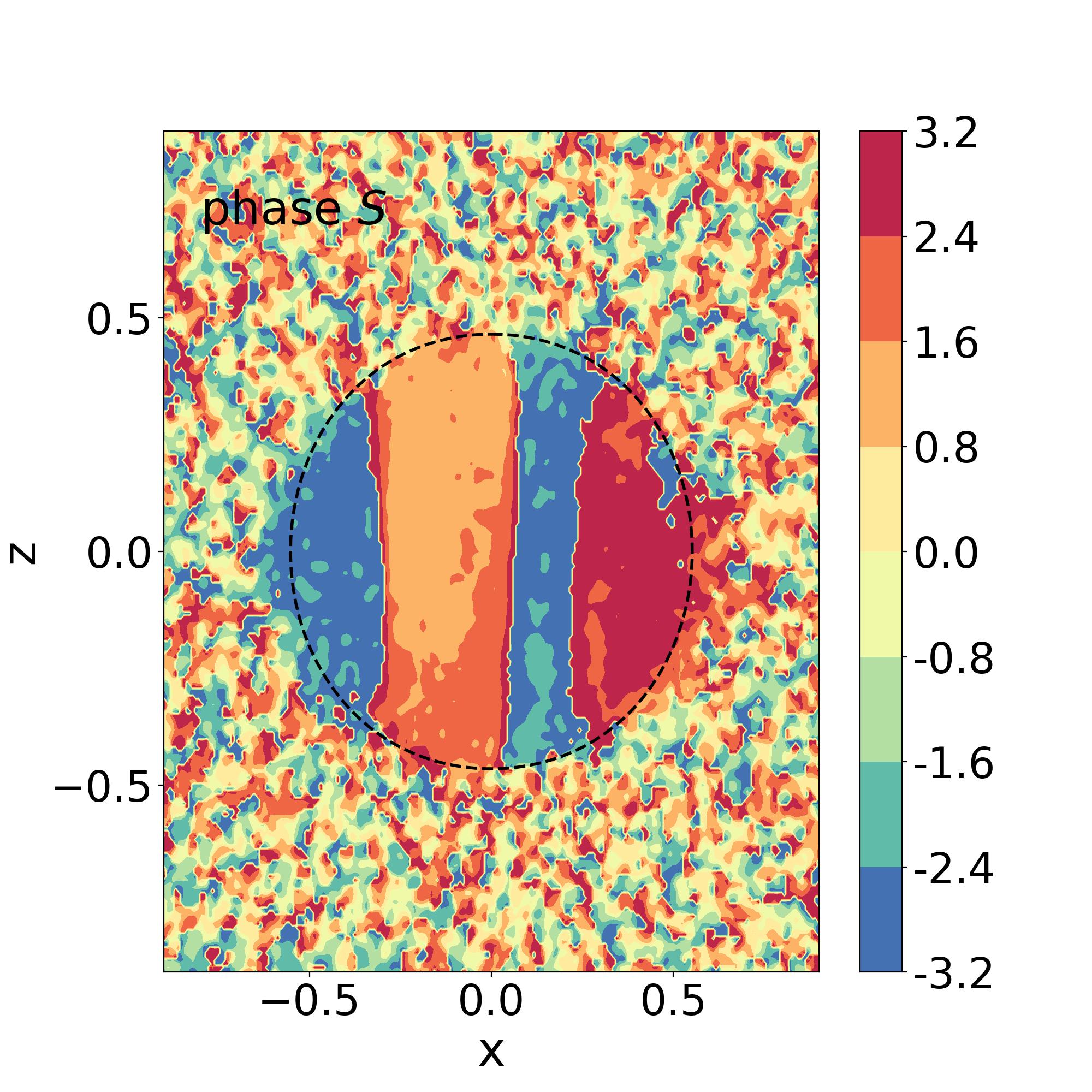}
\includegraphics[height=4.8cm,width=0.28\textwidth]{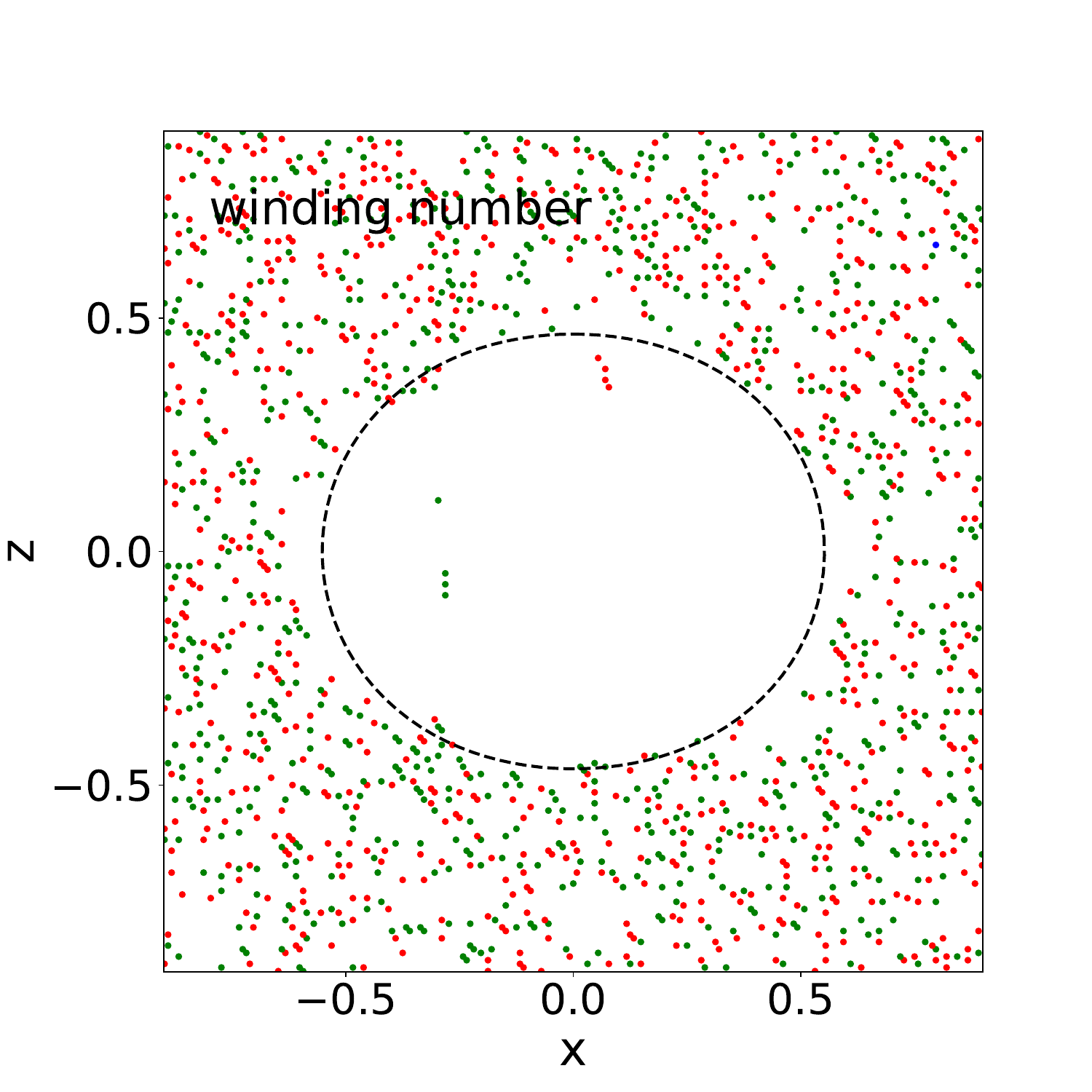}
\caption{
$[\epsilon=0.01, \alpha=1]$.
{\it Upper row, from left to right:} 2D maps in the plane $(x,y)$, at $t=85$, of the
density $\rho$, the phase $S$ and the winding number $w$
(green dots correspond to $w=1$ and red dots to $w=-1$).
The black dashed lines are the boundary of the soliton from Eq.(\ref{eq:Rsol-res}).
{\it Lower row:} 2D maps in the plane $(x,z)$.
}
\label{fig:2D-rho-mu1-0p01}
\end{figure*}

\subsubsection{2D density and phase maps}

We show in Fig.~\ref{fig:2D-rho-mu1-0p01} the density $\rho$, the phase $S$ and the winding number
$w$ in the $(x,y)$ and $(x,z)$ planes.
We also display the soliton boundary predicted by Eq.(\ref{eq:Rsol-res}).
The winding number in the plane $(x,y)$ is obtained as follows. Around each point of the numerical
grid within this plane, we draw a co-planar square of size made of three grid points
Then, we measure the phase difference $\Delta S$ found by moving clockwise along this loop,
$\Delta S = \oint \vec{d\ell} \cdot \vec \nabla S = \sum _i \Delta_i S$, where $\Delta_i S$ is the
phase difference between two successive points $i$ and $i+1$ along the loop.
Next, we define the winding number as $w=\Delta S/(2\pi)$.
By definition, $w$ is an integer as we recover the same value of the wave function $\psi$
at the identical starting- and end-point of the loop.
If the phase and its gradient $\vec\nabla S$ are regular, we have $\Delta S = 0$ and $w=0$.
However, if the loop encircles a vortex line of spin $\sigma = \pm 1$ we obtain $w=\pm 1$.
Therefore, the winding number map allows us to count the vortex lines that cross the equatorial
plane $(x,y)$, as shown in the third panel in the upper row.
We proceed in a similar fashion in the $(x,z)$ plane and the third panel in the lower row
shows the vortex lines that cross the $(x,z)$ plane.

We can clearly see the anisotropy of the system, as the equatorial and vertical planes $(x,y)$ and
$(x,z)$ show different behaviors. In agreement with the azimuthal symmetry, we checked that the
plane $(y,z)$ shows the same features as the plane $(x,z)$.

Let us first consider the equatorial plane $(x,y)$ in the upper row.
We recover a circle for the soliton boundary, with a radius that agrees with the prediction
(\ref{eq:Rxy-def}).
We can clearly see in the upper left panel the transition between the outer halo, with a low mean density
and strong fluctuations on the small de Broglie scale, and the soliton, with a smoother density profile
that rises towards the center.
Inside the soliton, in addition to the random fluctuations already apparent in the 1D profiles
shown in Fig.~\ref{fig:evol-mu1-0p01}, and also observed in the simulations of isotropic systems
without rotation in \cite{Garcia:2023abs} and in Fig.~\ref{fig:2D-rho-mu0-0p01} below, 
we distinguish a regular lattice of low-density troughs,
associated with the vortex lines in agreement with the vanishing density in Eq.(\ref{eq:rho-core}).

These two regimes are also clearly apparent in the upper middle panel, with a phase $S$ that shows
uncorrelated small-scale fluctuations in the outer halo and smoother and larger-scale features inside
the soliton.
Inside the soliton, the phase shows a characteristic branching structure also found in the 2D simulations
of rotating systems \cite{Brax:2025uaw}, spreading from the central region.
The start of each new branch corresponds to a vortex line, i.e. a singularity of the phase,
where a new line where $S$ jumps from $\pi$ to $-\pi$ appears in the equatorial plane.
These singular points coincide with the density troughs found in the upper left panel.

These two regimes are also manifested in the upper right panel. Whereas the outer halo shows a proliferation
of vortex lines with a spin of either sign, $\sigma = \pm 1$, due to the random interferences between
uncorrelated excited modes, with fluctuations on the de Broglie scale, the soliton shows a regular lattice
of vortex lines of spin $\sigma = 1$. This is due to the sign of the initial angular momentum,
$\langle J_z \rangle > 0$ (i.e. $\alpha > 0$), which dictates the sign of the rotation and of the vorticity
of the soliton.
Thus, after the formation and relaxation of the soliton, inner vortices of opposite signs have
mostly annihilated and we are left with vortices of the requisite sign to support the angular momentum
of the soliton.
Moreover, in agreement with the analysis presented in Sec.~\ref{sec:uniform-lattice},
these vortex lines self-organize to form a regular lattice, which is the discrete representation of the
uniform vortex density (\ref{eq:vortex-density}).
This is associated with the solid-body rotation (\ref{eq:v-solid-rotation}), as seen in (\ref{eq:vortex-density}).

\begin{figure*}
\centering
\includegraphics[height=4.8cm,width=0.27\textwidth]{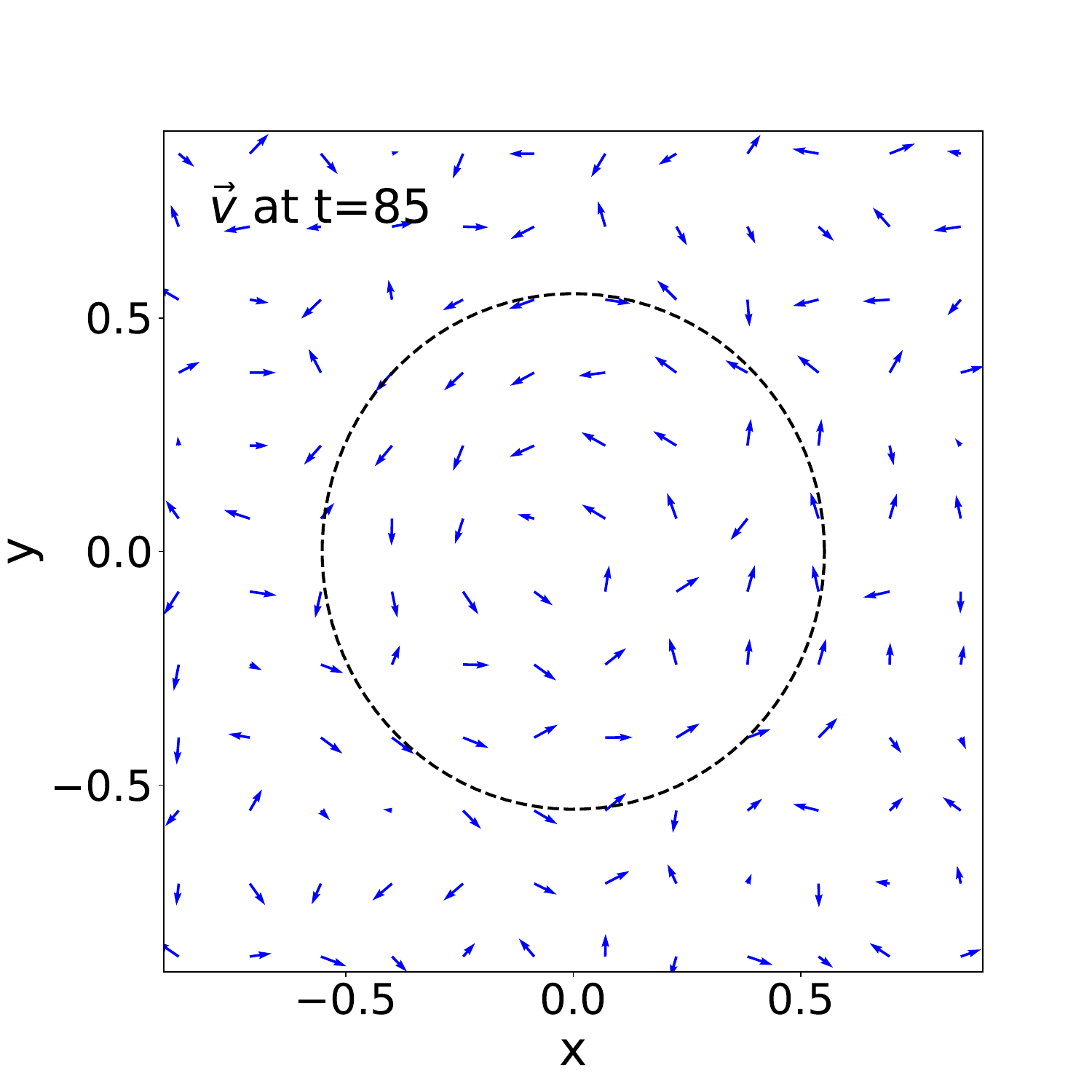}
\includegraphics[height=4.8cm,width=0.27\textwidth]{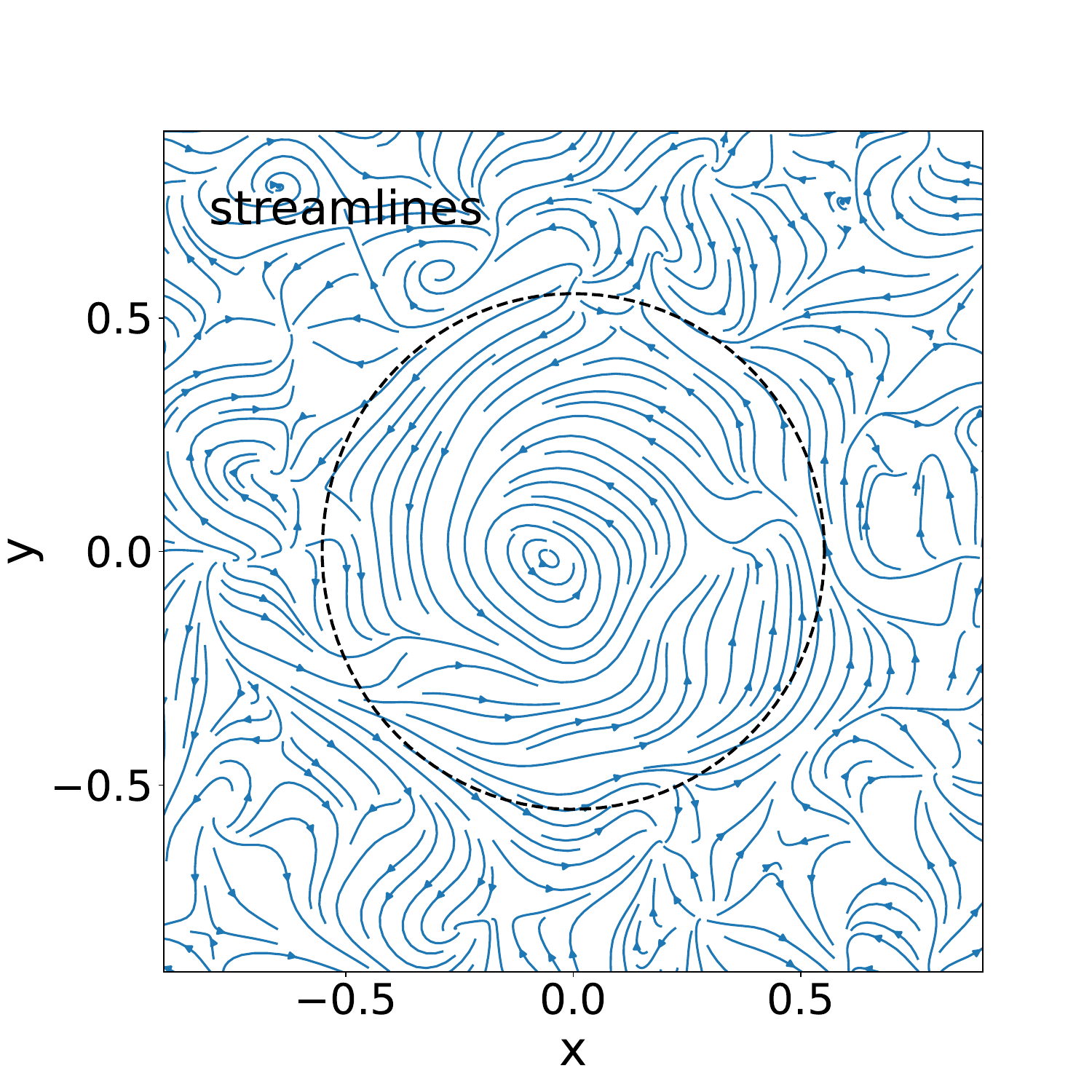}
\includegraphics[height=4.8cm,width=0.33\textwidth]{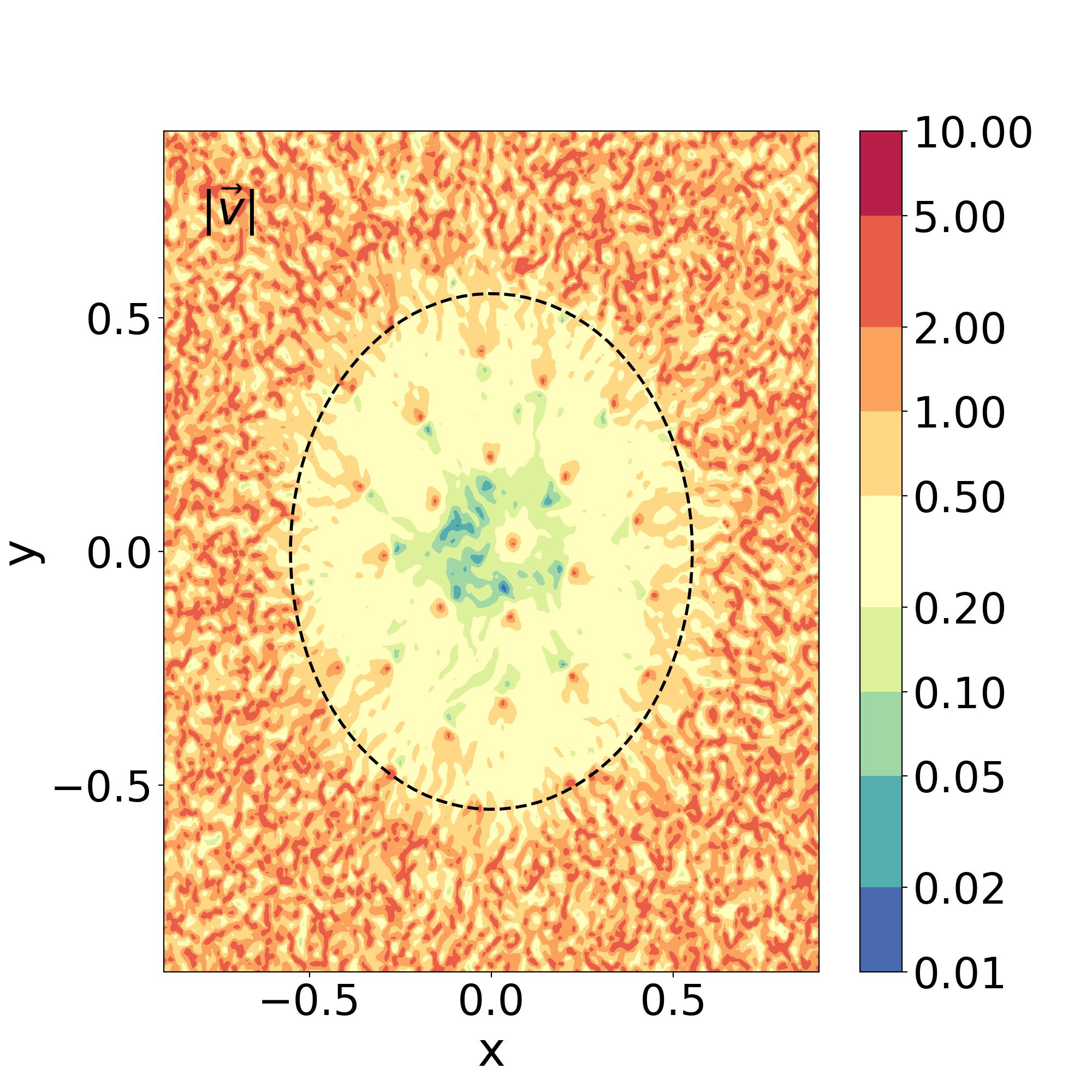}\\
\includegraphics[height=4.8cm,width=0.27\textwidth]{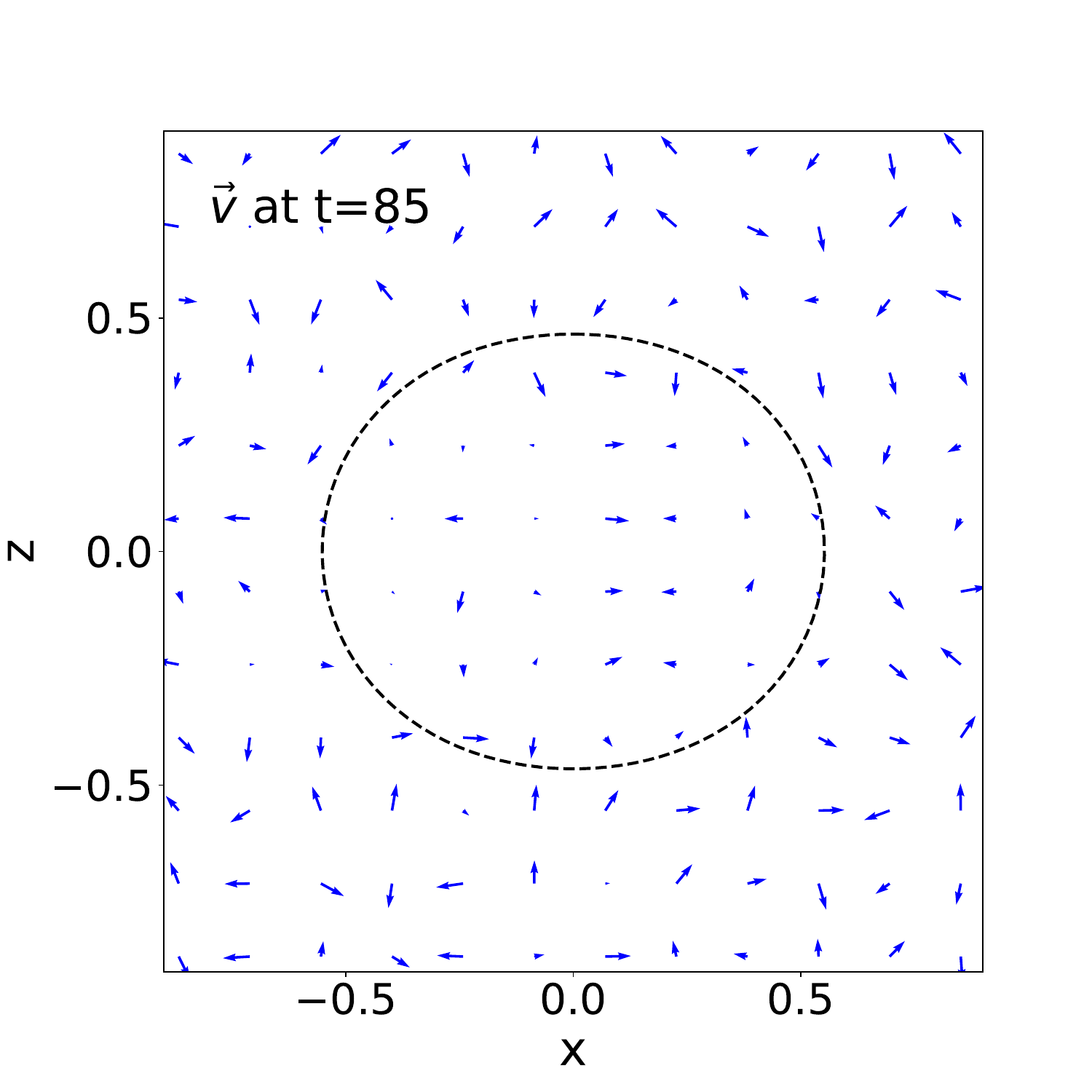}
\includegraphics[height=4.8cm,width=0.27\textwidth]{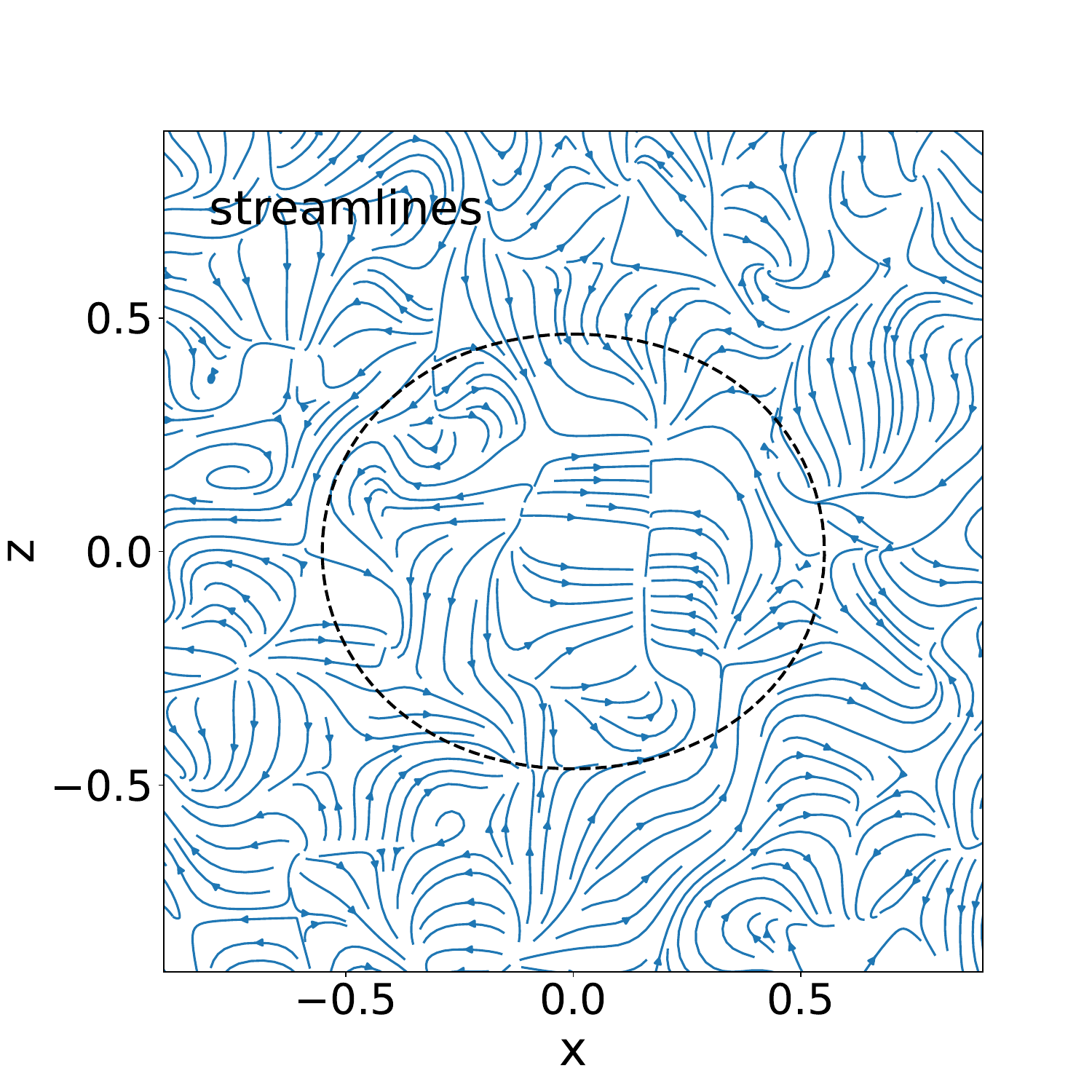}
\includegraphics[height=4.8cm,width=0.33\textwidth]{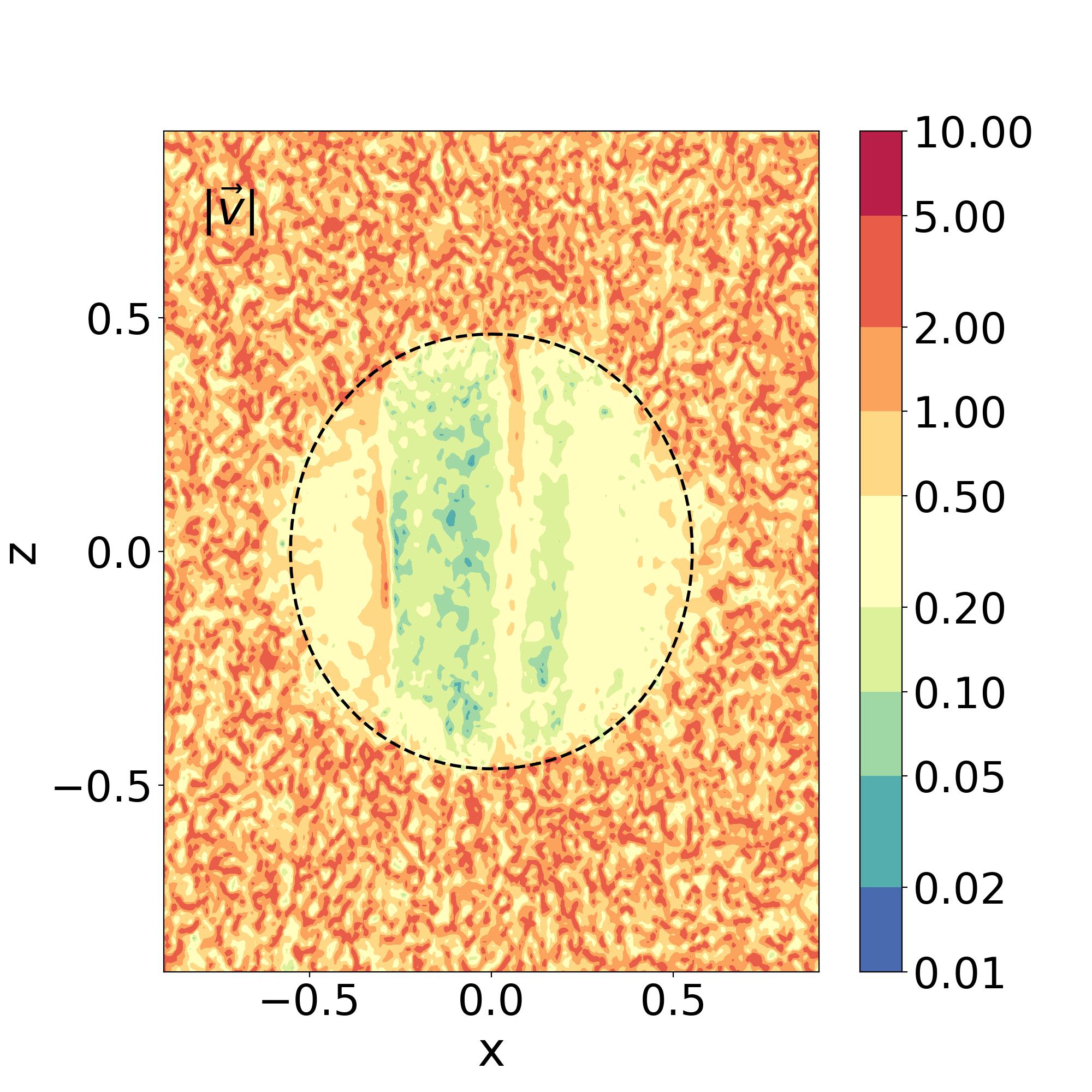}
\caption{
$[\epsilon=0.01, \alpha=1]$.
{\it Upper row, from left to right:} 2D maps in the plane $(x,y)$, at $t=85$, of the
normalized velocity field $\vec v/|\vec v|$ projected onto the plane $(x,y)$,
the streamlines of this projected velocity field,
and the amplitude $|\vec v(\vec r)|$ of the 3D velocity.
{\it Lower row:} 2D maps in the plane $(x,z)$.
}
\label{fig:2D-v-mu1-0p01}
\end{figure*}

As explained in Sec.~\ref{sec:rotating-soliton}, this solid-body rotation arises as the minimum
of the energy at fixed mass and angular momentum. Thus, this is the analog of the static soliton 
profile (20) with vanishing velocity obtained for a system at constant mass and zero angular momentum.
Therefore the mechanism leading to this configuration is the same as that for the usual static
soliton. In turn, this solid-body rotation implies a uniform density of vortex lines, as shown
in Sec.~\ref{sec:uniform-lattice} in the continuum limit.
For a finite number of vortices, this corresponds to a regular lattice with roughly equidistant
spacing between neighbouring vortices.
Thus, it is easier to understand this process in terms of the relaxation of the velocity field,
within the hydrodynamical picture associated with the continuum limit.
Then, for a nonzero $\epsilon$, that is, a finite number of vortices, the vortex lattice is simply
dictated by this equilibrium velocity profile. 
In principle, one could directly look for the lattice that minimizes the energy at fixed
angular momentum, but this is a much more difficult discrete problem.
In some systems, such as superconductors or superfluids, it is possible to show that the
lowest-energy lattices are square or triangular \cite{Abrikosov:1956sx,Tkachenko1966}.
They are known as Abrikosov lattices in the condensed matter context.
In this paper we do not investigate the detailed shape of the lattice cells as we find that the
location of the vortices is not perfectly stationary because of the perturbations
due to the wave packets passing by from the outer envelope and the fluctuating forces
from the other vortices.
However, in very well-relaxed solitons it could be interesting to study the exact shape of these lattices
and their oscillation modes.

The characteristic radial branching pattern found in the upper middle panel in 
Fig.~\ref{fig:2D-rho-mu1-0p01} is also due to the solid-body rotation. Indeed, 
to recover $\vec v = \epsilon \vec \nabla S$ in the continuum limit
with $\vec v = v_\varphi \vec e_{\varphi}$, the phase $S$ must be independent
of $r$ and $z$. 
The independence on $z$ is clearly seen in the lower middle panel and the independence on $r$
in the upper middle panel, with mostly constant-phase lines or sectors that emanate from the center
of the soliton. However, a phase $S$ that only depends on the azimuthal angle $\varphi$ would
give a decaying velocity $v_\varphi = \epsilon S'(\varphi)/r$ instead of the linear growth
$v_\varphi = r \Omega$ with radius $r$ associated with solid-body rotation.
The contradiction is resolved by the singularities due to the vortex lines, which generate new radial 
branch cuts (where the phase jumps from $\pi$ to $-\pi$) as we move towards larger radii.
In this manner, within an annulus of width $\Delta r$ around radius $r$, the phase only depends
on $\varphi$, leading to $\vec v = v_\varphi \vec e_{\varphi}$, but as we move to larger distance
the winding number $\Delta S/(2\pi)$ grows as $r^2$, that is, the number of cuts or ``branches''
grows as $r^2$.

The lower row in Fig.~\ref{fig:2D-rho-mu1-0p01} shows that the vertical $(x,z)$ plane exhibits
very different features.
The soliton boundary now displays an oblate shape and agrees with the prediction (\ref{eq:Rsol-res}).
Because the vortex lines are roughly vertical, aligned with the rotation axis $\vec\Omega$,
we no longer see a regular lattice of density troughs inside the soliton in the lower left panel.
Indeed, the vortex lines are aligned with the vertical axis and parallel to the $(x,z)$ plane,
which generically they do not cross.
We mostly see the smooth rise towards the center of the soliton density profile.
Nevertheless, we can see the vertical trace of two vortex lines that happen to be close to the
$(x,z)$ plane, shown by two vertical density troughs.
In particular, we recover the vortex line at $\vec r_{\perp} \simeq (-0.3,0)$ that was already
found in Figs.~\ref{fig:evol-mu1-0p01} and \ref{fig:v-mu1-0p01}.

The vertical structure is clearly apparent in the lower middle panel: the phase is mostly smooth
inside the soliton with vertical bands associated with the rotation axis of the system.
Note that the vertical lines where the phase jumps from $\pi$ to $-\pi$ are not singular lines.
These jumps are merely due to our definition of the phase in the interval $]-\pi,\pi]$.
Within a strip around such a vertical line, we can locally redefine the phase to the interval
$[0,2\pi[$ (i.e., we add a factor $2\pi$ to the phase if it is negative). This gives a strip where
the phase is regular, with small fluctuations around $\pi$ and small velocities.
In contrast, in the outer halo the $(x,z)$ plane looks similar to the $(x,y)$ plane, as it is dominated
by the interferences between uncorrelated modes that lead to a proliferation of vortex lines
of either sign. They are not related to the global angular momentum of the system, which is hidden
by these many incoherent vortices.

The lower right panel, where we display the winding number through the $(x,z)$ plane, also shows the
dichotomy between the outer halo, where we can see the signature of the many vortex lines of random
direction that cross any plane with equal probability, and the soliton which is mostly devoid of vortex
crossings.
The few points inside the soliton are again the traces of the two vertical vortex lines at
$\vec r_{\perp} \simeq (-0.3,0)$ and $\vec r_{\perp} \simeq (0.1,0)$ found in the density map
in the lower left panel. Indeed, because the vortex lines are not perfectly vertical they may happen
to cross the $(x,z)$ plane along a finite vertical interval (because of the finite resolution and of the
fluctuations and bending of the vortex lines).

\subsubsection{2D velocity maps}

\begin{figure}
\centering
\includegraphics[height=5cm,width=0.35\textwidth]{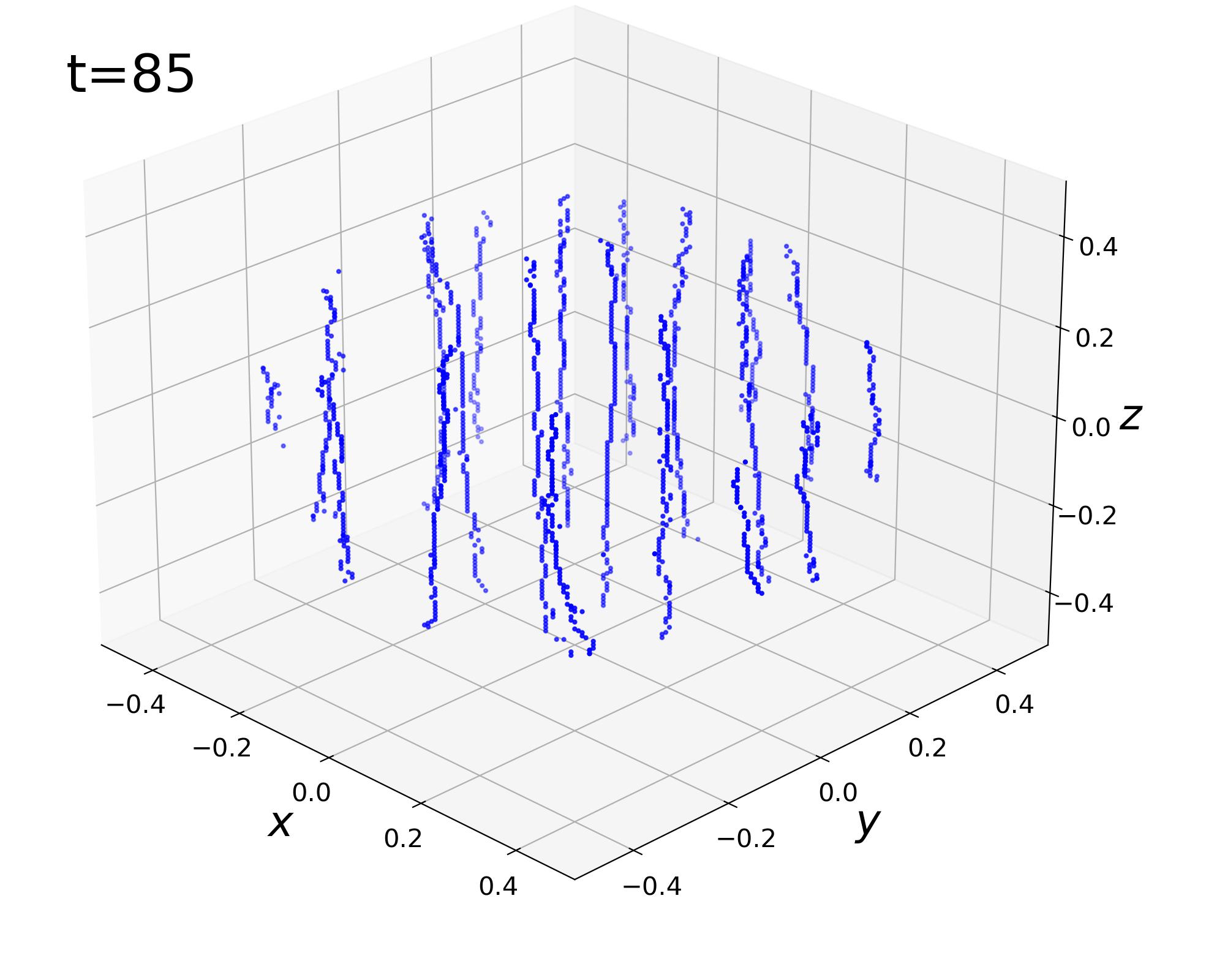}\\
\includegraphics[height=4.5cm,width=0.237\textwidth]{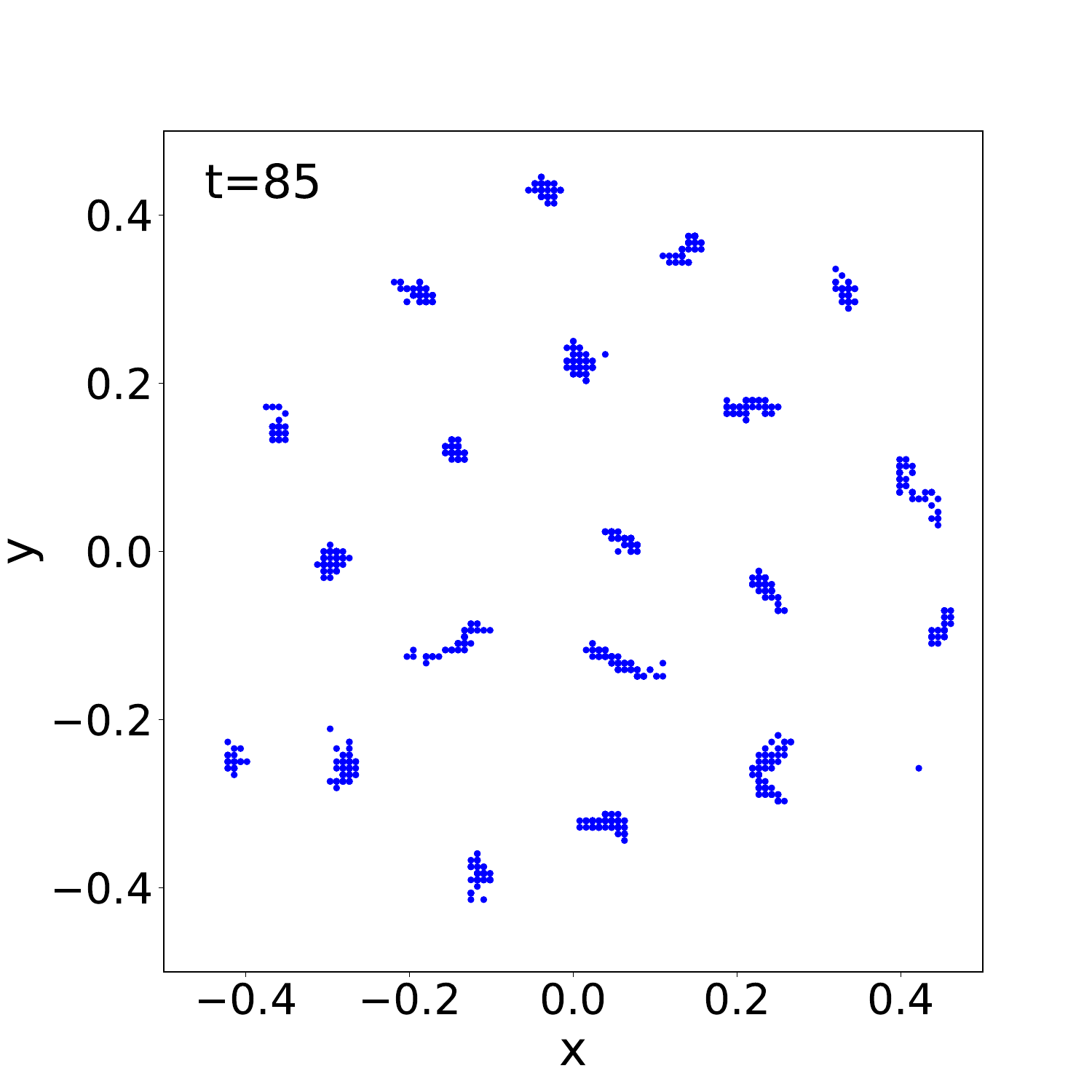}
\includegraphics[height=4.5cm,width=0.237\textwidth]{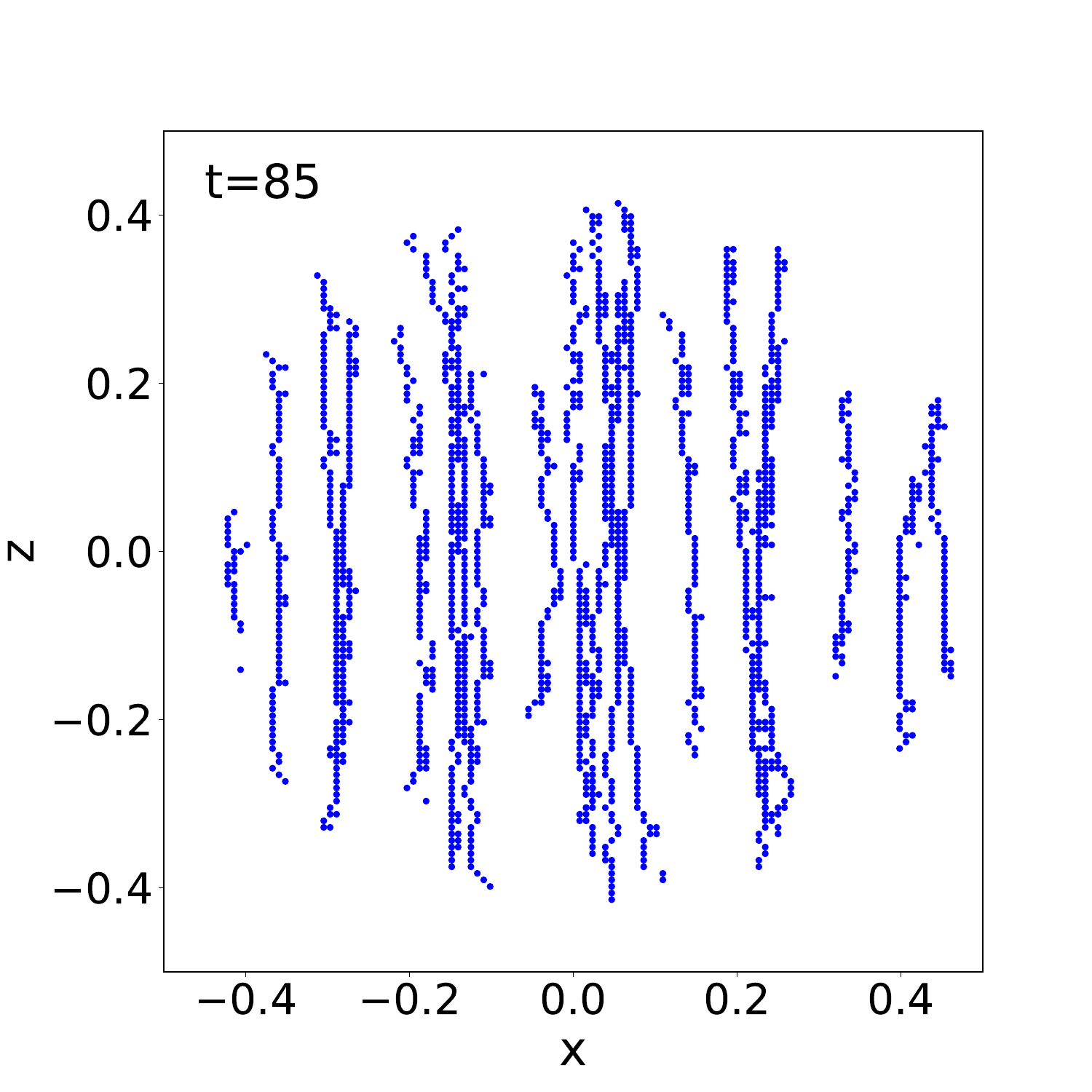}
\caption{
$[\epsilon=0.01, \alpha=1]$.
{\it Upper panel:} singularities of the phase (i.e., vortex lines) inside the soliton
at $t=85$.
{\it Lower left panel:} projection of these vortex lines onto the plane $(x,y)$.
{\it Lower right panel:} projection of these vortex lines onto the plane $(x,z)$.
}
\label{fig:3D-vortices-mu1-0p01}
\end{figure}

\begin{figure}
\centering
\includegraphics[height=4.4cm,width=0.237\textwidth]{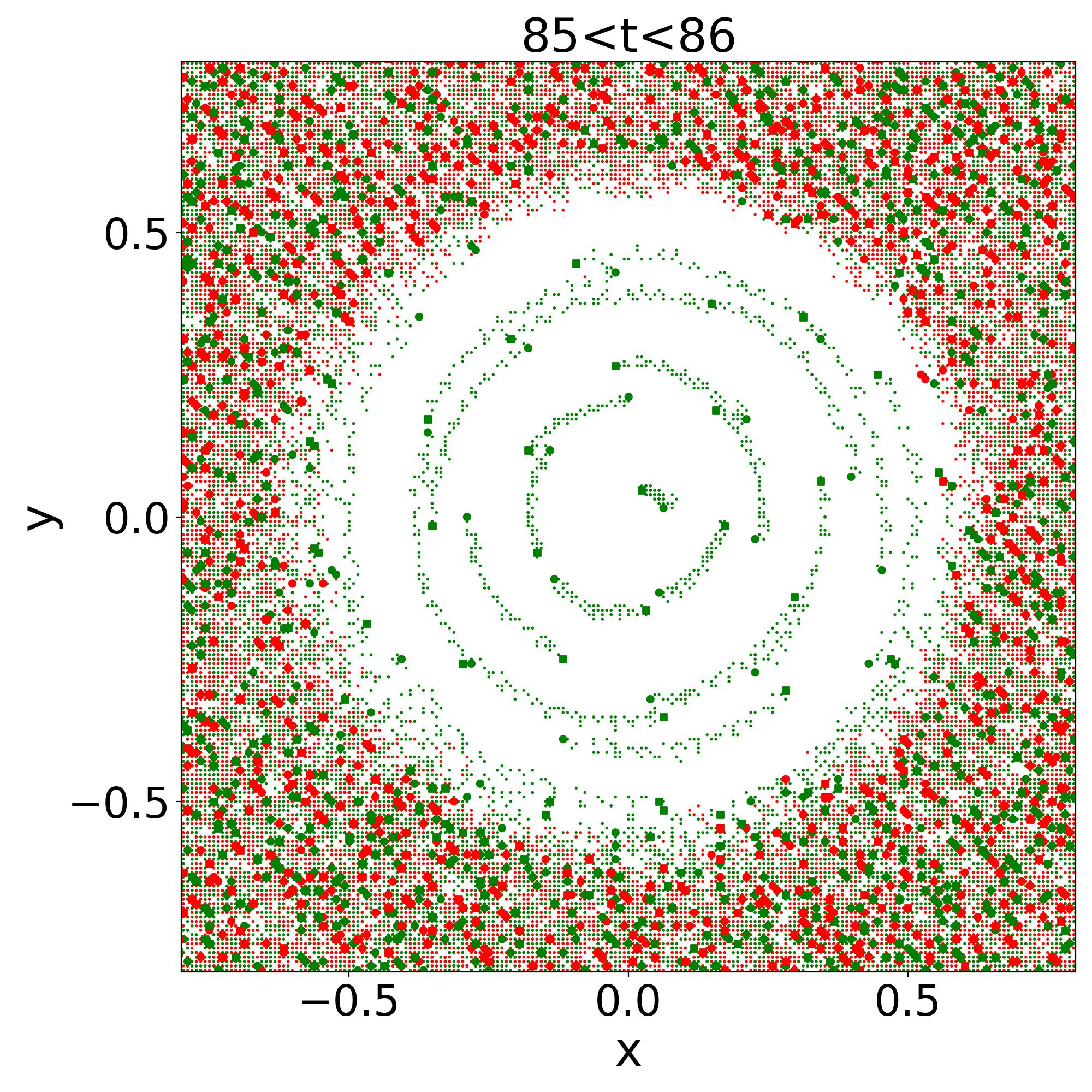}
\includegraphics[height=4.4cm,width=0.237\textwidth]{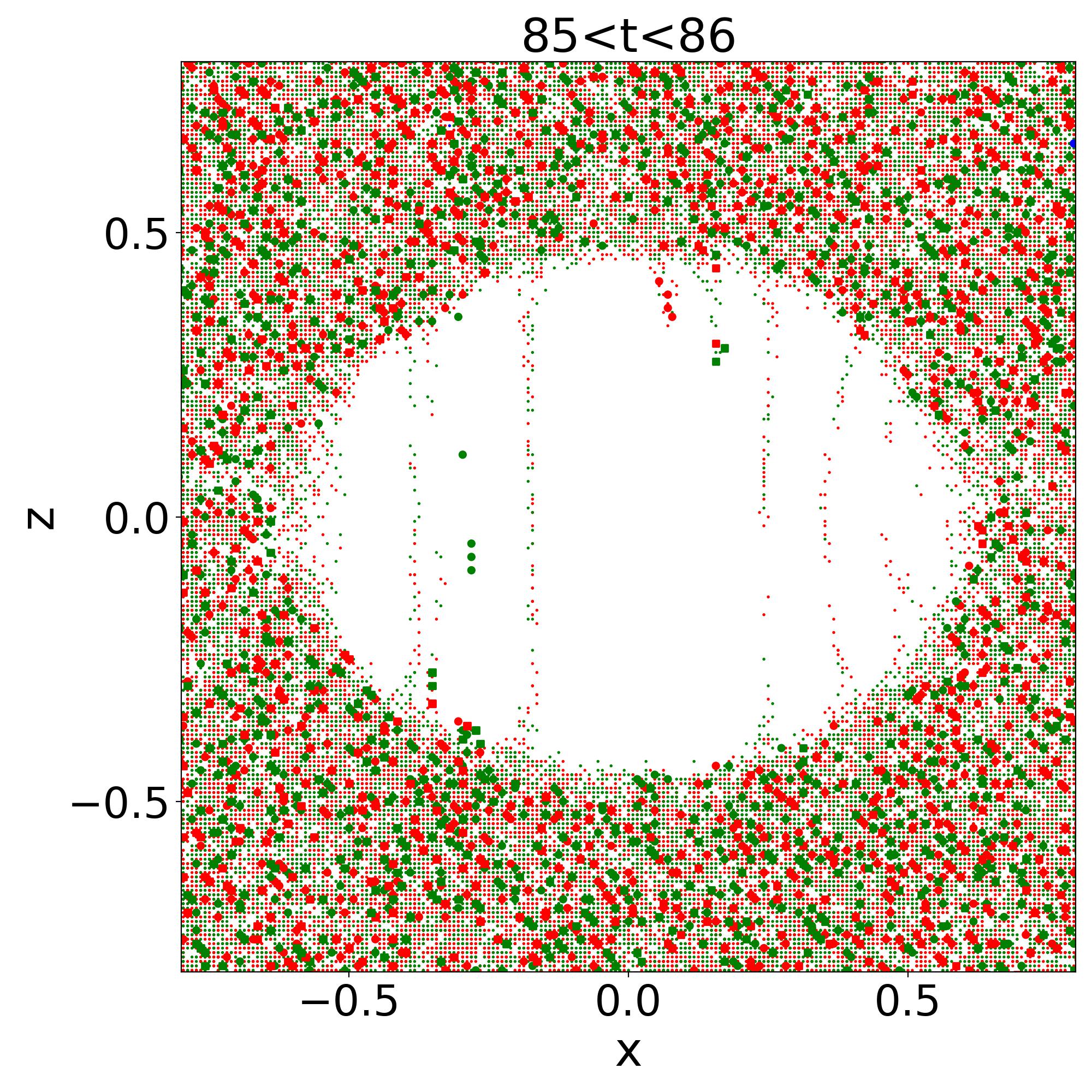}
\caption{
$[\epsilon=0.01, \alpha=1]$.
{\it Left panel:} superposition of many snapshots between times $t=85$ and $t=86$ of the maps
of the vortices in the plane $(x,y)$, as in the upper right panel in Fig.~\ref{fig:2D-rho-mu1-0p01}.
The filled circles/squares are the positions at the initial/final time.
{\it Right panel:} superposition of vortices snapshots in the plane $(x,z)$.
}
\label{fig:3D-vortex-motion-mu1-0p01}
\end{figure}

\begin{figure*}
\centering
\includegraphics[height=4.8cm,width=0.33\textwidth]{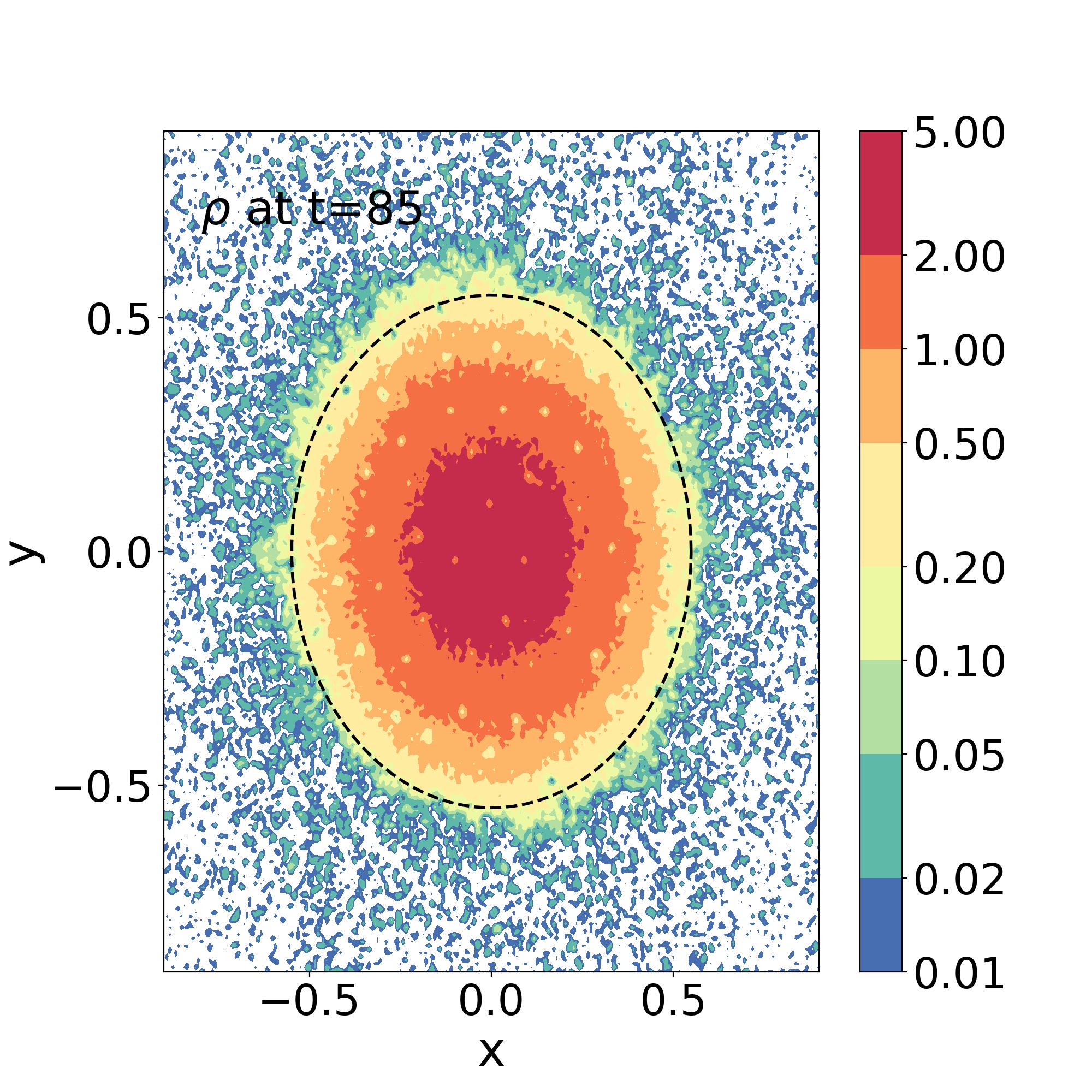}
\includegraphics[height=4.8cm,width=0.33\textwidth]{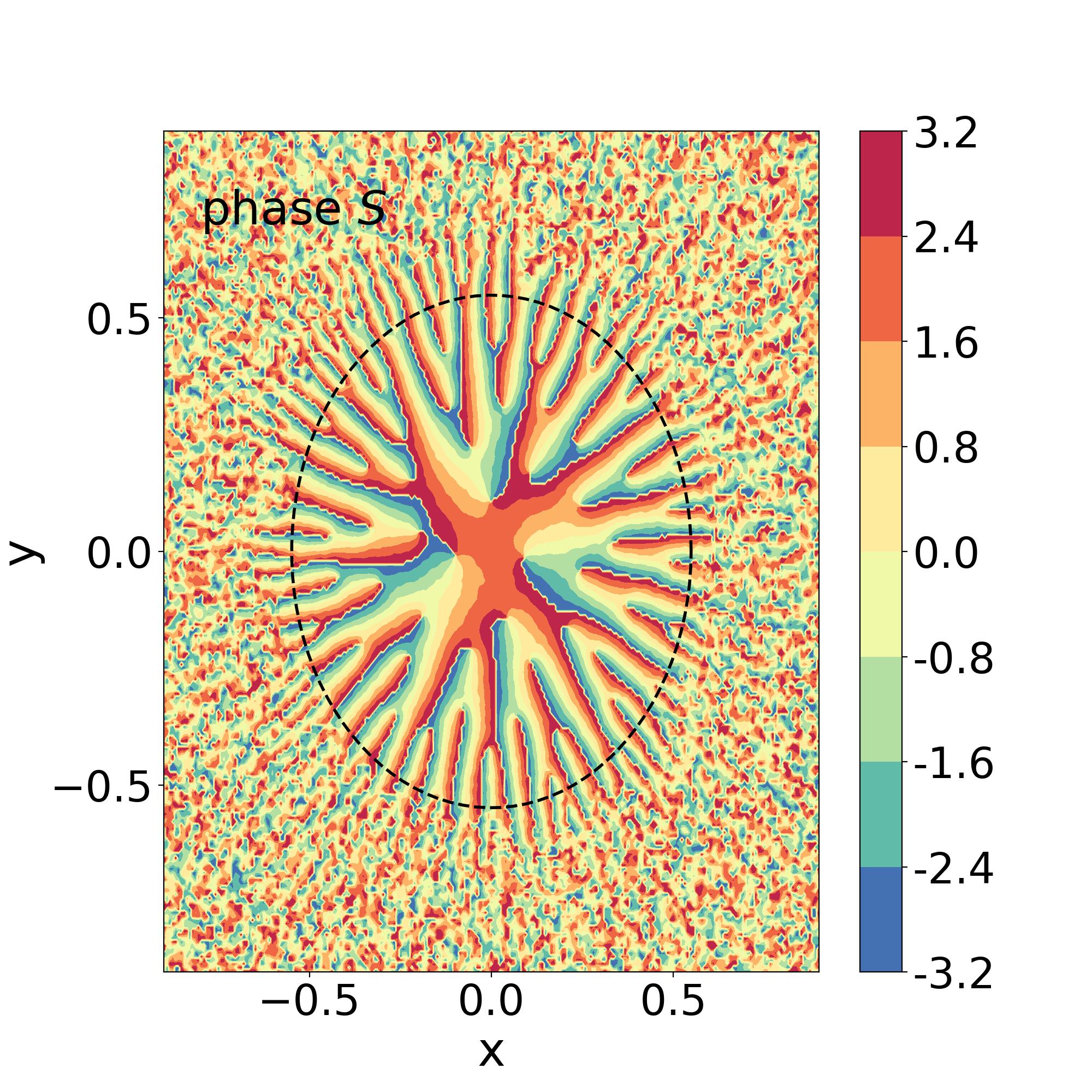}
\includegraphics[height=4.8cm,width=0.28\textwidth]{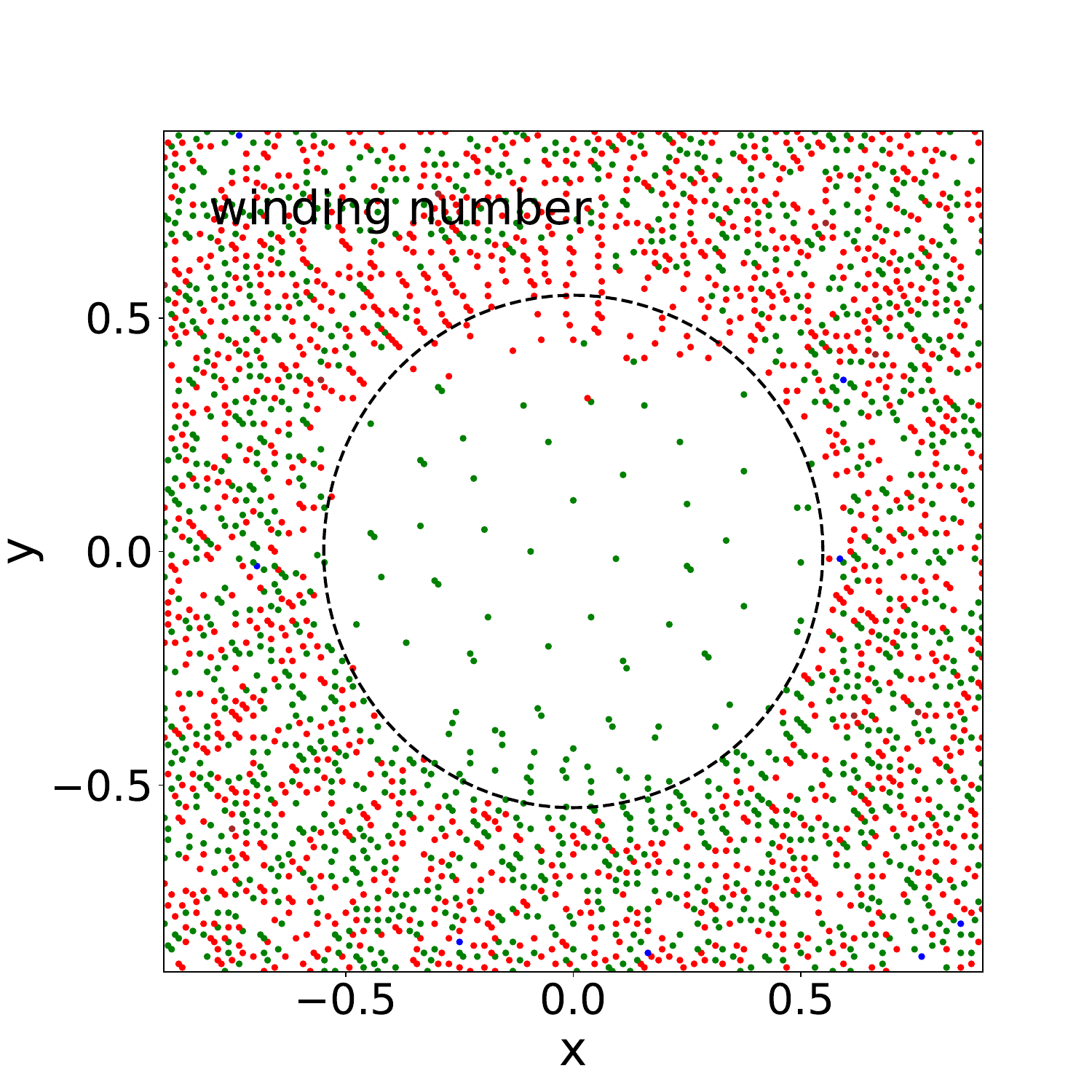}\\
\includegraphics[height=4.8cm,width=0.33\textwidth]{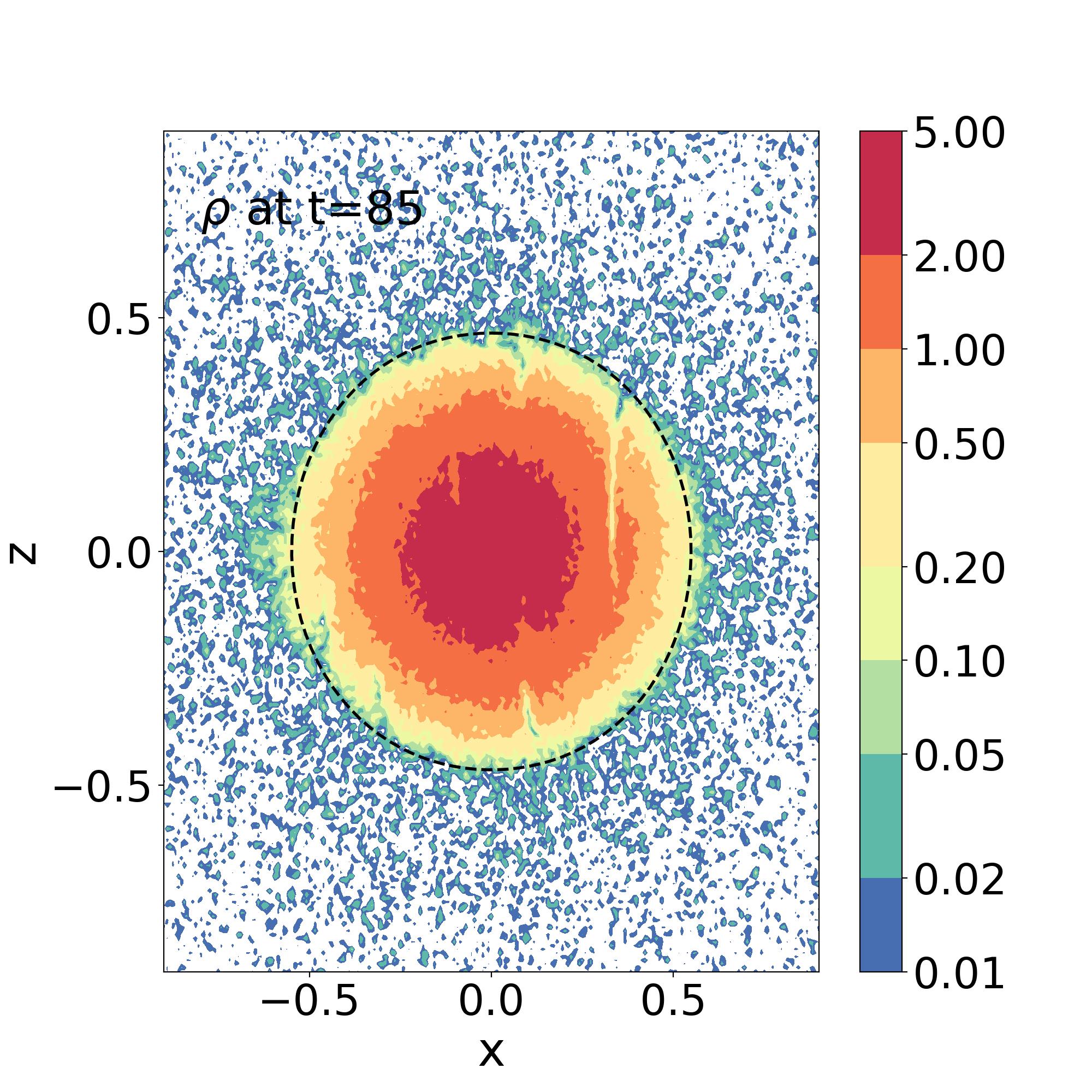}
\includegraphics[height=4.8cm,width=0.33\textwidth]{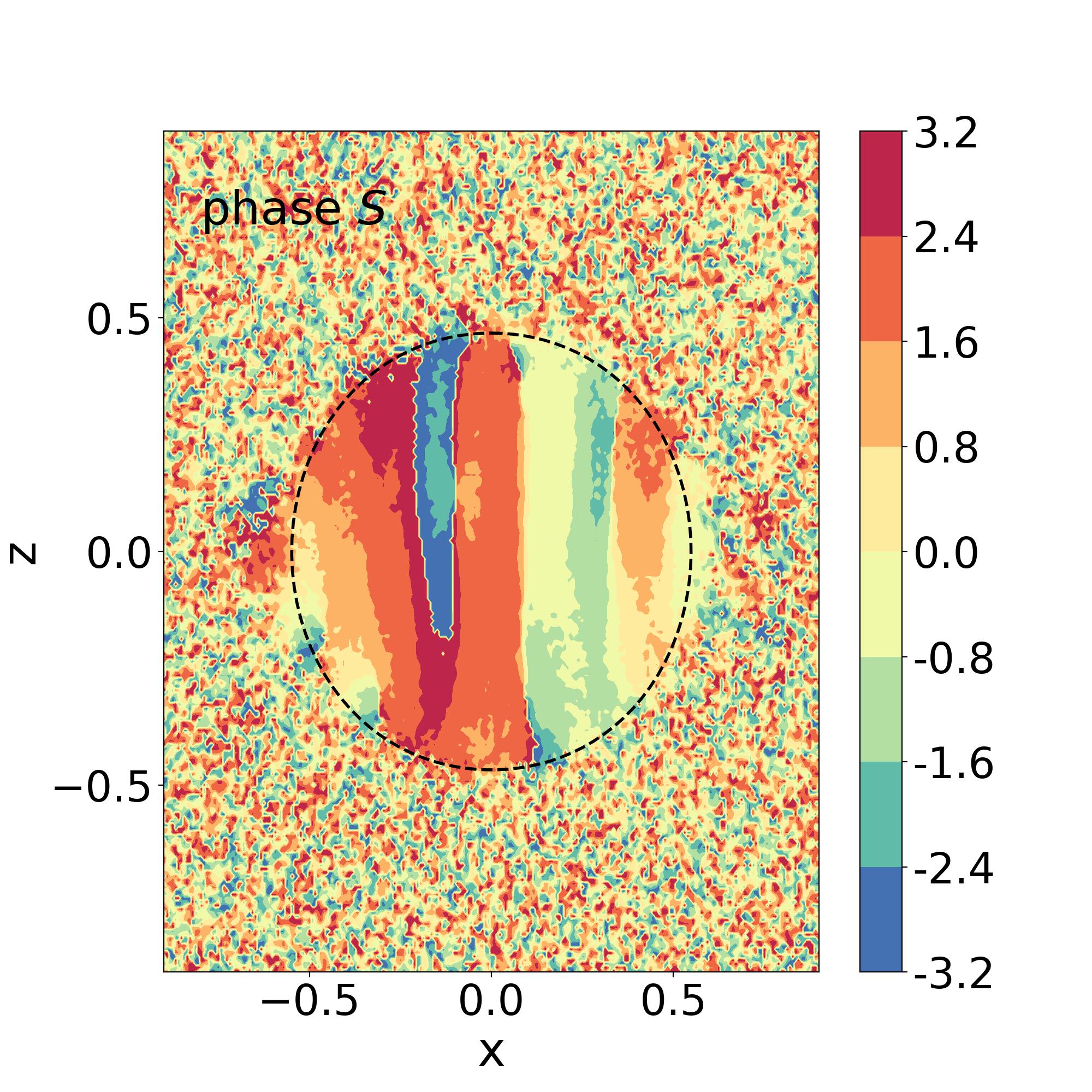}
\includegraphics[height=4.8cm,width=0.28\textwidth]{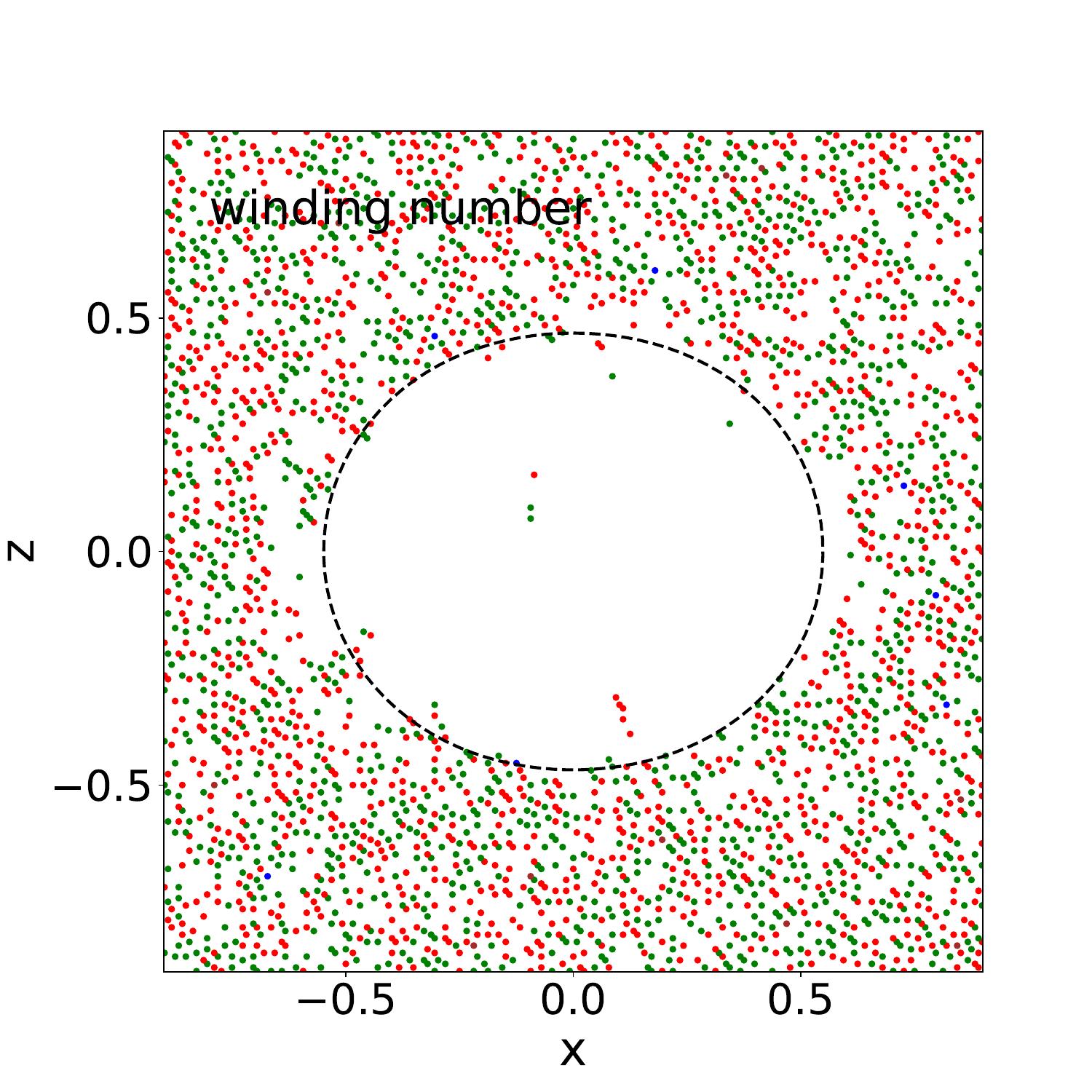}
\caption{
$[\epsilon=0.005, \alpha=1]$.
2D maps at $t=85$ of the density $\rho$, the phase $S$ and the winding number $w$,
as in Fig.~\ref{fig:2D-rho-mu1-0p01}.
}
\label{fig:2D-rho-mu1-0p005}
\end{figure*}

We show in Fig.~\ref{fig:2D-v-mu1-0p01} the velocity field in the $(x,y)$ and $(x,z)$ planes.
In the upper left and middle panels we can clearly see the rotation pattern inside the soliton
in the equatorial plane, in agreement with the prediction (\ref{eq:v-solid-rotation}).
Outside the soliton, the velocity field is disordered, as for a collisionless classical system
mostly supported by its velocity dispersion, in agreement with the results obtained in the previous
figures.
In the upper right panel, we again see the large fluctuations of the velocity on the small de Broglie scale.
Inside the soliton, we mostly find a smooth pattern with a decreasing velocity towards the center,
in agreement with the solid-body rotation (\ref{eq:v-solid-rotation}), $|\vec v| \propto r_\perp$.
On top of this regular trend we recover the lattice of vortex lines, already seen in
Fig.~\ref{fig:2D-rho-mu1-0p01}, that generates large velocity spikes in agreement with the divergence
(\ref{eq:v-single-vortex}).

In the lower row, we can see that the velocity field inside the soliton in the $(x,z)$ plane is disordered,
that is, the components $(v_x,v_z)$ are irregular. Indeed, for an exact solid-body rotation
(\ref{eq:v-solid-rotation}) we would have $v_x=v_z=0$ in the $(x,z)$ plane, so that the
nonzero values that we find in Fig.~\ref{fig:2D-v-mu1-0p01} are due to discrete effects
and perturbations from the incomplete relaxation and the outer halo modes.
Again, in the lower right panel we can see the vertical structure associated with the axis of rotation.
We mostly see the decrease of the 3D velocity amplitude towards the rotation axis,
$|\vec v| \propto r_\perp$.
Nevertheless, as in Fig.~\ref{fig:2D-rho-mu1-0p01} , we can distinguish the trace of two vortex
lines at $\vec r_{\perp} \simeq (-0.3,0)$ and $\vec r_{\perp} \simeq (0.1,0)$ that happen to be close 
to the $(x,z)$ plane. They appear as vertical lines where $|\vec v|$ is large.

\subsubsection{Vortices}

We show in Fig.~\ref{fig:3D-vortices-mu1-0p01} a 3D plot of the vortex lines inside the soliton
at $t=85$ and their projections onto the planes $(x,y)$ and $(x,z)$.
In the upper panel, we plot the 3D locations where the phase $S$ is singular, using the same
procedure as for the right panels in Fig.~\ref{fig:2D-rho-mu1-0p01}.
We restrict these points to the soliton volume, as defined by Eq.(\ref{eq:Rsol-res}) which
we multiply by a factor $0.9$, so as to focus on the interior of the soliton and to avoid
hiding the inner vortex lines by the tangle of random vortex lines in the outer halo.
We can check that the vortex lines inside the soliton are mostly aligned with the vertical axis,
up to small perturbations, with a regular spacing.
This is also clearly apparent in their projections onto the $(x,y)$ and $(x,z)$ planes
shown in the lower panels.

We show in Fig.~\ref{fig:3D-vortex-motion-mu1-0p01} the motions of the vortex lines by
stacking many snapshots of the vortices inside the $(x,y)$ and $(x,z)$ planes over the
time interval $85 \leq t \leq 86$.
The left panel shows that these vortex lines rotate around the central vertical axis, in agreement
with the solid-body rotation and the result (\ref{eq:dot-rj}), which states that the vortices follow the
velocity field, whence the solid-body rotation.
In the right panel we can see the vertical traces of vortex lines which have crossed the plane
$(x,z)$ during the time interval $85<t<86$.
Because the vortex lines are not perfectly vertical, when they move through the plane $(x,z)$ they
can generate a nonzero winding number in this plane.
As the perturbation from the vertical direction is random, these winding numbers can take either
sign, $w = \pm 1$, in contrast with the positive winding numbers $w=1$ found in the plane $(x,y)$
that are related to the sign of the angular momentum of the soliton.

\subsection{Dependence on $\epsilon$: case $\epsilon=0.005$ with $\alpha=1$}
\label{sec:eps-0p005}

\begin{figure*}
\centering
\includegraphics[height=4.8cm,width=0.27\textwidth]{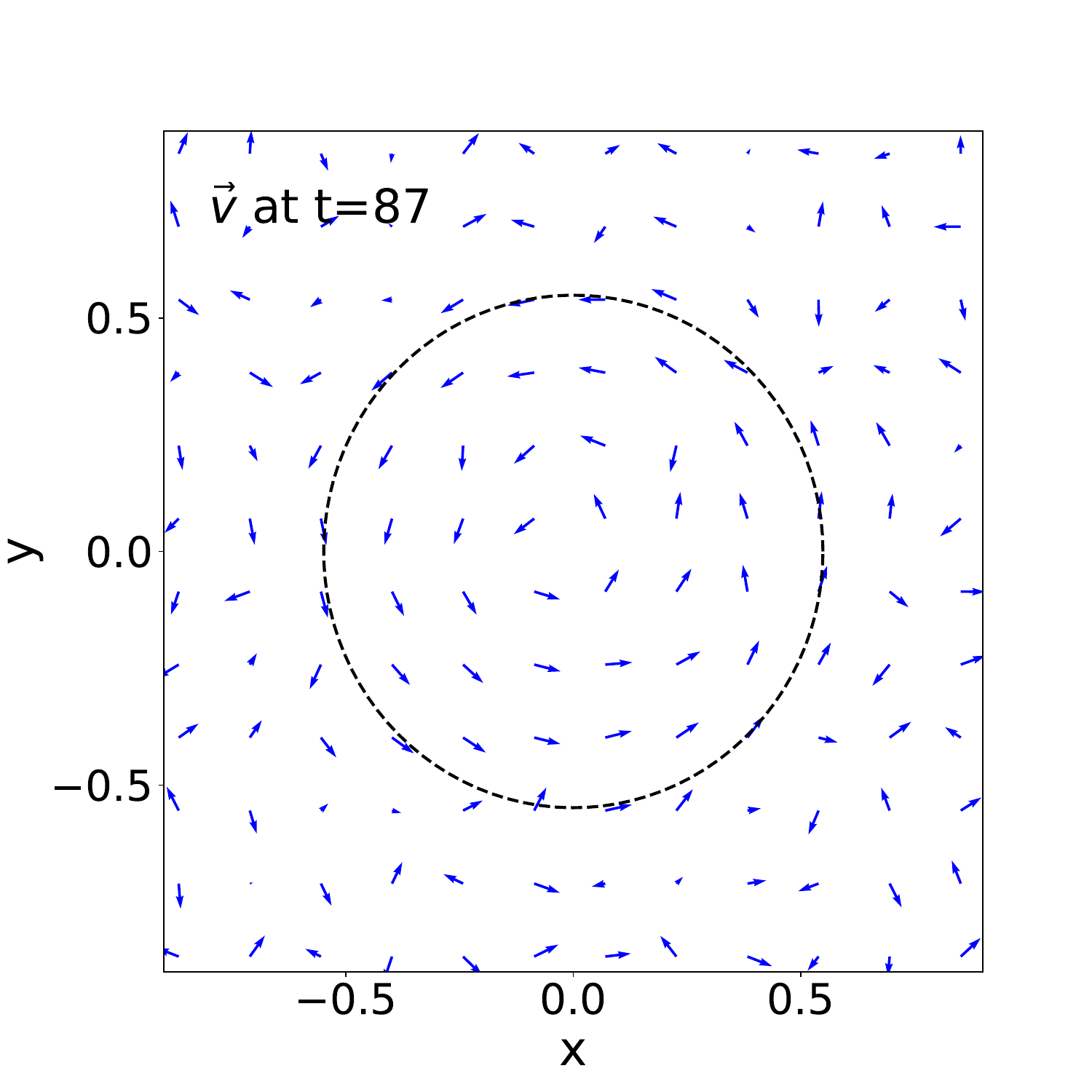}
\includegraphics[height=4.8cm,width=0.27\textwidth]{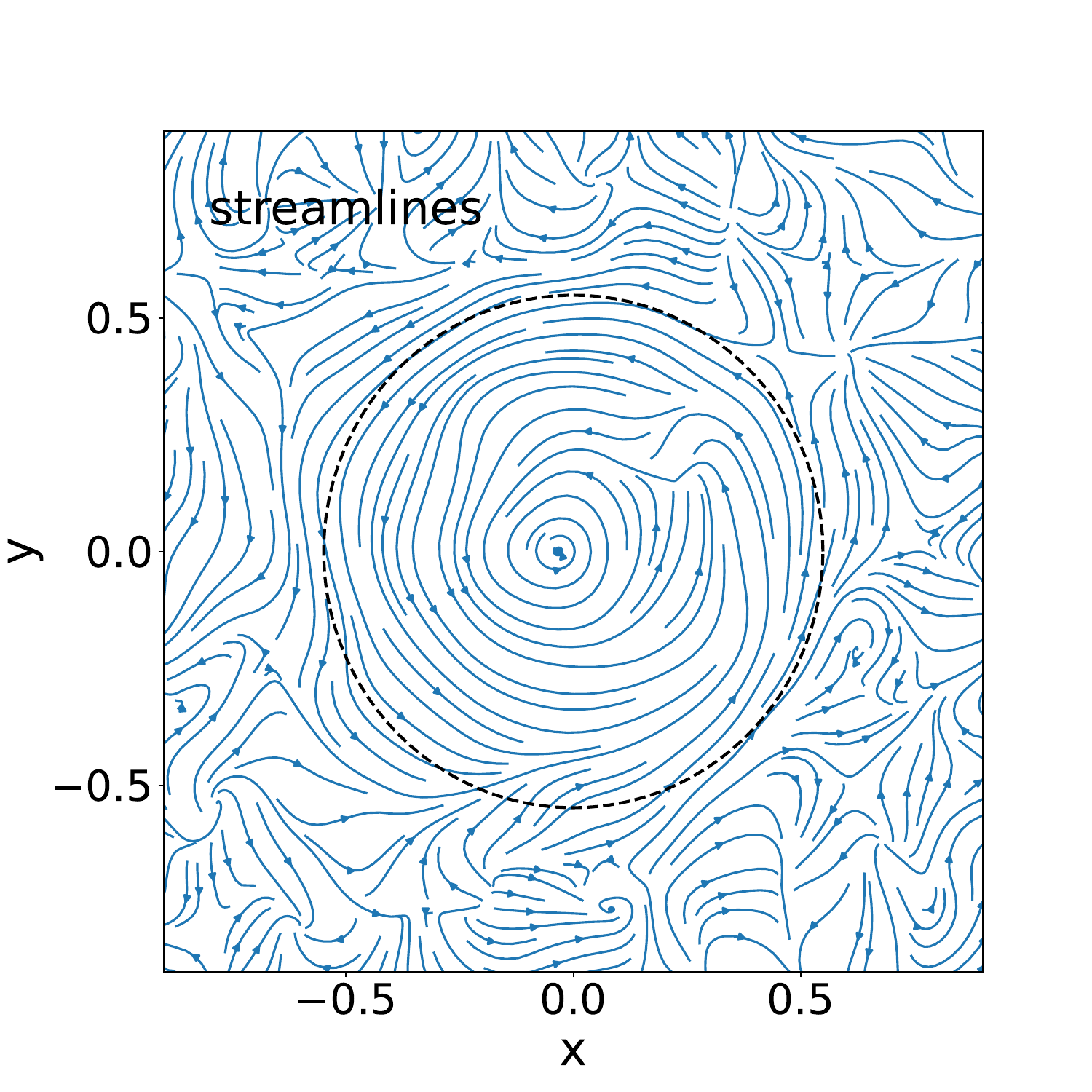}
\includegraphics[height=4.8cm,width=0.33\textwidth]{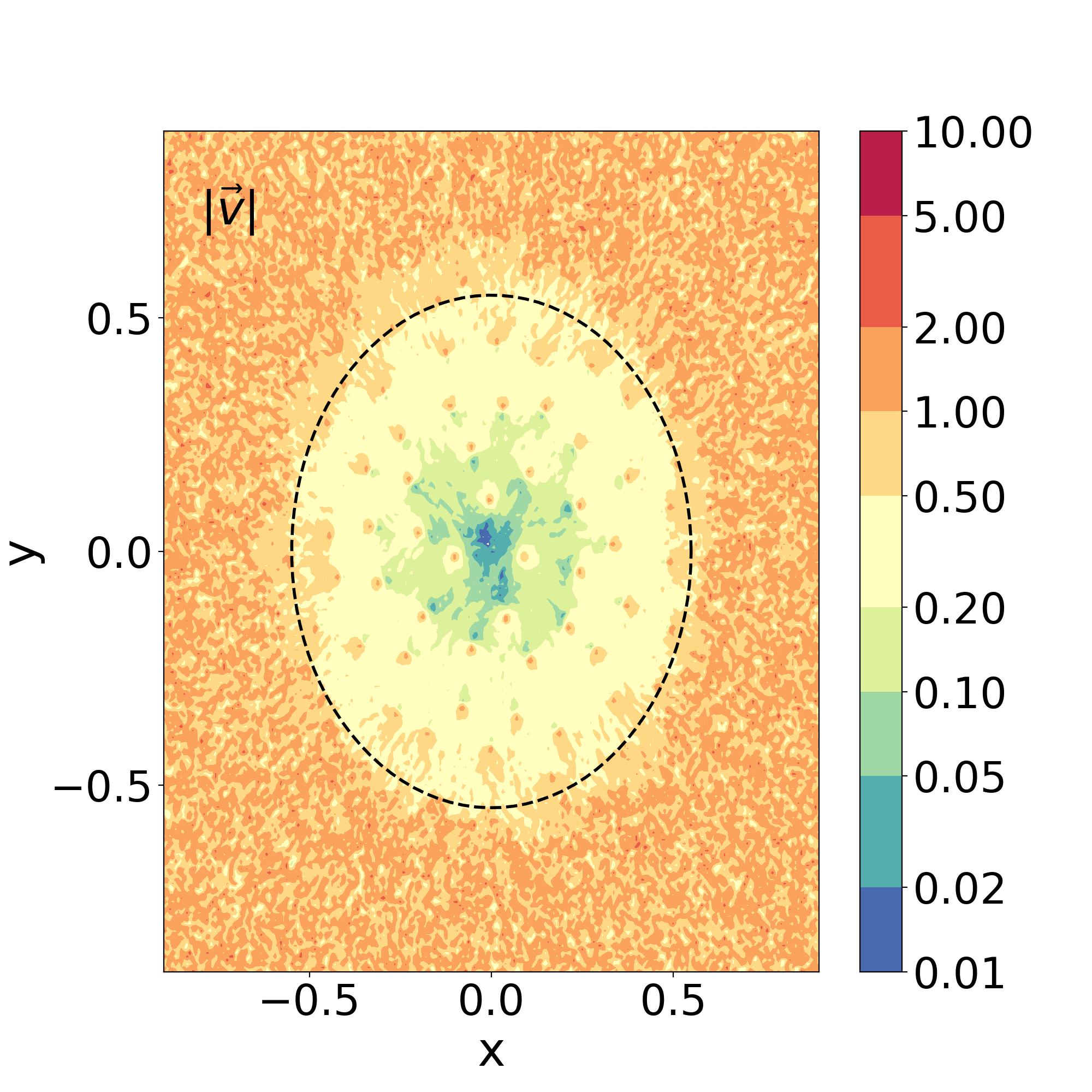}\\
\includegraphics[height=4.8cm,width=0.27\textwidth]{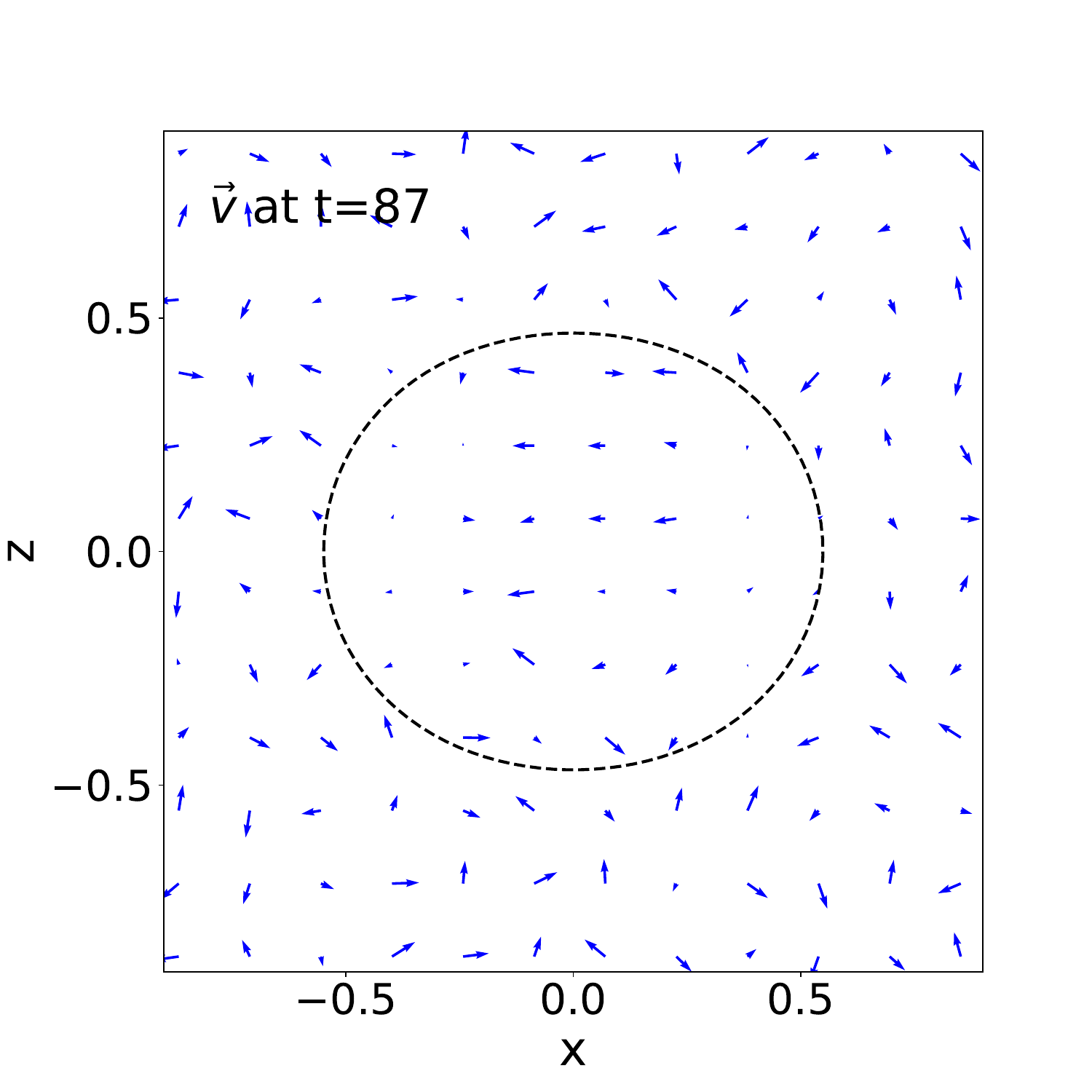}
\includegraphics[height=4.8cm,width=0.27\textwidth]{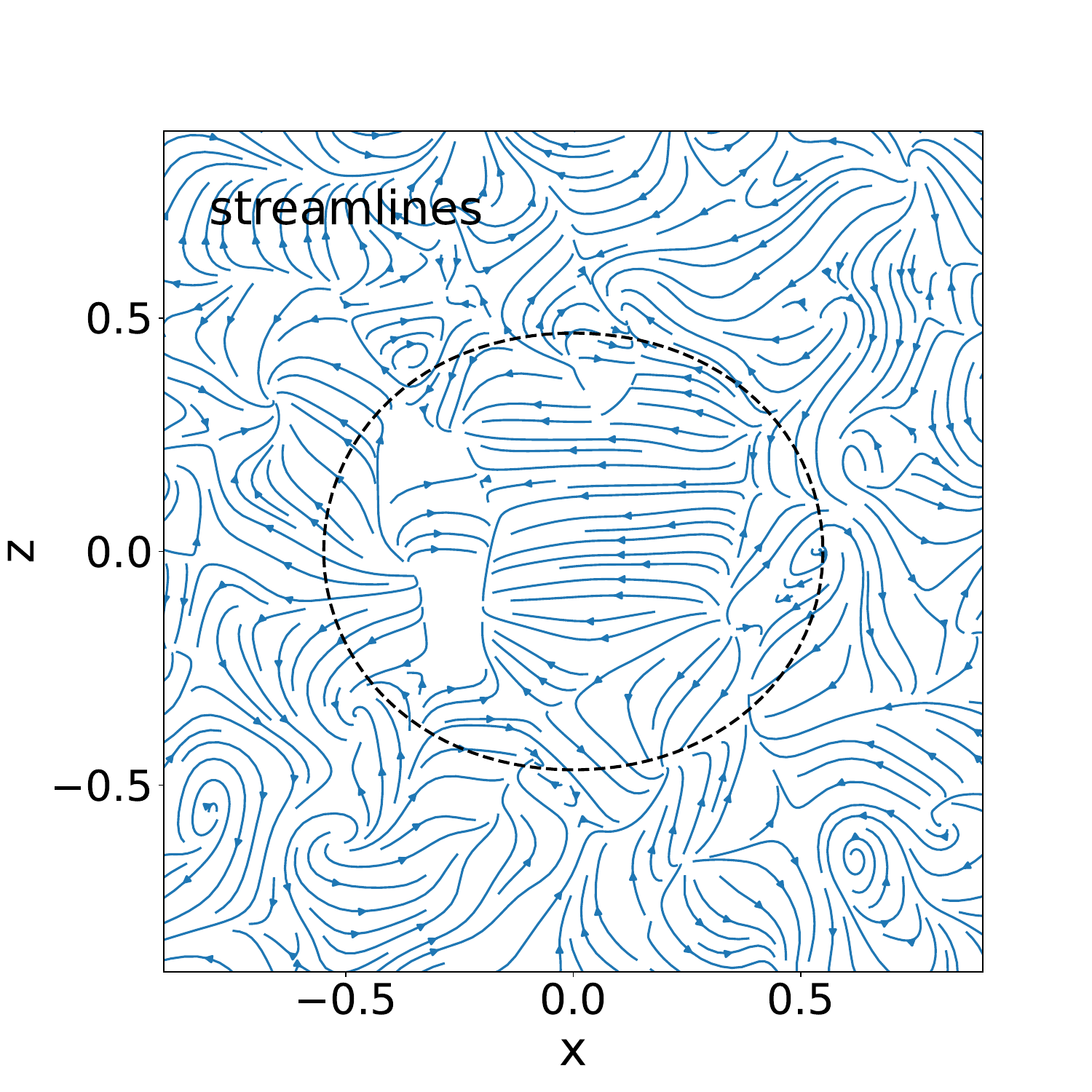}
\includegraphics[height=4.8cm,width=0.33\textwidth]{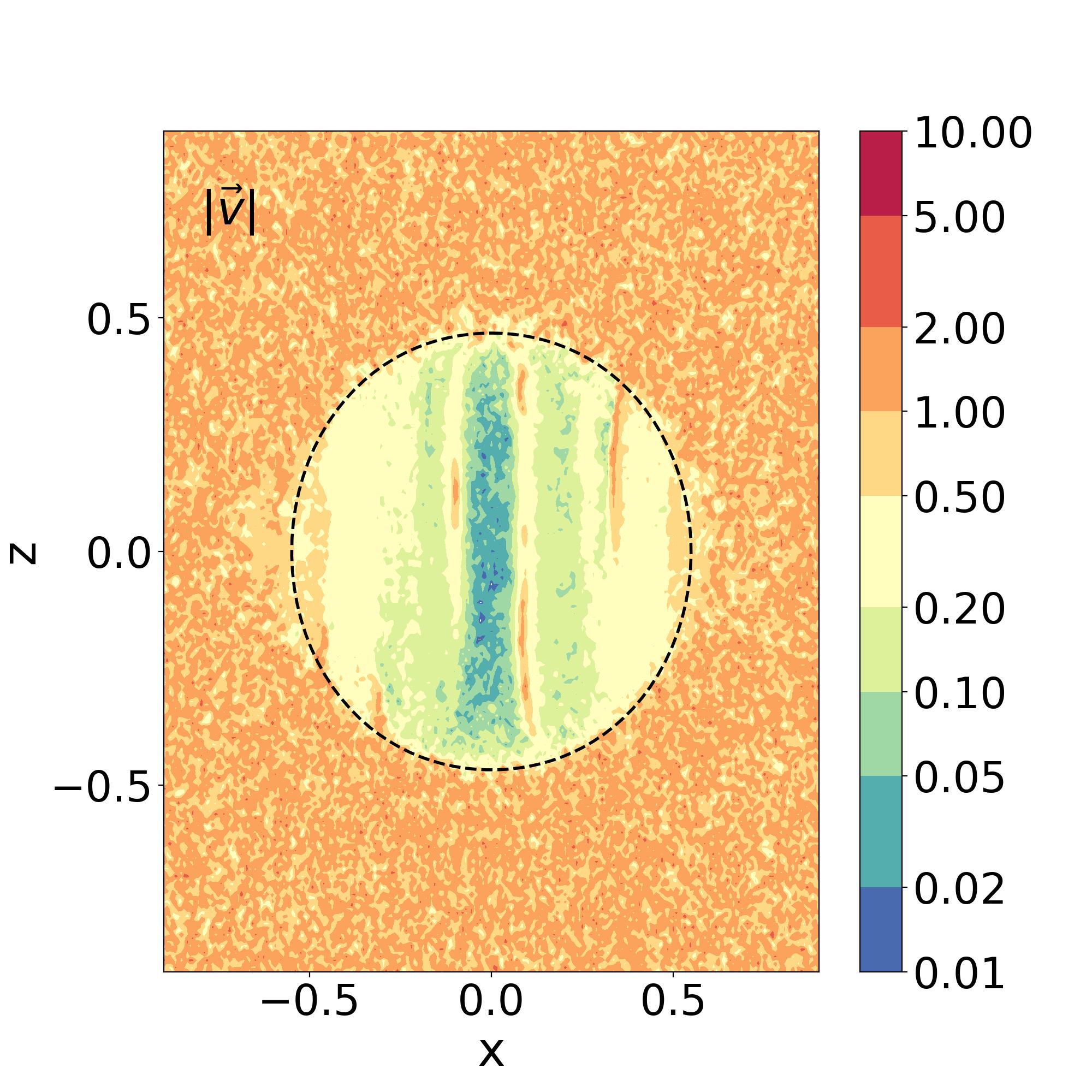}
\caption{
$[\epsilon=0.005, \alpha=1]$.
2D maps of the velocity, as in Fig.~\ref{fig:2D-v-mu1-0p01}.
}
\label{fig:2D-v-mu1-0p005}
\end{figure*}

\begin{figure*}
\centering
\includegraphics[height=5cm,width=0.35\textwidth]{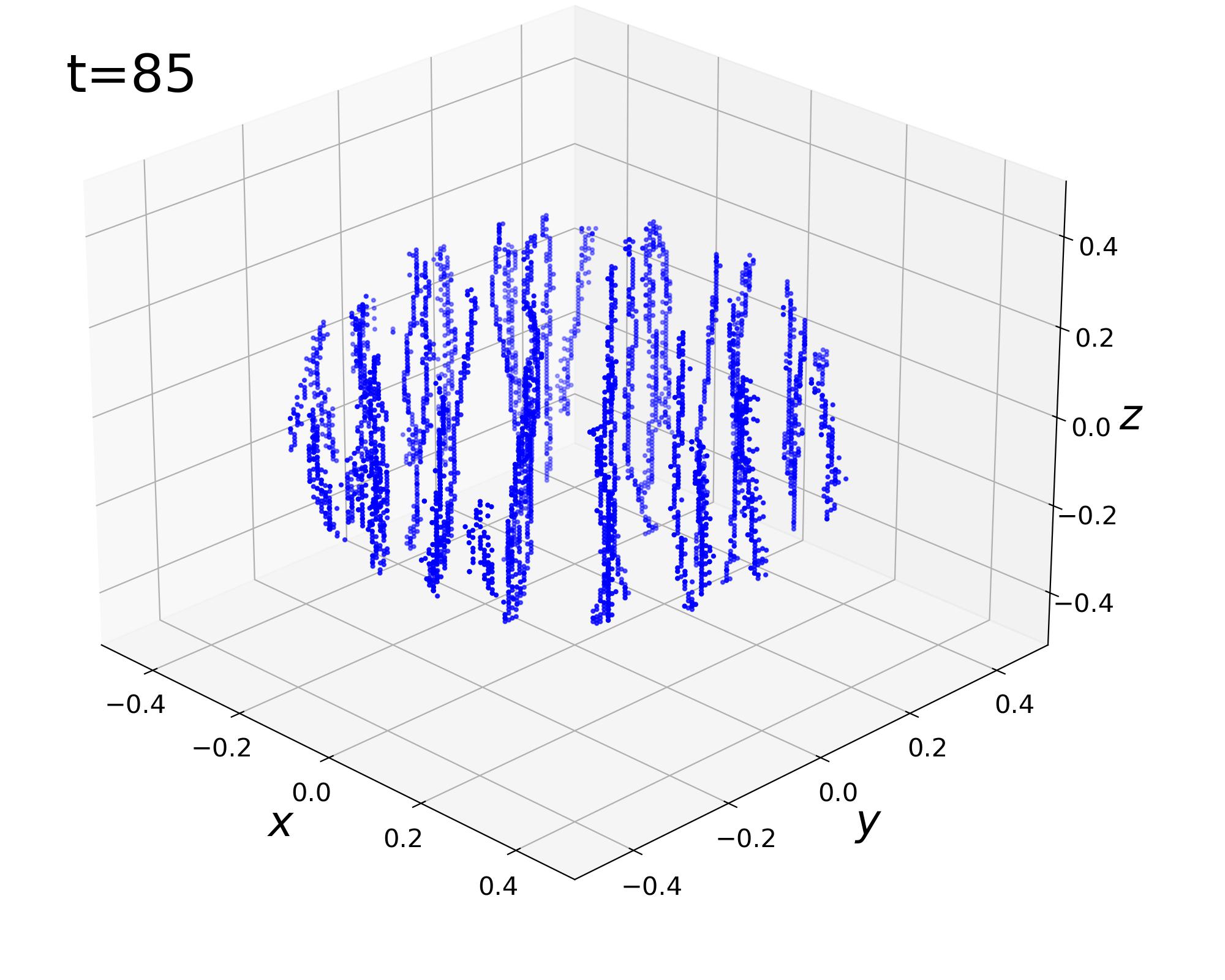}
\includegraphics[height=5cm,width=0.28\textwidth]{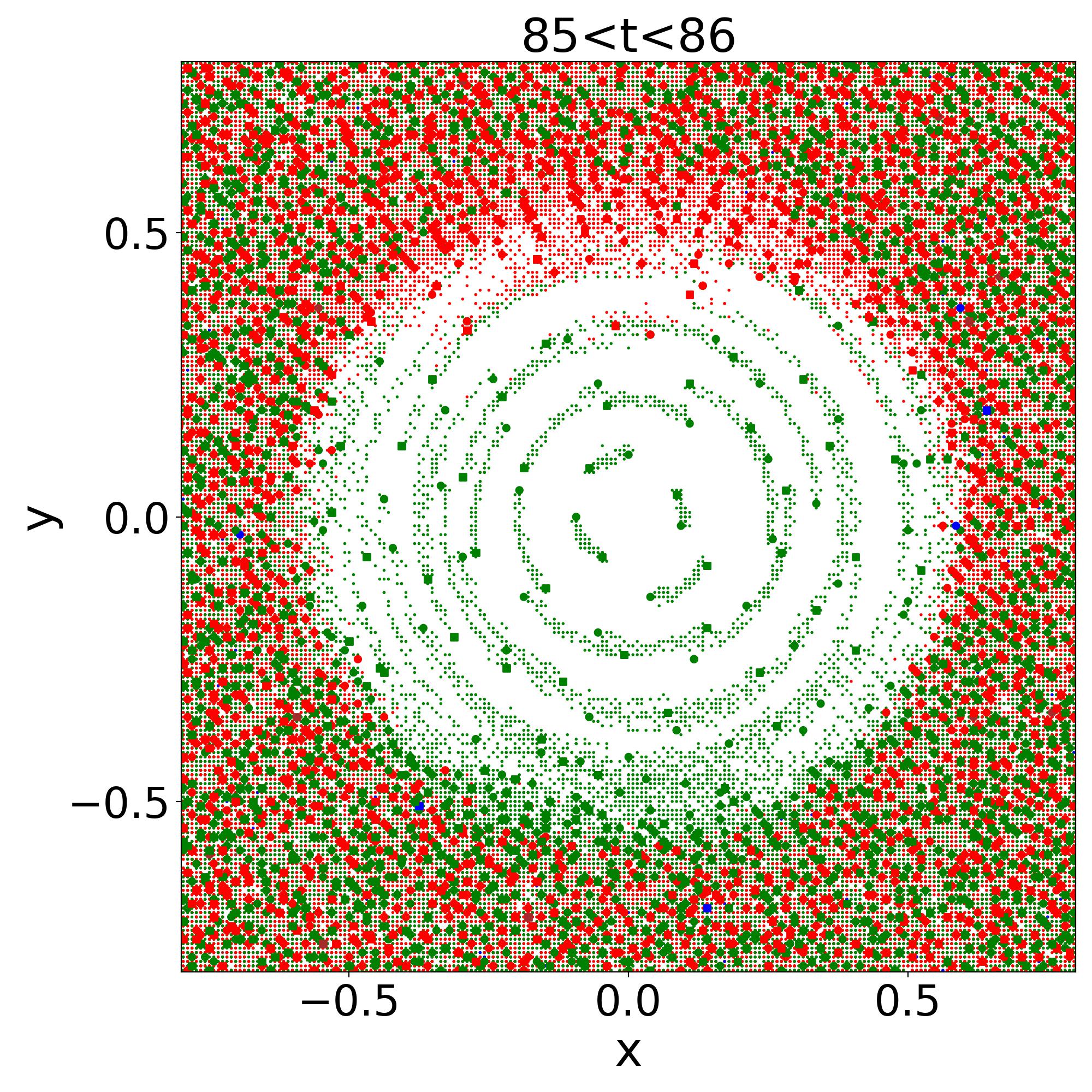}
\includegraphics[height=5cm,width=0.28\textwidth]{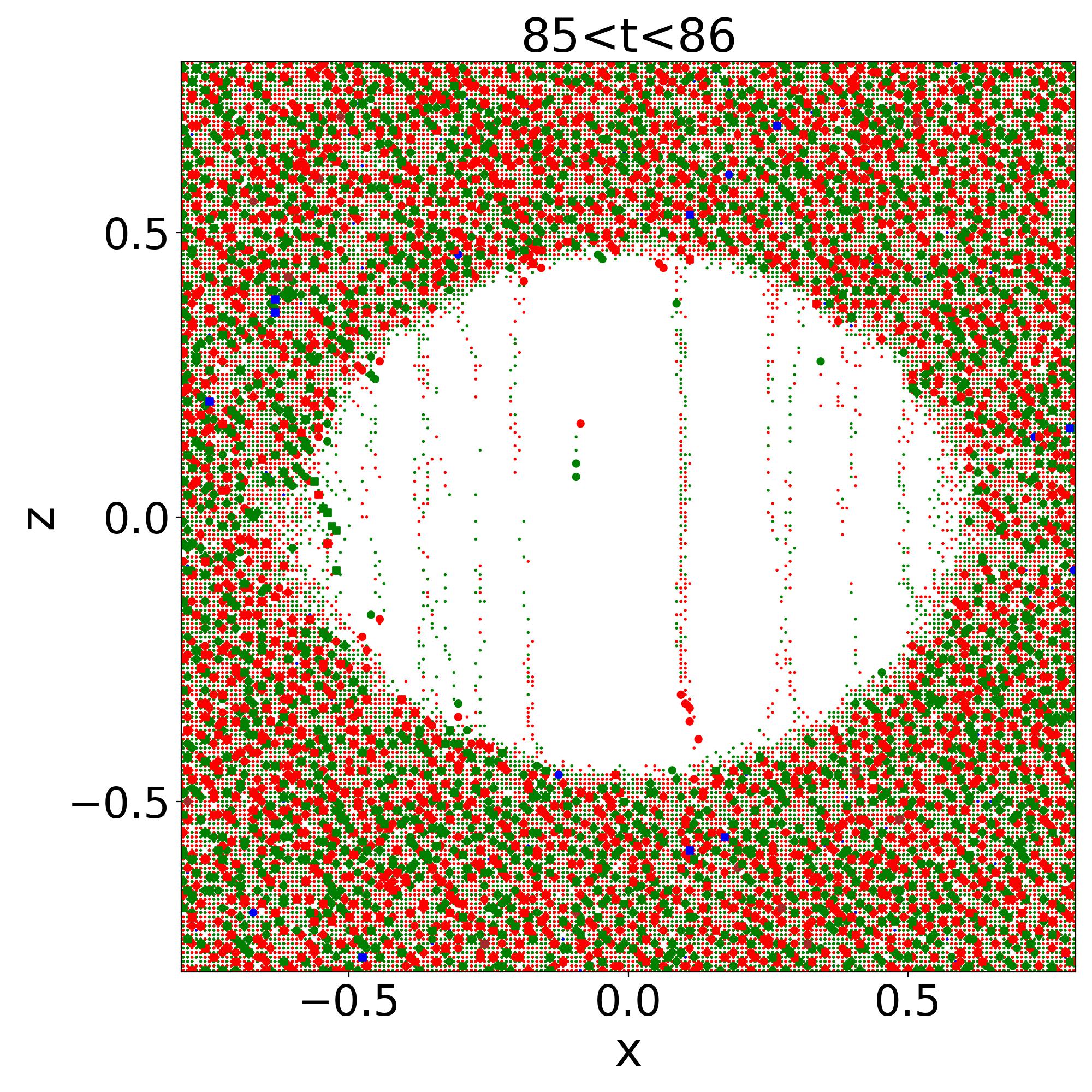}
\caption{
$[\epsilon=0.005, \alpha=1]$.
vortices as in Fig.~\ref{fig:3D-vortices-mu1-0p01}.
}
\label{fig:3D-vortices-mu1-0p005}
\end{figure*}

We now study how the properties of the system vary with $\epsilon$, by running
a simulation with a value $\epsilon=0.005$ that is twice smaller than that of the simulation
analyzed in the previous section \ref{sec:eps-0p01-alpha-1}.
The qualitative properties remain the same as for $\epsilon=0.01$ but,
in agreement with the linear scaling over $\epsilon$ of the vorticity quantum (\ref{eq:vorticity-vortex}),
the same global angular momentum of the soliton requires twice more vortices.
The de Broglie wavelength also scales linearly with $\epsilon$, which implies that random
density and phase fluctuations in the outer halo exhibit a characteristic scale that is twice smaller.
These changes are most apparent in the 2D and 3D maps. Therefore, we only show our results for the 2D
and 3D maps in Figs.~\ref{fig:2D-rho-mu1-0p005}, \ref{fig:2D-v-mu1-0p005} and
\ref{fig:3D-vortices-mu1-0p005}.

We can check in Fig.~\ref{fig:2D-rho-mu1-0p005} that the scale of the random fluctuations in the
outer halo is indeed twice smaller than in Fig.~\ref{fig:2D-rho-mu1-0p01}, while the number of vortices
in the soliton within the equatorial plane is twice greater.
Thanks to these finer details, we can see even more clearly the good agreement with the
prediction (\ref{eq:Rxy-def}) for the shape of the soliton.

The smaller value of $\epsilon$ also means that the continuum limit analyzed in Sec.~\ref{sec:continuum}
provides an even better approximation of the system.
Thus, the rotation pattern inside the soliton is even more apparent and regular in the upper row
in Fig.~\ref{fig:2D-v-mu1-0p005} as compared with Fig.~\ref{fig:2D-v-mu1-0p01}.
We can also see better in the lower right panel the solid-body rotation inside the soliton,
$|\vec v| = \Omega r_\perp$, which is independent of $z$ and grows linearly with $r_\perp$.
Thus, we can clearly see the low-velocity region in a central column around the rotation axis
and the symmetric growth of $|\vec v|$ on both sides with $r_\perp = x$.
We can also see the narrow traces of several vertical vortex lines, associated with high-velocity lines.

We recover the verticality of these vortex lines as well as their rotational motion
in Fig.~\ref{fig:3D-vortices-mu1-0p005}.

\subsection{Dependence on $\alpha$: cases $\alpha=0.5$ and $0$ with $\epsilon=0.01$}
\label{sec:alpha-0p5-0}

\begin{figure*}
\centering
\includegraphics[height=4.8cm,width=0.33\textwidth]{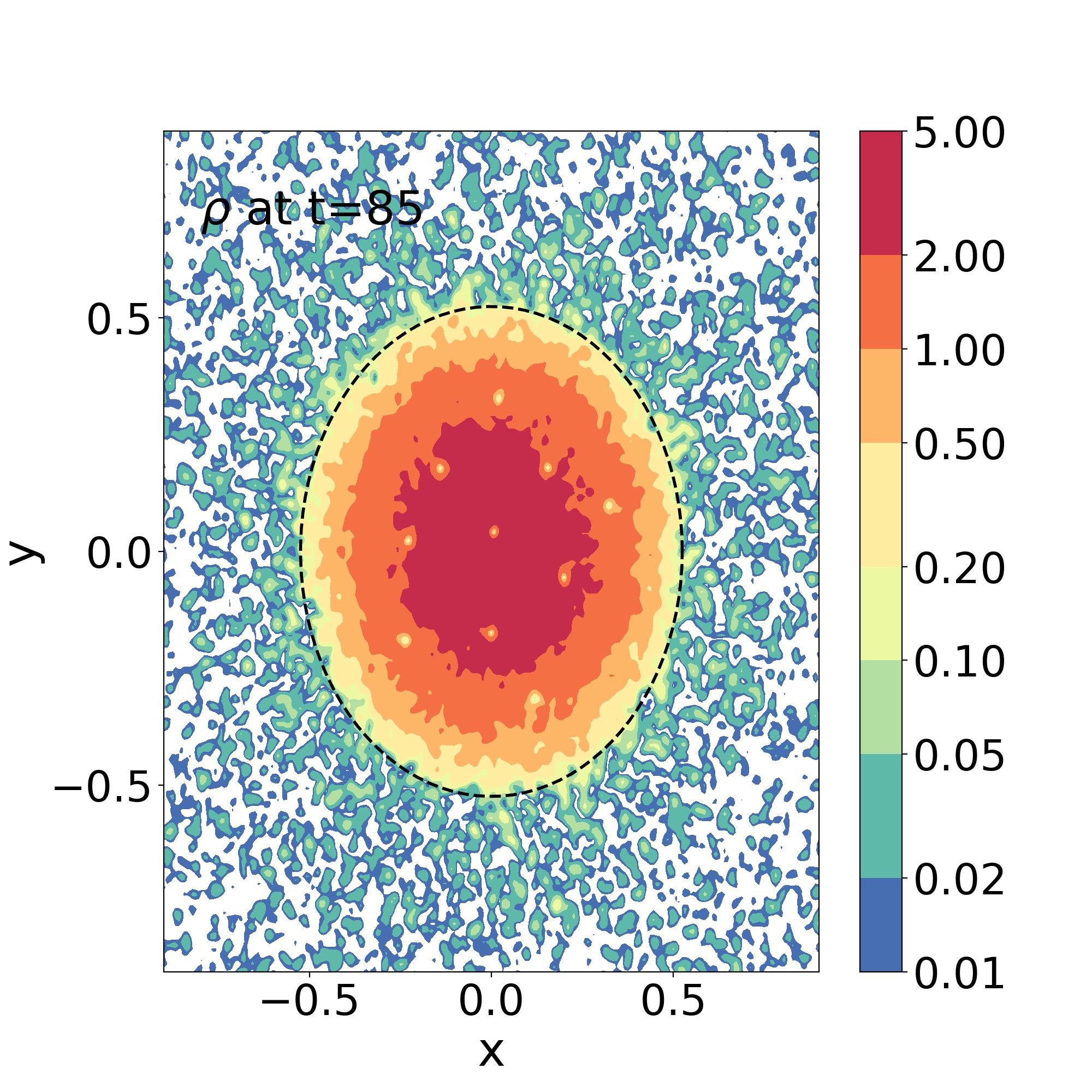}
\includegraphics[height=4.8cm,width=0.33\textwidth]{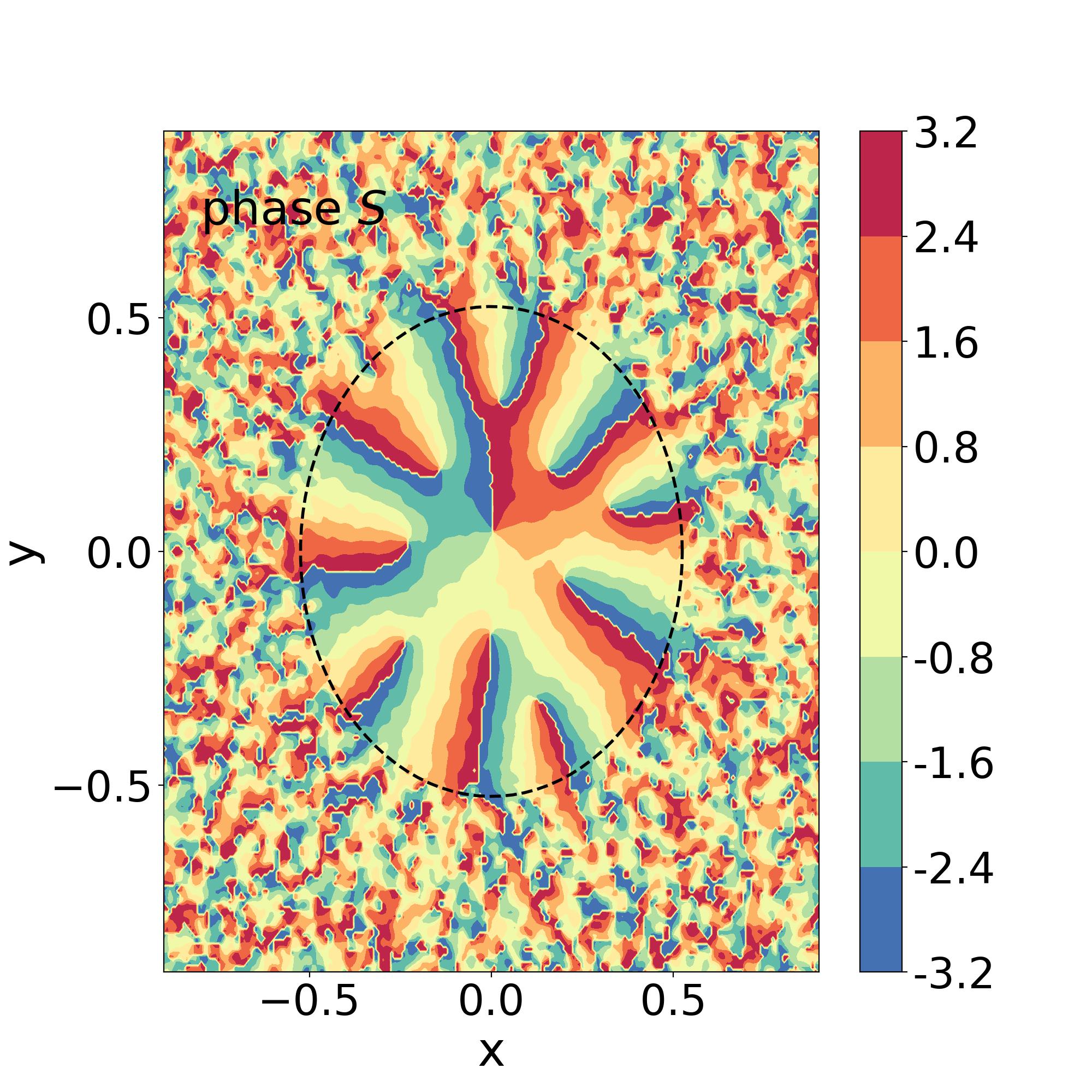}
\includegraphics[height=4.8cm,width=0.28\textwidth]{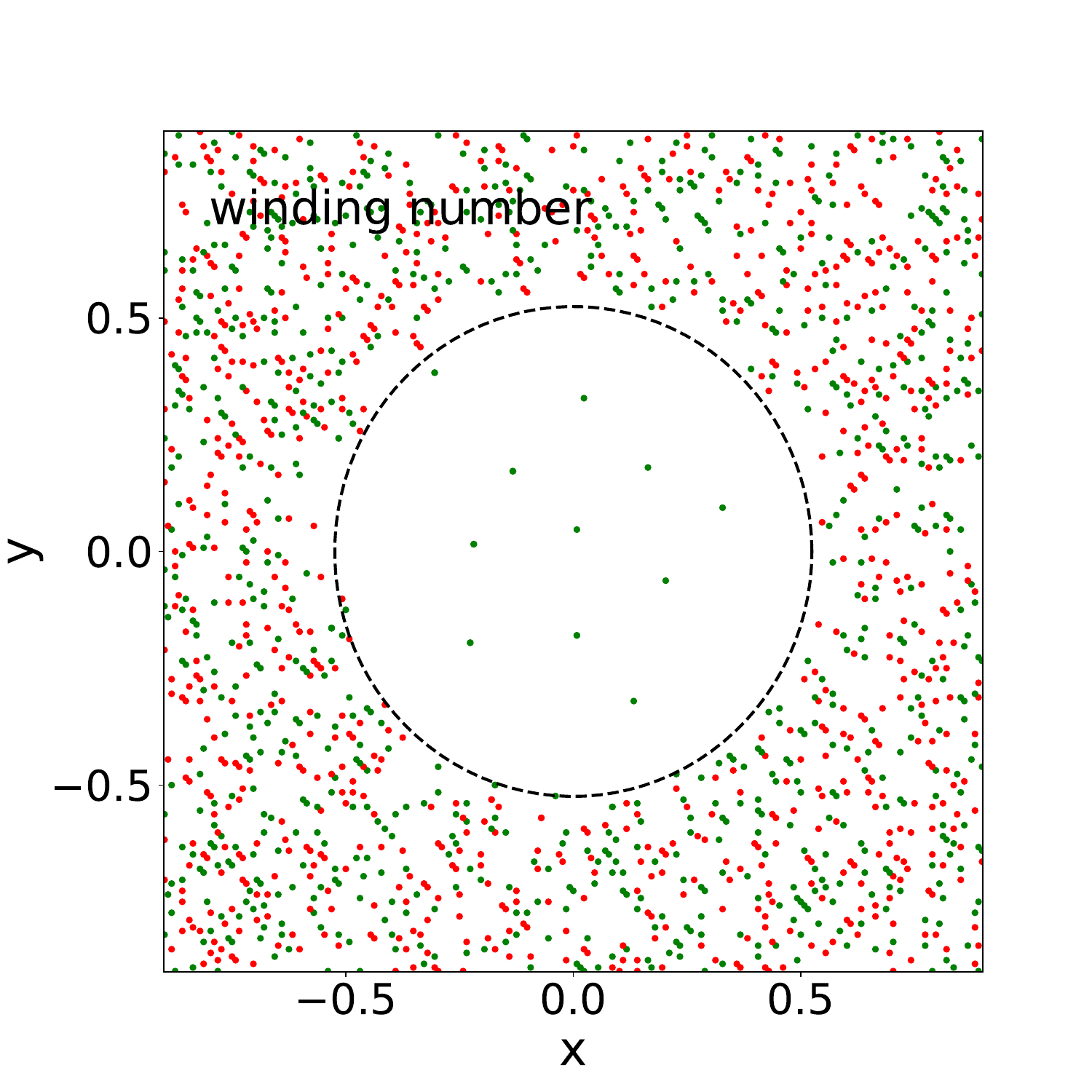}\\
\includegraphics[height=4.8cm,width=0.27\textwidth]{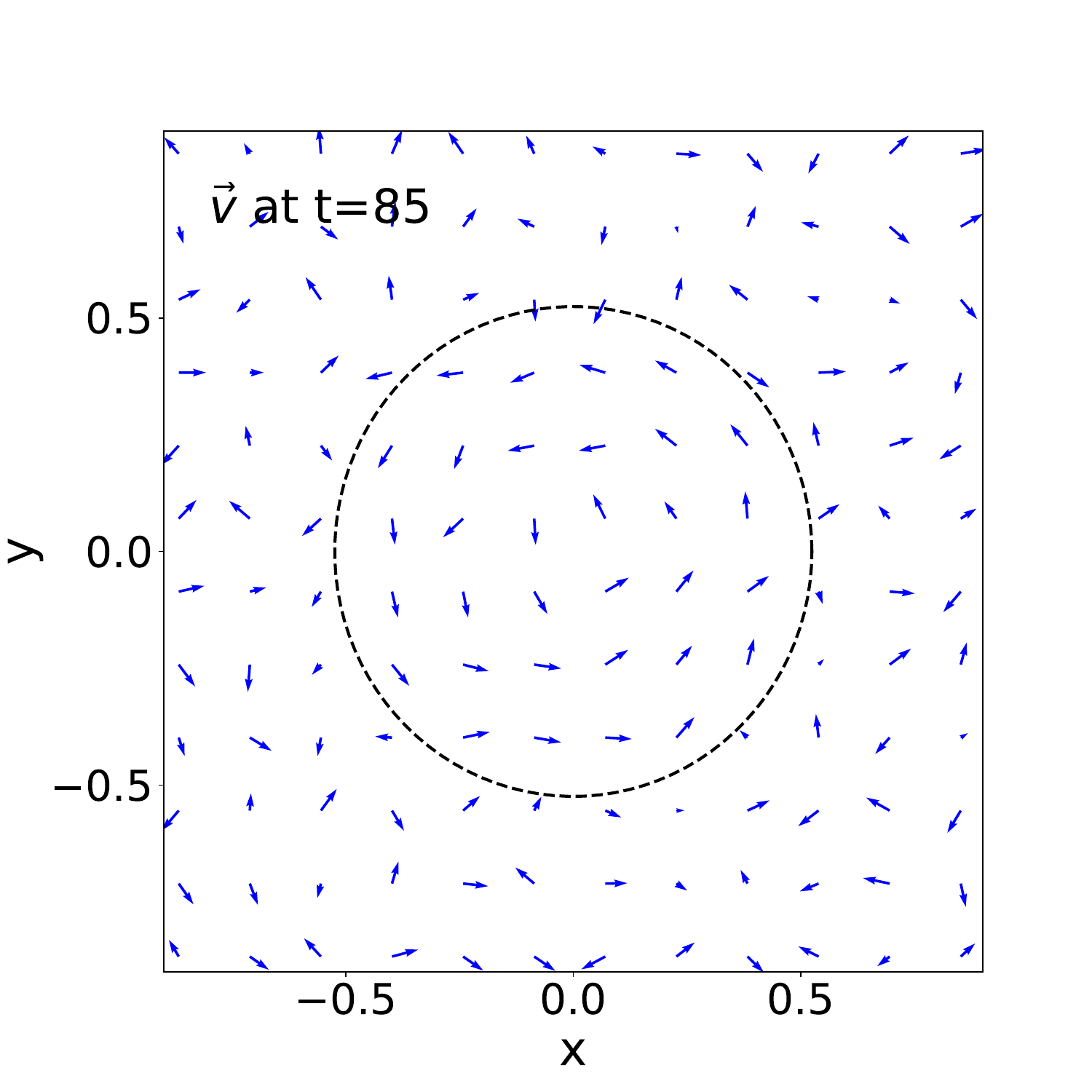}
\includegraphics[height=4.8cm,width=0.27\textwidth]{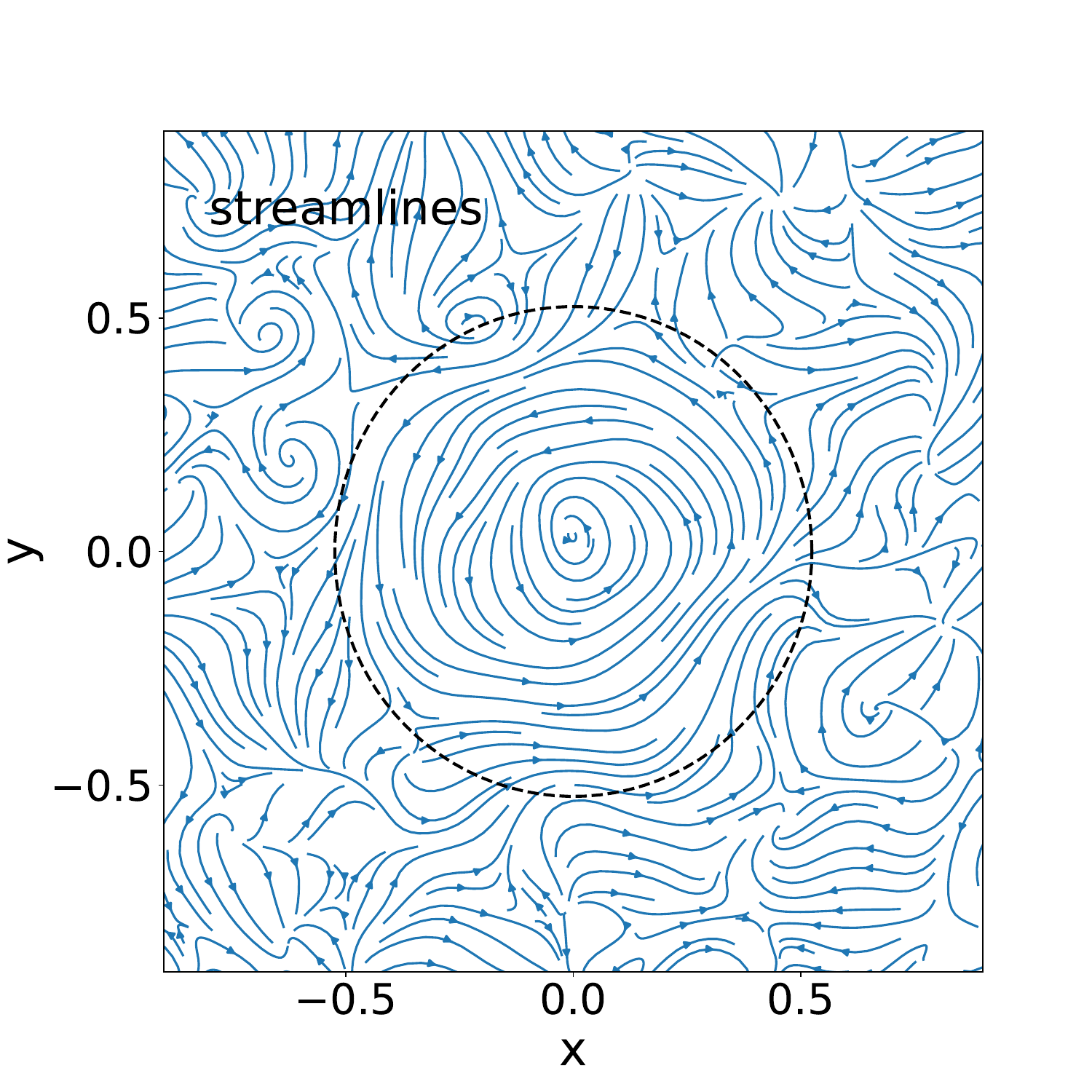}
\includegraphics[height=4.8cm,width=0.33\textwidth]{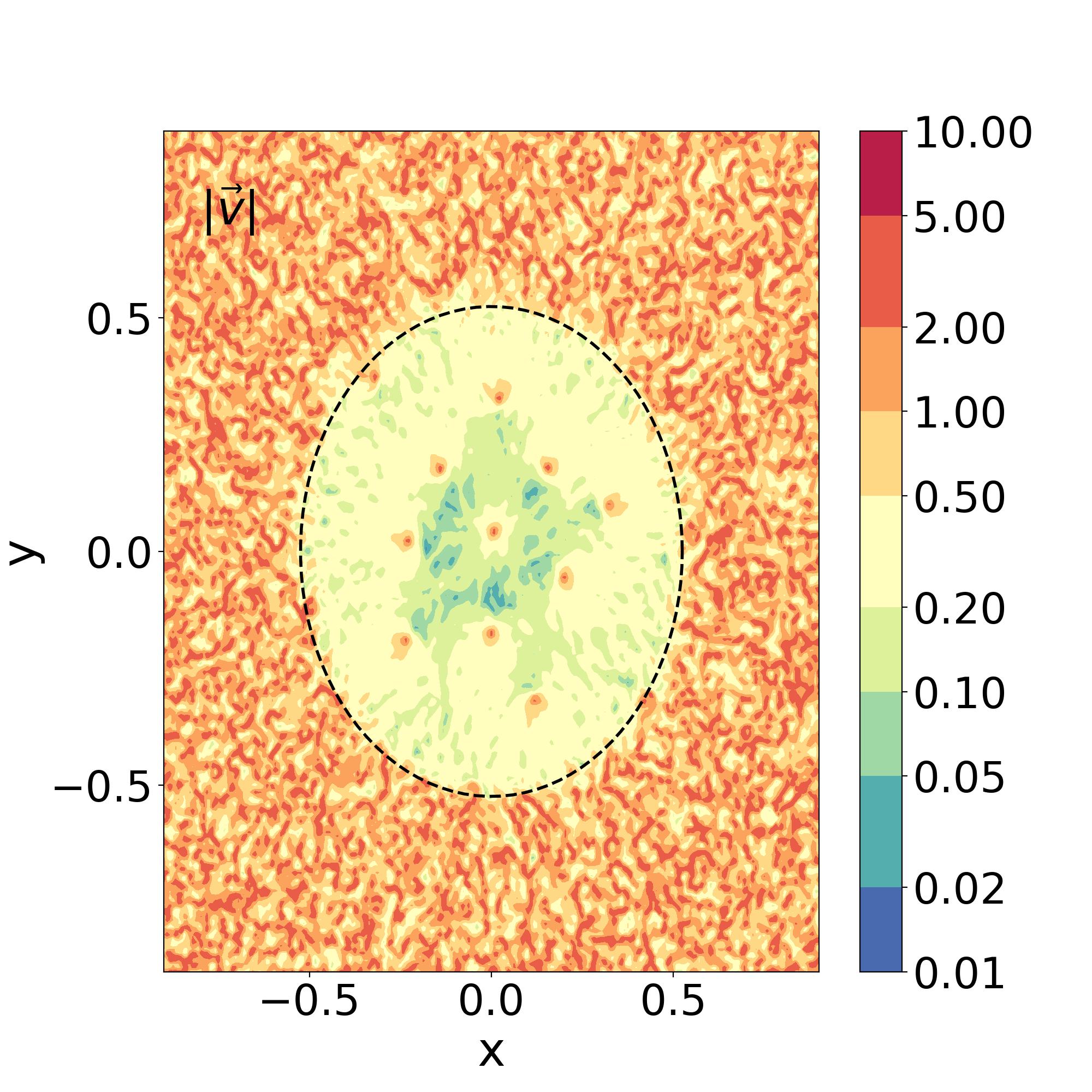}\\
\caption{
$[\epsilon=0.01, \alpha=0.5]$.
{\it Upper row:} 2D maps at $t=85$ of the density $\rho$, the phase $S$ and the winding number $w$,
in the $(x,y)$ plane, as in the upper row in Fig.~\ref{fig:2D-rho-mu1-0p01}.
{it Lower row:} 2D maps of the velocity in the $(x,y)$ plane, as in the upper row
in Fig.~\ref{fig:2D-v-mu1-0p01}.
}
\label{fig:2D-rho-mu0p5-0p01}
\end{figure*}

\begin{figure*}
\centering
\includegraphics[height=5cm,width=0.35\textwidth]{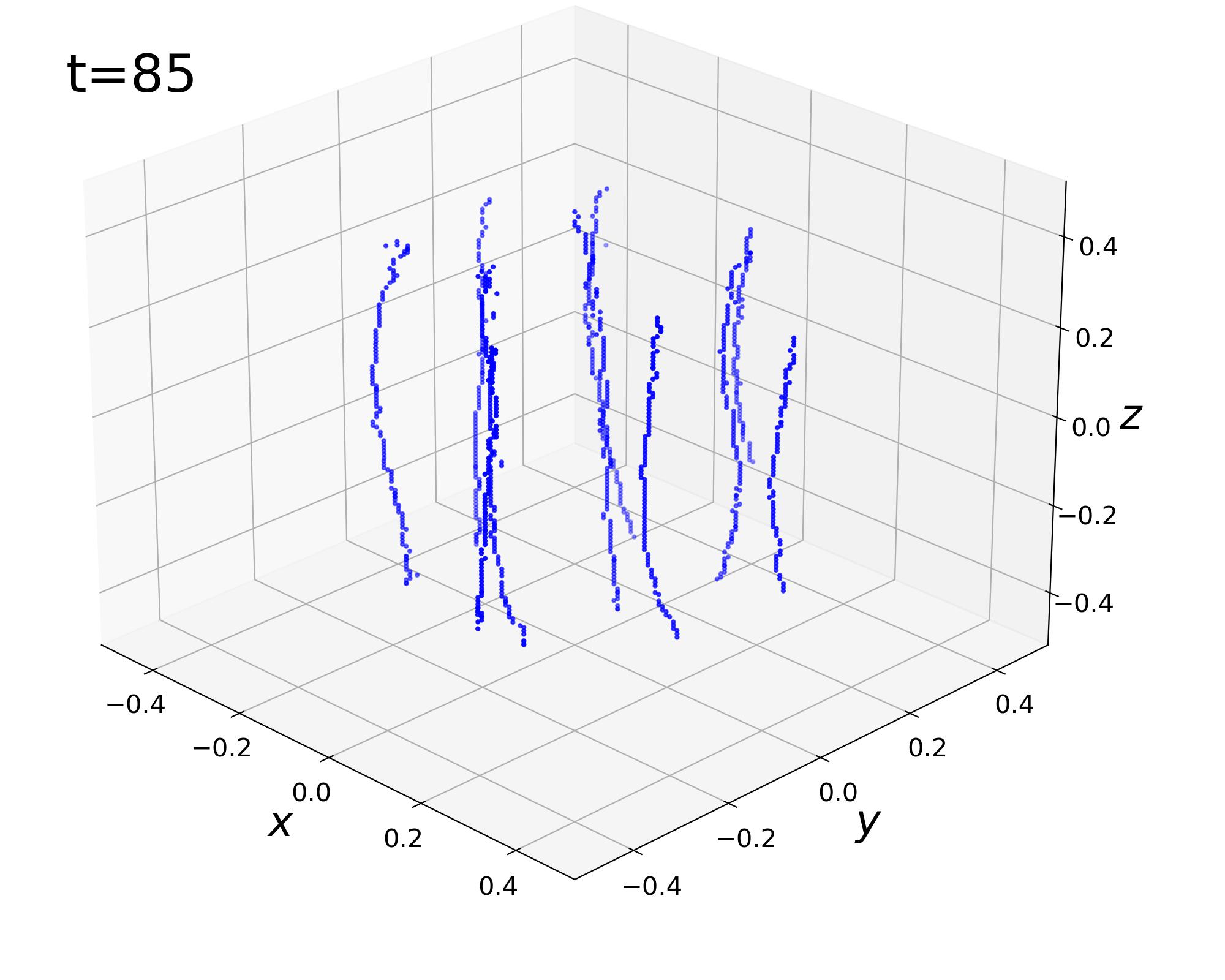}
\includegraphics[height=5cm,width=0.28\textwidth]{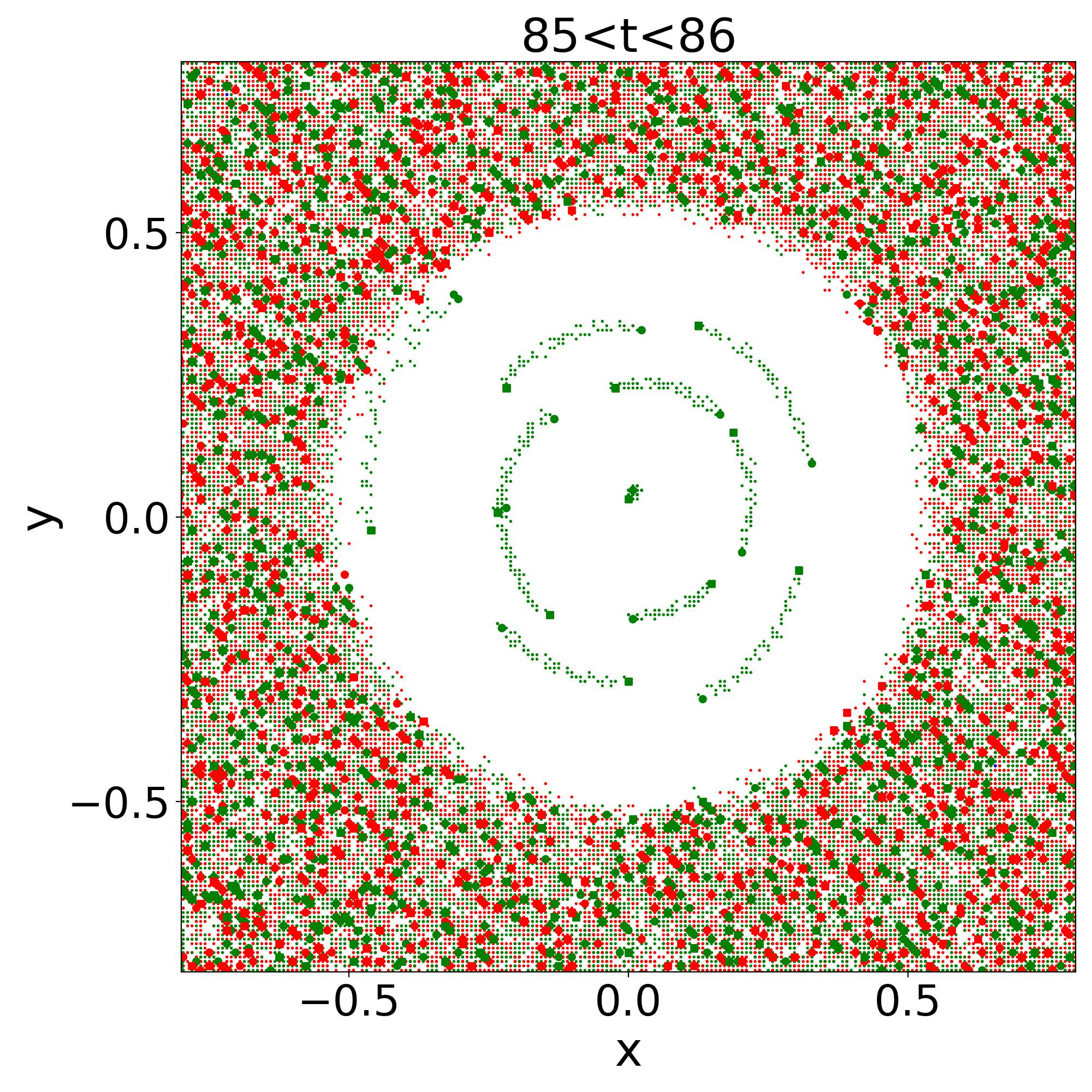}
\includegraphics[height=5cm,width=0.28\textwidth]{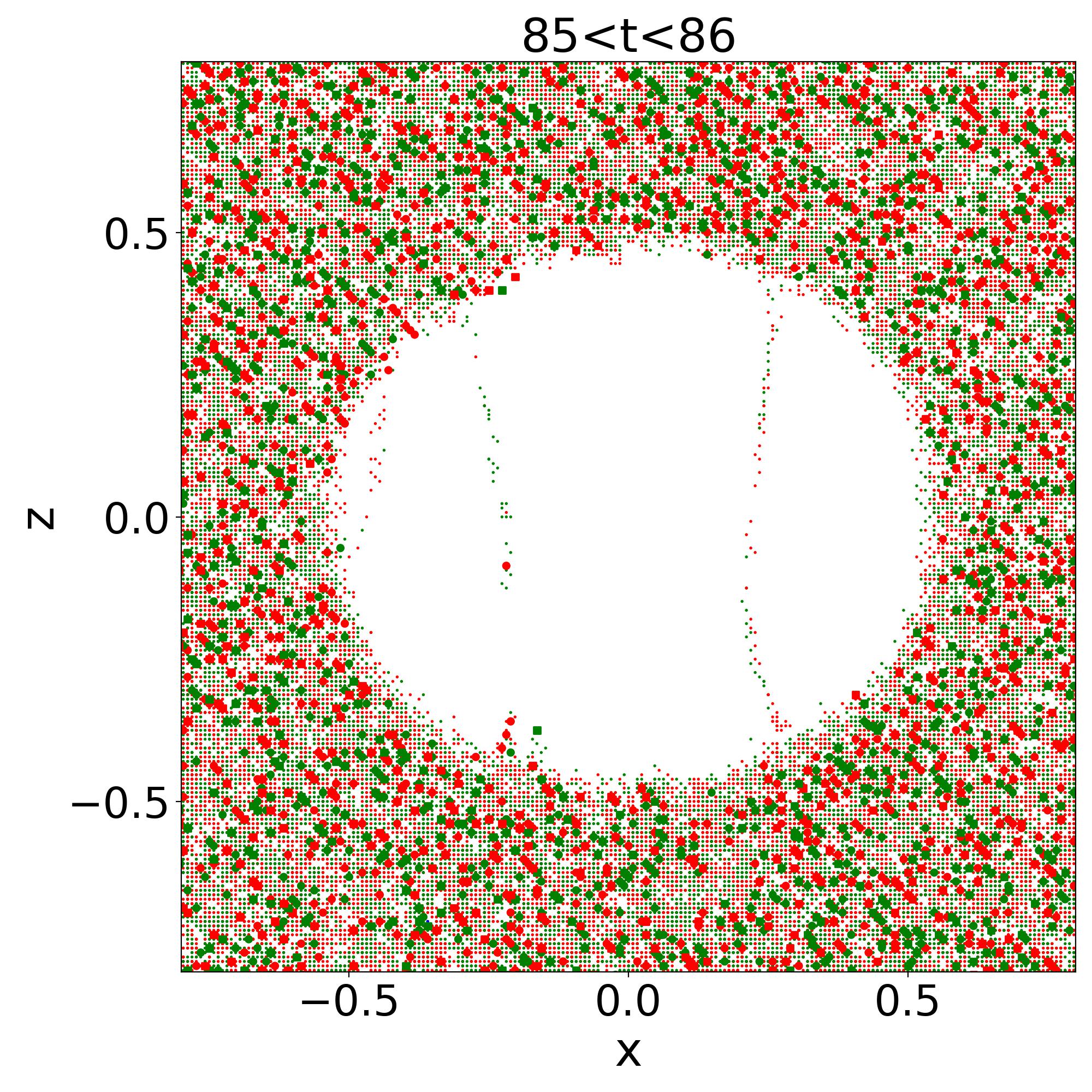}
\caption{
$[\epsilon=0.01, \alpha=0.5]$.
vortices as in Fig.~\ref{fig:3D-vortices-mu1-0p01}.
}
\label{fig:3D-vortices-mu0p5-0p01}
\end{figure*}

We now study how the properties of the system vary with $\alpha$ by running
simulations with the values $\alpha=0.5$ and $\alpha=0$, keeping $\epsilon=0.01$
as in the reference case presented in Sec.~\ref{sec:eps-0p01-alpha-1}.

\begin{figure*}
\centering
\includegraphics[height=4.8cm,width=0.33\textwidth]{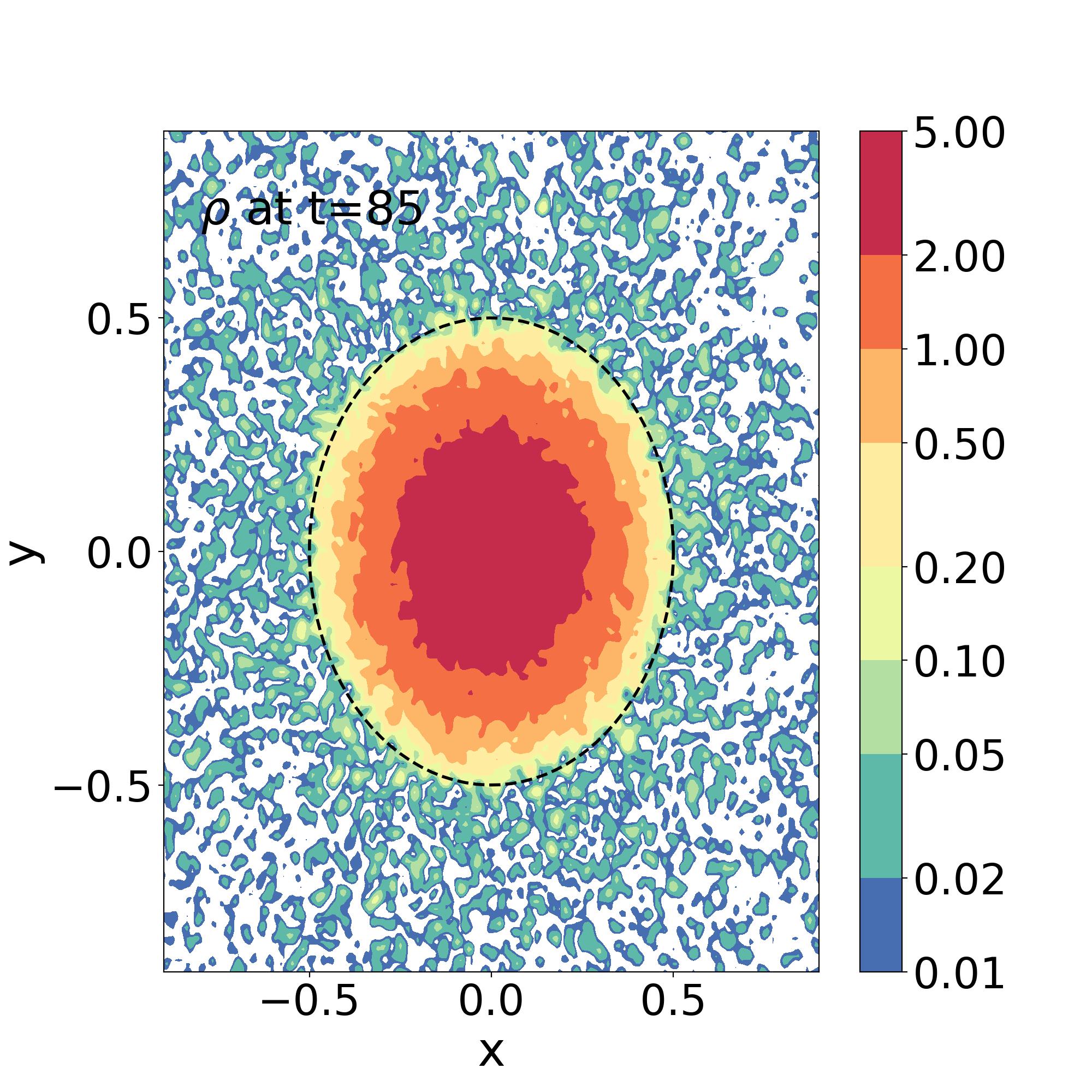}
\includegraphics[height=4.8cm,width=0.33\textwidth]{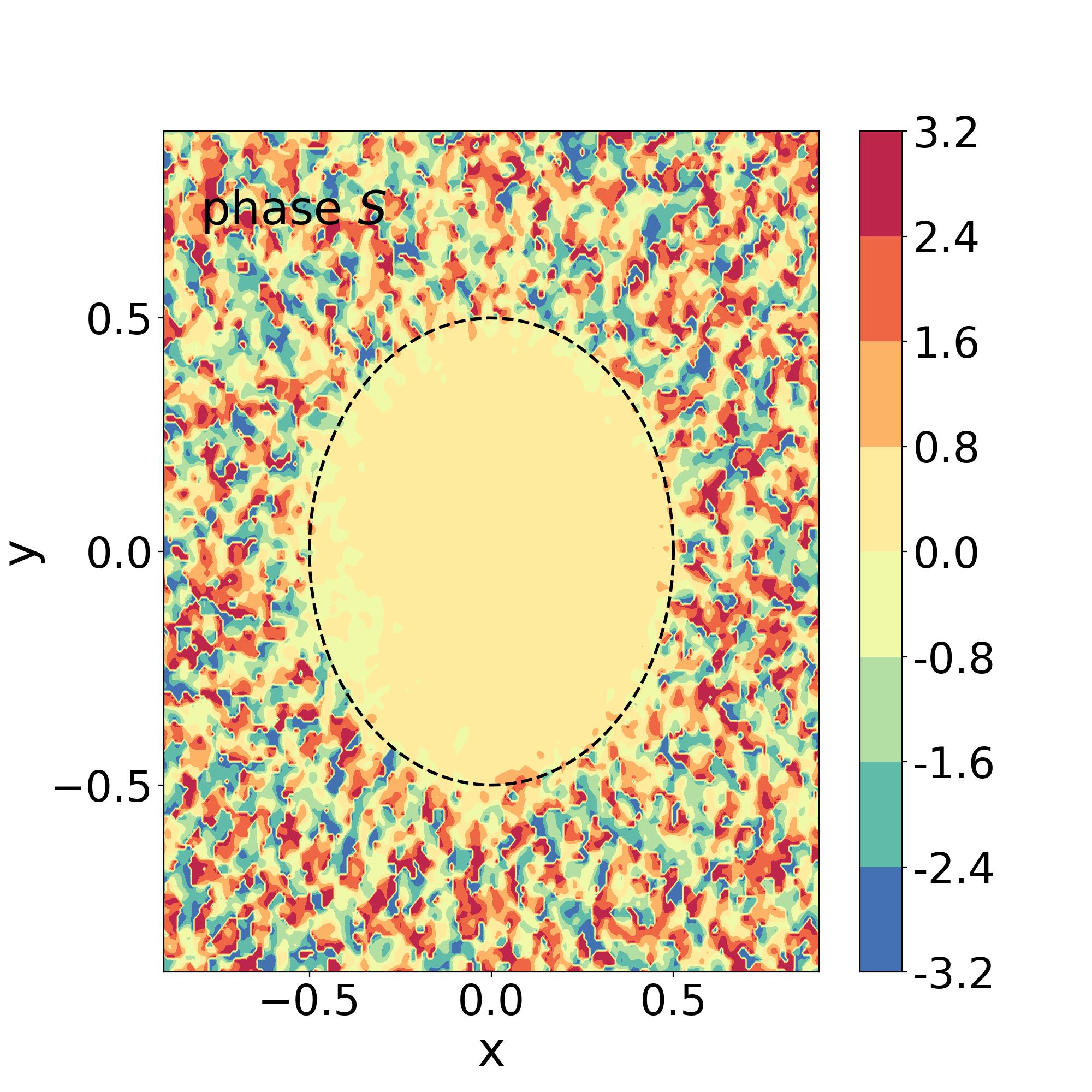}
\includegraphics[height=4.8cm,width=0.28\textwidth]{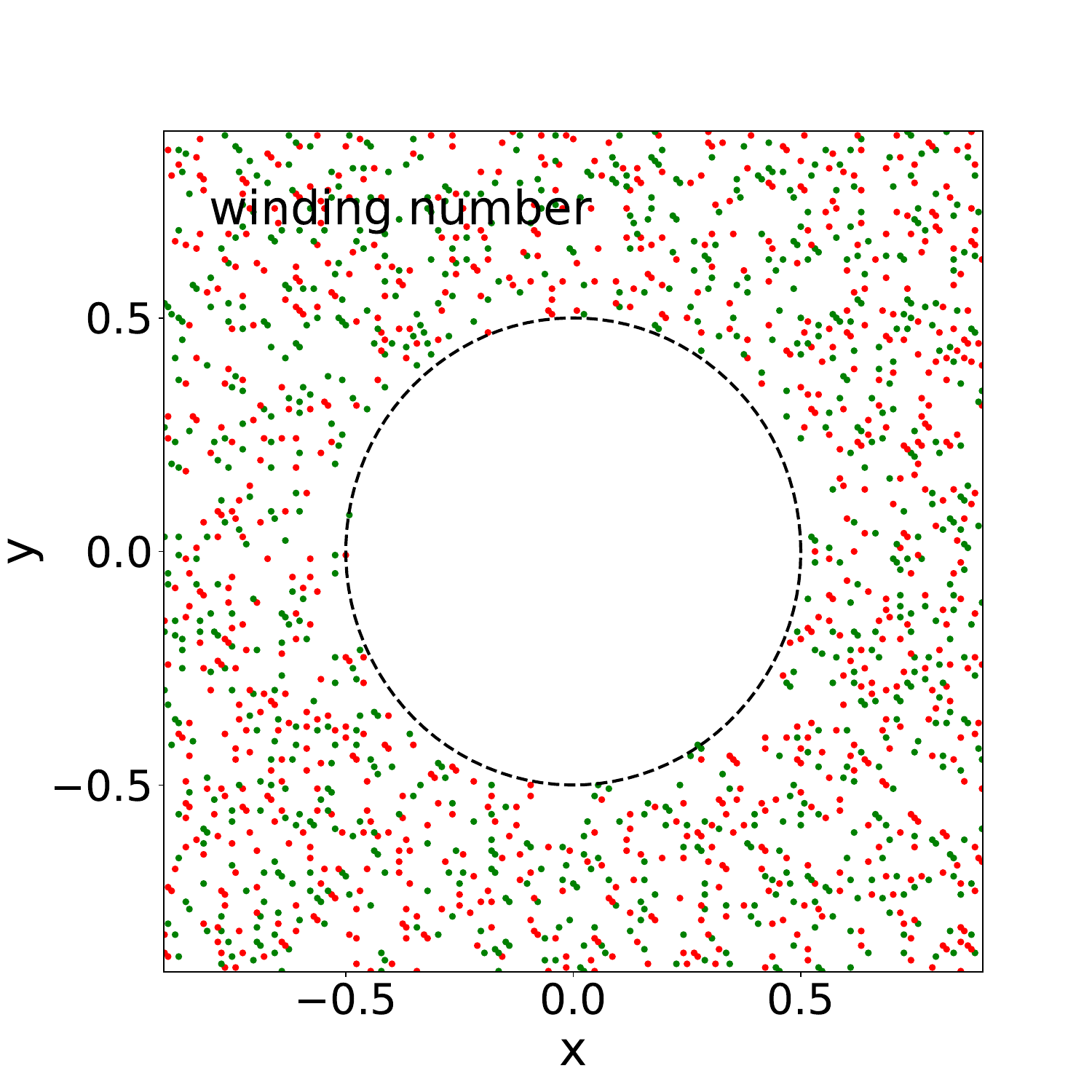}\\
\includegraphics[height=4.8cm,width=0.27\textwidth]{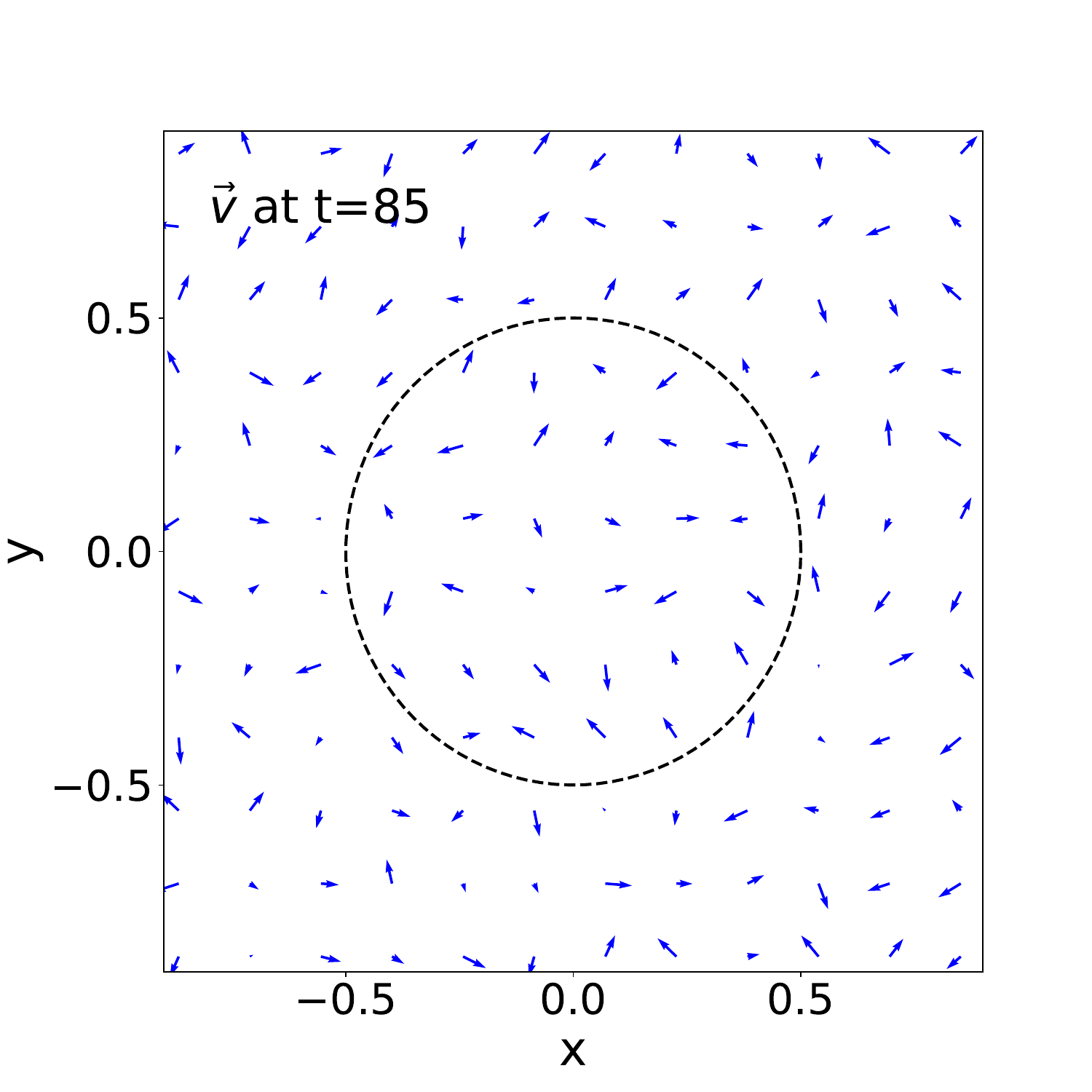}
\includegraphics[height=4.8cm,width=0.27\textwidth]{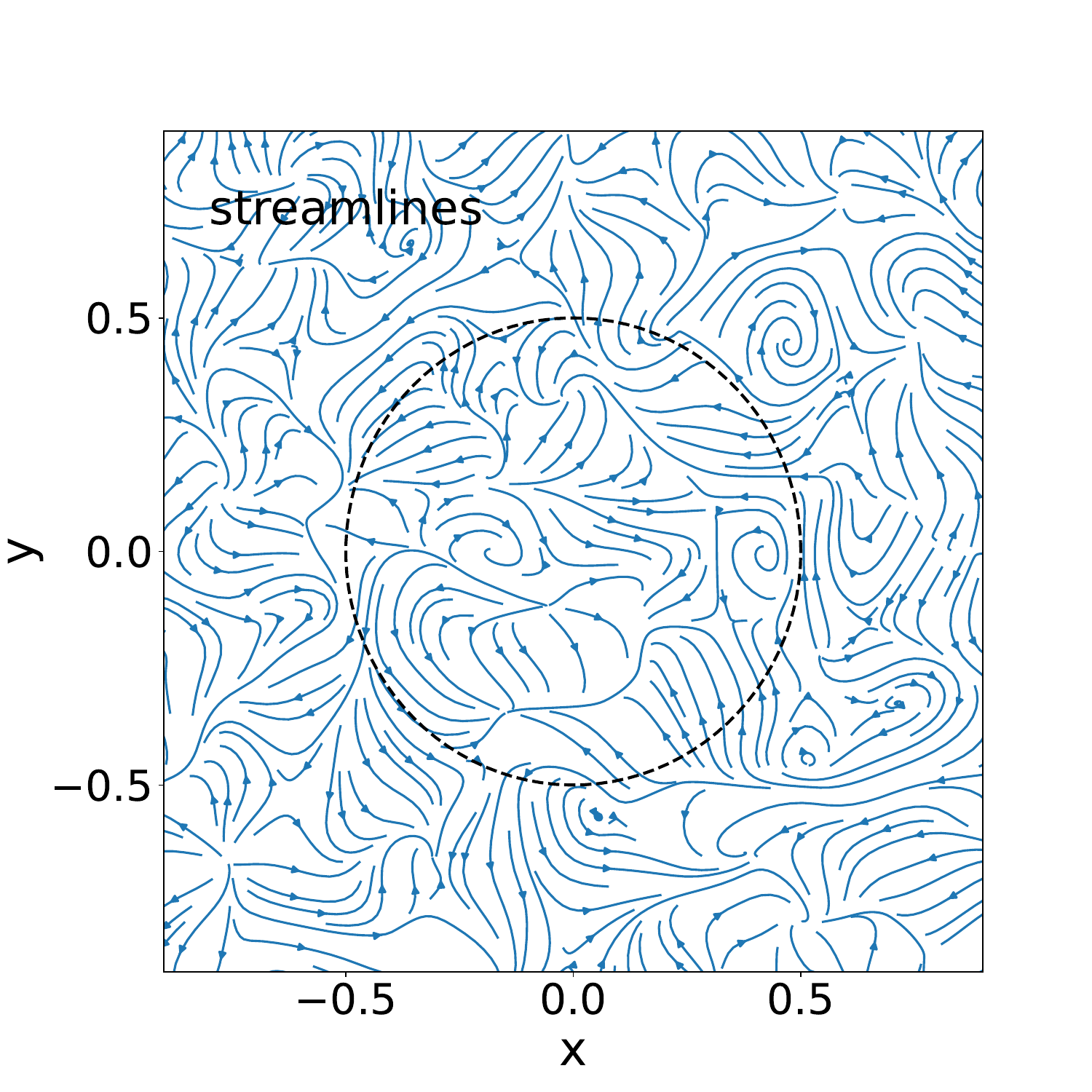}
\includegraphics[height=4.8cm,width=0.33\textwidth]{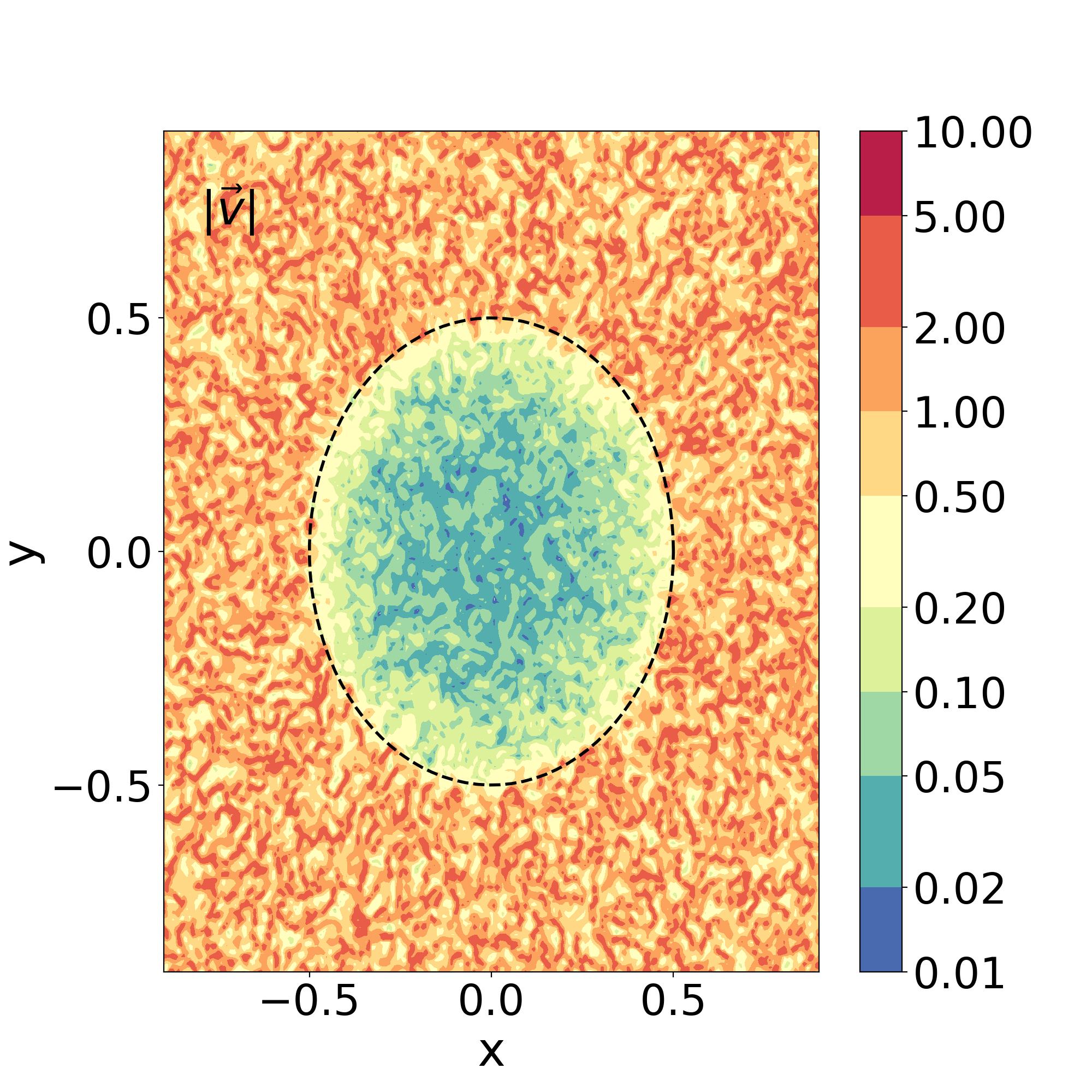}
\caption{
$[\epsilon=0.01, \alpha=0]$.
{\it Upper row:} 2D maps at $t=85$ of the density $\rho$, the phase $S$ and the winding number $w$,
in the $(x,y)$ plane, as in the upper row in Fig.~\ref{fig:2D-rho-mu1-0p01}.
{it Lower row:} 2D maps of the velocity in the $(x,y)$ plane, as in the upper row
in Fig.~\ref{fig:2D-v-mu1-0p01}.
}
\label{fig:2D-rho-mu0-0p01}
\end{figure*}

Let us first consider the case $\alpha=0.5$.
The qualitative behaviors are similar to those of the reference case $\alpha=1$
analyzed in Sec.~\ref{sec:eps-0p01-alpha-1}, therefore we only show the 2D maps in the
equatorial plane $(x,y)$.
We recover the same features for the density and velocity fields in
Fig.~\ref{fig:2D-rho-mu0p5-0p01} as in
Figs.~\ref{fig:2D-rho-mu1-0p01} and \ref{fig:2D-v-mu1-0p01}, except that the number of
vortex lines inside the soliton is about twice smaller, in agreement with the fact that the
initial angular momentum of the system is twice smaller.
In fact, we obtain a rotation rate $\Omega(t) \simeq 0.6$, slightly above half the value
$\Omega(t) \simeq 0.9$ obtained in Fig.~\ref{fig:Om-mu1-0p01}.
This means that a somewhat larger fraction of the initial angular momentum ends up inside the
soliton.
This could be due to the fact that we are deeper inside the stability domain
(\ref{eq:Omega-stable}), so that the initial violent relaxation associated with the formation
of the soliton proceeds in a somewhat more ordered manner with a lower redistribution of matter,
energy and angular momentum within the system.

We recover in Fig.~\ref{fig:3D-vortices-mu0p5-0p01} the vertical vortex lines inside the soliton,
with their rotation around the $z$ axis.
Even though the number of vortices is rather small, their combined velocity field (\ref{eq:dot-rj})
already mimics rather well the solid-body rotation (\ref{eq:v-solid-rotation}) derived in the continuum limit
and the orbits of the vortex lines are roughly circular circles, as those
displayed in Fig.~\ref{fig:3D-vortex-motion-mu1-0p01}.

We show in Fig.~\ref{fig:2D-rho-mu0-0p01} our results for the isotropic case $\alpha=0$,
where there is no initial angular momentum.
This case was already studied in \cite{Garcia:2023abs}, but only from the point of view of the
density field. As the planes $(x,y)$ and $(x,z)$ are now statistically equivalent, as we checked
in our numerical simulation, we only display our results in the plane $(x,y)$ in
Fig.~\ref{fig:2D-rho-mu0-0p01}.
We recover the spherical static soliton profile (\ref{eq:rho-TF-0}) and its radius $R_0$,
which agrees with Eq.(\ref{eq:Rsol-res}) for $\Omega=0$.
As the angular momentum of the system vanishes, the soliton is not rotating and it does not
contain vortices. The phase is roughly constant, without singularities. This gives a low amplitude
$|\vec v|$ for the velocity inside the soliton, with random velocity fluctuations due to the
incomplete relaxation and the impact of the outer halo modes.
In the outer halo, which is supported by its velocity dispersion, we recover many random vortex
lines of any direction, as in the simulations analyzed in the previous sections.

\section{Conclusion}
\label{sec:conclusion}

This paper deals with the creation of vortices and rotating solitons in scalar dark matter models,
focusing on scenarios with non-negligible repulsive self-interactions.
There is somehow an apparent paradox in this issue as the hydrodynamical formulation of scalar dark
matter is usually described by an irrotational  fluid, together with a so-called quantum pressure and 
an additional self-interaction pressure with a polytropic equation of state. 
As is well-known, this picture which emerges after a Madelung transform of the Gross-Pitaevskii equation
is only valid as long as the density of the associated fluid does not vanish. 
However, interferences between different eigenmodes generically lead to the vanishing of the wave function
at many locations inside a virialized halo, which give rise to vortices in 2D and vortex lines in 3D.
For a fuzzy dark matter model, or in our case in the outer virialized halo, these stochastic interferences 
generate strong velocity fluctuations and a maze of disordered vortices that mimic a classical system
of collisionless particles supported by its velocity dispersion 
\cite{Widrow:1993qq,Mocz:2018ium,GalazoGarcia:2022nqv,Garcia:2023abs,Liu:2024pjg}.
On the other hand, any nonzero total angular momentum $\vec J$ of the system, which is conserved
by the dynamics, breaks the statistical isotropy and generates an ordered set of vortex lines aligned with
$\vec J$. These vortices no longer arise from random interferences. Instead they provide the means
for the system to support a macroscopic angular momentum and rotation, which is the focus of this paper.

We have presented an analytic description of these vortex lines and of the rotating solitons that
can form inside such systems. We have retrieved the fact that vortices are advected by the flow created
by the other vortices. 
In the Thomas-Fermi limit where the de Broglie wavelength is tiny as compared with the system size,
the width of the vortex tubes is much smaller than their nearest-neighbor distance, which itself is also
much smaller than the system size, even though their number becomes very large.
Then, the system approximates a continuum limit where the vorticity is non-vanishing at each point.
This resolves the apparent paradox alluded to above. Thanks to these vortices, the Gross-Pitaevskii
equation can still be mapped to hydrodynamical equations of motion but the velocity field no longer
needs to be curl-free. 
We have shown that taking into account the conservation of angular momentum generalizes the well-known
static and spherically symmetric solitons to rotating and axisymmetric solitons with an oblate shape.
Moreover, these oblate spheroids obey a solid-body rotation and display an uniform density
of vortex lines.

In fuzzy dark matter scenarios or models with attractive self-interactions, so-called
rotating boson stars have been shown to be unstable \cite{Dmitriev:2021utv} 
(but a central vortex can be stabilized by the external gravitational potential of a central black hole \cite{Glennon:2023oqa}).
However, the objects studied in these works are quite different from the rotating solitons
obtained in this paper. They correspond to a single vortex of possibly large spin $\sigma$, 
as in Eq.(\ref{eq:single-vortex-psi}) but with a non-uniform envelope $\rho(r)$ that describes
a compact object. 
In contrast, in our case the rotating soliton or boson star is made of a lattice of many vortices
of unit spin and its central density does not vanish. Instead of a torus, it shows an oblate
spheroidal shape, with a maximum density at the center.
We have also shown that these rotating solitons are dynamically stable, as long as the rotational
energy is below the self-interaction and gravitational energies.

We have checked that these analytical calculations are confirmed by numerical simulations.
To simplify the analysis, we have considered a single isolated halo with stochastic initial conditions.
Using the WKB approximation to make the correspondence between the initial scalar field $\psi(\vec x)$
and the phase-space distribution $f(\vec x,\vec v)$ of a classical system of particles,  
we can set initial conditions with a prescribed averaged density profile and total angular momentum.
Starting with such virialized halos without a central soliton, we have found that a soliton forms in a few 
dynamical times, as for the isotropic initial conditions presented in \cite{Garcia:2023abs}.
For a nonzero initial angular momentum, the soliton shows a solid-body
rotation with a uniform lattice of vortex lines and an oblate shape, in good agreement with the analytical
predictions.
We have checked that the number of vortices scales roughly linearly with the inverse of the
de Broglie wavelength and with the total angular momentum.

The dynamics of these solitons and of the vortex lines are extremely interesting and this paper
is only a first step in their study.
In particular, whereas in a 2D system the vortices are merely isolated point vortices, in a 3D
system they are extended vortex lines. Because of the nonzero winding number around these
topological defects, the vortex lines cannot terminate inside the fluid. Within the Thomas-Fermi
limit and for a compact object, the vortex lines might terminate at the boundary of the object, where
the density vanishes, but going beyond this approximation the density no longer vanishes at the 
border and it shows instead extended exponential tails. 
In laboratory experiments of Bose-Einstein condensates, the vortex lines can terminate at the metallic
container, where the Gross-Pitaevskii equation is no longer a good description and electrons can
for instance regain their individuality.
In our case, we do not have such an external boundary.
However, because of the periodic boundary conditions, the few extra positive vortex lines associated
with the nonzero angular momentum might run through the outer halo maze and extend to infinity
through the series of periodic boxes along the vertical axis.
In a more realistic cosmological context, we note that the angular momentum of virialized halos
arises from the torques exerted by neighbouring structures
\cite{Peebles:1969jm,Doroshkevich-1970,White-1984}.
Then, the conservation of angular momentum means that there is a causal connection between
the spins of neighboring halos inside the cosmic web. In the case of ultralight dark matter models,
this connection would be physically associated with actual vortex lines running from one halo to the
other. 
One possible picture would thus be that as the angular momentum of various proto-halos grows
within the cosmic web, a network of vortex lines would also develop along the dark matter filaments.
Then, in contrast with BEC laboratory experiments, there would be no need to terminate the
vortex lines by going beyond the Gross-Pitaevskii formulation.
Rather, because of the conservation of the total angular momentum and of the hierarchical
process that led to the formation of the cosmic web and of distinct spinning halos,
a continuous network of vortex lines would simultaneously grow and link the various halos,
along cosmic filaments, while forming ordered lattices inside the central solitons at the nodes.

A connection between spinning halos and filaments was considered for CDM scenarios
in \cite{Codis:2012ep,Xia:2020znt}. 
In the case of ultralight dark matter, this connection would thus be materialized
by physical vortex lines. 
These cosmic vortex lines could explain the recent observation \cite{Wang:2021axr}
of the spin of cosmic filaments, as explored in \cite{Alexander:2021zhx}.
A detailed study of such a mechanism is beyond the scope of this paper and would require
dedicated simulations. A difficulty for numerical computations would be to distinguish the 
extended vortex lines that link distant halos (if they exist) from the chaotic maze of 
vortices generated by the random interferences between excited modes in the virialized outer halos
and filaments, outside of the relaxed solitons.
We leave such a study to future works.

Another topic worthy of investigation is the interplay with baryons. Whether through gravitational
effects or through additional couplings to baryons (such as the coupling to electromagnetism for axions),
rotating solitons or vortex lines could have some observational signatures.
They might impact the dynamics of the gas or of stars and, in turn, lead to radiative processes
(e.g. synchrotron emission of accelerated gas).
On the other hand, the rotation of the soliton could be related to the galactic rotation curves
\cite{Kain:2010rb}.
Finally, it would be useful to study in detail whether dark matter substructures such as those
vortex lines could be detected by gravitational lensing.

\appendix

\section{Rotating soliton profile}
\label{app:rotating-soliton}

In this appendix we detail our computation of the rotating soliton profile.

We can exactly solve the linear differential equations (\ref{eq:PhiN-in}) and (\ref{eq:PhiN-out})
by expanding in Legendre polynomials, as we are looking for an axisymmetric solution that does
not depend on $\varphi$.
Inside the soliton, we obtain
\ba
&& \Phi_{N\rm in} = \sum_{\ell = 0}^{\infty} \phi_\ell \, j_\ell\!\left(\frac{\pi r}{R_0}\right)
P_\ell(\cos\theta ) + \frac{\Omega^2 r_\perp^2}{2} + \mu - \frac{\lambda \Omega^2}{2\pi} , \nonumber \\
&& \rho_{\rm in} = -\frac{1}{\lambda} \sum_{\ell = 0}^{\infty} \phi_\ell \,
j_\ell\!\left(\frac{\pi r}{R_0}\right) P_\ell(\cos\theta) + \frac{\Omega^2}{2\pi}  ,
\label{eq:PhiN-in-general}
\ea
where $j_\ell$ and $P_\ell$ are the spherical Bessel function and Legendre polynomial of order $\ell$.
Here we used the Poisson equation to obtain the density and the requirement that the density be finite
at the center to discard the spherical Bessel functions of the second kind.
Outside of the soliton we can write
\be
\Phi_{N\rm out} = \sum_{\ell = 0}^{\infty} \psi_\ell \, r^{-\ell-1} P_\ell(\cos\theta) , \;\;\;
\rho_{\rm out} = 0 ,
\label{eq:PhiN-out-general}
\ee
as we require the gravitational potential to vanish at infinity.
As the solution is even in $z$ and $\cos\theta$, odd-order coefficients vanish,
$\phi_{2\ell+1}=0$ and $\psi_{2\ell+1}=0$.
At zeroth order over $\Omega^2$ we have the usual static solution (\ref{eq:rho-TF-0})-(\ref{eq:PhiN-TF-0}).
For a slow rotation we can use a perturbative expansion over powers of $\Omega^2$ and
up to first order we can write
\ba
&& \Phi_{N\rm in}(r,\theta) = - \lambda \left( \rho_0 - \frac{\Omega^2}{2\pi} \right) \, j_0\left(\frac{\pi r}{R_0}\right)
+ \frac{\Omega^2 r_\perp^2}{2} \nonumber \\
&& + \Omega^2 \sum_{\ell = 2}^{\infty} \phi_\ell  \, j_\ell\left(\frac{\pi r}{R_0}\right) P_\ell(\cos\theta)
-\lambda \rho_0 + \Omega^2 \nu , \;\;\;
\label{eq:PhiN-in-1}
\ea
\ba
\rho_{\rm in}(r,\theta) & = & \left( \rho_0 - \frac{\Omega^2}{2\pi} \right) \, j_0\left(\frac{\pi r}{R_0}\right)
+ \frac{\Omega^2}{2\pi}
\nonumber \\
&& - \frac{\Omega^2}{\lambda} \sum_{\ell = 2}^{\infty} \phi_\ell \, j_\ell\left(\frac{\pi r}{R_0}\right)
P_\ell(\cos\theta) ,
\label{eq:rho-in-1}
\ea
and
\be
\Phi_{N\rm out}(r,\theta) = - \lambda \rho_0 R_0/r + \Omega^2 \sum_{\ell = 0}^{\infty} \psi_\ell r^{-\ell-1}
P_\ell(\cos\theta)  .
\label{eq:PhiN-out-1}
\ee
Here we renormalized the coefficients $\phi_\ell$ and $\psi_\ell$ by a factor $\Omega^2$ and
the coefficient of the $j_0$ term in $\rho_{\rm in}$ and $\Phi_{N\rm in}$ is obtained from the
normalization of the central density to a given fixed value $\rho_0$.
The coefficient $\nu$ is related to the Lagrange multiplier $\mu$ by
\be
\mu = -\lambda \rho_0 + \Omega^2 \left( \nu + \frac{\lambda}{2\pi} \right) .
\ee

The inner and outer expressions of the gravitational potential and of its gradient must be matched
on the surface of the soliton, $r=R_{\Omega}(\theta)$, which we write up to first order over $\Omega^2$ as
\be
R_{\Omega}(\theta) = R_0 + \Omega^2 \sum_{\ell = 0}^{\infty} q_\ell P_\ell(\cos\theta) .
\label{eq:R-theta-1}
\ee
This surface is defined by the locations where the soliton density vanishes, that is,
$\rho_{\rm in}(R_{\Omega}(\theta),\theta)=0$.
Expanding this condition up to first order over $\Omega^2$, using Eqs.(\ref{eq:rho-in-1})
and (\ref{eq:R-theta-1}), gives
\be
q_0 = \frac{R_0}{2\pi\rho_0} , \;\;\; \ell \geq 2: \; q_\ell = - \frac{R_0 j_\ell(\pi)}{\lambda\rho_0} \phi_\ell.
\ee
Next, the matching of the gravitational potentials (\ref{eq:PhiN-in-1}) and (\ref{eq:PhiN-out-1}) at the
surface of the soliton,
$\Phi_{N\rm in}(R_{\Omega}(\theta),\theta)=\Phi_{N\rm out}(R_{\Omega}(\theta),\theta)$, gives
\ba
&& \psi_0 = \frac{R_0^3}{3} + \nu R_0 , \;\;
\psi_2 = \frac{3 R_0^3}{\pi^2} \phi_2 - \frac{R_0^5}{3} , \nonumber \\
&& \ell \geq 4 : \;\; \psi_\ell = j_\ell(\pi) R_0^{\ell+1} \phi_\ell .
\ea
Finally, matching the gradient $\partial\Phi_N/\partial r$ of the gravitational potentials gives
\be
\nu = \frac{2-\pi^2}{\pi^2} R_0^2 , \;\; \ \phi_2 = \frac{5 R_0^2}{3} , \;\;
\ell \geq 4 : \; \phi_\ell = 0 .
\ee
Collecting these results we obtain the expressions (\ref{eq:rho-in-res})-(\ref{eq:mu-res})
for the density and gravitational potentials, the soliton surface and the Lagrange multiplier $\mu$.

\section{Initial conditions}
\label{app:initial-conditions}

In this appendix we detail the construction of the initial conditions of our simulations, based
on the expansion (\ref{eq:psi-halo-a_nlm}) over eigenfunctions.

\subsection{WKB approximation}
\label{app:WKB}

Using the WKB approximation, the radial part ${\cal R}_{n\ell}(r)$ of the eigenfunctions
(\ref{eq:Rnl-def}) reads
\be
r_1 \! < \! r \! < \! r_2 : \; {\cal R}_{n\ell}(r) \simeq \frac{N_{n\ell}}{r\sqrt{k_{n\ell}(r)}}
\sin \! \left[ \frac{1}{\epsilon} \! \int_{r_1}^r \!\! dr' k_{n\ell}(r') \! + \! \frac{\pi}{4} \! \right]
\ee
and ${\cal R}_{n\ell}\simeq 0$ outside of this radial range,
where $r_1$ and $r_2$ are the turning points of the classical orbit, $N_{n\ell}$ is the
normalization factor and we introduced
\be
k_{n\ell}(r) = \sqrt{2 ( E_{n\ell} - \Phi_N(r) ) - \frac{\epsilon^2 \ell (\ell+1)}{r^2} } .
\label{eq:k-n-ell-def}
\ee
The normalization condition (\ref{eq:norm-R}) gives
\be
N_{n\ell} = \left( \int_{r_1}^{r_2} \frac{dr}{2 k_{n\ell}(r)} \right)^{-1/2} ,
\ee
while the quantization of the energy levels is given by
\be
\frac{1}{\epsilon} \int_{r_1}^{r_2} dr k_{n\ell}(r) = \left( n + \frac{1}{2} \right) \pi .
\label{eq:levels}
\ee
From Eqs.(\ref{eq:k-n-ell-def}) and (\ref{eq:levels}) we obtain at fixed $\ell$ the relation
between the energy level and the energy
\be
\left. \frac{\partial n}{\partial E} \right|_{\ell} = \frac{1}{\pi\epsilon} \int_{r_1}^{r_2} \frac{dr}{k_{n\ell}(r)} .
\label{eq:dn-dE}
\ee
With the change of variables $L=\epsilon\ell$ and $L_z=\epsilon m$ in
Eq.(\ref{eq:rho-psi-squared}), the average density reads in the continuum limit
\be
\langle \rho \rangle = \frac{1}{\pi\epsilon^3 r^2} \int_0^{\infty} dL \int_{-L}^L dL_z | Y_\ell^m |^2
\int dE \frac{| a_{n\ell m} |^2}{k_{n\ell}} .
\label{eq:rho-average}
\ee
For the total angular momentum $J_z$ of the halo along the vertical axis,
defined by Eq.(\ref{eq:Jz-def}), we obtain from Eq.(\ref{eq:j-phi-psi-squared})
\be
\langle J_z \rangle = \frac{1}{\pi\epsilon^3} \int dr \int_0^\infty dL \int_{-L}^L dL_z L_z
\int dE \frac{| a_{n\ell m} |^2}{k_{n\ell}} .
\label{eq:Jz-average}
\ee

\subsection{Collisionless classical system}
\label{app:collisionless}

For a classical system of collisionless particles, we have for the angular momentum and the
energy of a particle
\be
L_z= r \sin\theta v_\varphi , \; L^2 = r^2 ( v_\theta^2 + v_\phi^2) , \;
E = \frac{v_r^2}{2} + \frac{L^2}{2 r^2} + \Phi_N .
\ee
Then, for a phase-space distribution function $f(E,L,L_z)$ the density
$\rho = \int d\vec v \, f$ also reads
\ba
\rho & = & \frac{2}{r_\perp r} \int_{\Phi_N}^{\infty} dE \int_0^{2 r^2 (E-\Phi_N)}
\frac{d L^2}{\sqrt{2 (E-\Phi_N) - L^2/r^2}} \nonumber \\
&& \times \int_{-L r_\perp/r}^{L r_{\perp}/r}
\frac{d L_z}{\sqrt{L^2 - L_z^2 r^2/r_{\perp}^2}} f(E,L,L_z) ,
\ea
where $r_\perp = r \sin\theta$ is the distance from the vertical axis.
If the distribution function only depends on $E$ and $L_z$ we can integrate over $L$,
which gives
\be
\rho = \frac{2\pi}{r_\perp} \int_{\Phi_N}^{\infty} dE
\int_{-r_\perp \sqrt{2 (E-\Phi_N)}}^{r_\perp \sqrt{2 (E-\Phi_N)}} dL_z f(E,L_z) ,
\label{eq:rho-f}
\ee
while the azimuthal current reads
\be
\rho v_\varphi = \frac{2\pi}{r_\perp^2} \int_{\Phi_N}^{\infty} dE
\int_{-r_\perp \sqrt{2 (E-\Phi_N)}}^{r_\perp \sqrt{2 (E-\Phi_N)}} dL_z L_z f(E,L_z) .
\label{eq:rho-v_phi-f}
\ee
From the definition (\ref{eq:Jz-def}), this gives for the total angular momentum
along the vertical axis
\ba
J_z & = & 8 \pi^2 \int dr \, r \int_{\Phi_N}^{\infty} dE
\int_{-r \sqrt{2 (E-\Phi_N)}}^{r \sqrt{2 (E-\Phi_N)}} dL_z \, L_z  \nonumber \\
&& \times {\rm Arccos}\left( \frac{| L_z |}{r \sqrt{2 (E-\Phi_N)}} \right)  f(E,L_z)  ,
\label{eq:Jz-f}
\ea
where we integrated over the angles on the sphere.

\subsection{Semi-classical limit}
\label{app:semi-classical}

For phase-space distributions of the form (\ref{eq:f-f0-fm})-(\ref{eq:fm-Lz-def}),
where the symmetric part over $L_z$ only depends on $E$,
we can see from Eq.(\ref{eq:rho-f}) that the antisymmetric part does not contribute to the density
and integrating over $L_z$ we obtain
\be
\rho = 4\pi \int_{\Phi_N}^{\infty} dE \sqrt{2 (E-\Phi_N)} f_0(E) .
\label{eq:rho-f0}
\ee
On the other hand, the symmetric part does not contribute to the total angular momentum
(\ref{eq:Jz-f}), which reads
\ba
J_z & = & 16 \pi^2 \int dr \, r \int_{\Phi_N}^{\infty} dE
\int_0^{r \sqrt{2 (E-\Phi_N)}} dL_z \, L_z  \;\;\; \nonumber \\
&& \times {\rm Arccos}\left( \frac{L_z}{r \sqrt{2 (E-\Phi_N)}} \right)  f_-(E,L_z) .
\label{eq:Jz-fm}
\ea

In a similar fashion, let us consider coefficients $a_{n\ell m}$ of the expansion
(\ref{eq:psi-halo-a_nlm}) of the form
\be
| a_{n\ell m} |^2 = S(E_{n\ell}) + T(E_{n\ell}, L_z) ,
\ee
where again $L_z = \epsilon m$, and
\be
T(E,-L_z) = - T(E,L_z) .
\ee
Then, the antisymmetric component $T$ does not contribute to the average density
(\ref{eq:rho-average}).
Using $\sum_{m=-\ell}^\ell | Y_\ell^m |^2 = (2\ell+1)/(4\pi)$ and integrating over $L$
we obtain
\be
\langle \rho \rangle = \frac{1}{2\pi^2 \epsilon^3} \int_{\Phi_N}^{\infty} dE \sqrt{2 (E-\Phi_N)} S(E) .
\label{eq:rho-S}
\ee
On the other hand, the symmetric part $S$ does not contribute to the total angular momentum
(\ref{eq:Jz-average}) and integrating over $L$ we obtain
\ba
\langle J_z \rangle & = & \frac{2}{\pi\epsilon^3} \int dr \, r \int_{\Phi_N}^\infty dE
\int_0^{r \sqrt{2 (E-\Phi_N)}} dL_z \, L_z  \nonumber \\
&& \times {\rm Arccos}\left( \frac{L_z}{r \sqrt{2 (E-\Phi_N)}} \right) T(E,L_z)  .
\label{eq:Jz-T}
\ea
Comparing Eqs.(\ref{eq:rho-f0})-(\ref{eq:Jz-fm}) with Eqs.(\ref{eq:rho-S})-(\ref{eq:Jz-T}),
we can see that we recover the density and angular momentum of the classical system
if we take
\be
S(E) = (2\pi\epsilon)^3 f_0(E) , \;\; T(E,L_z) = (2\pi\epsilon)^3 f_-(E,L_z) ,
\ee
which gives the expression (\ref{eq:a_nellm-f_E-Lz}) for the amplitude of the coefficients
$a_{n\ell m}$.

\acknowledgments

This work was granted access to the CCRT High-Performance Computing (HPC) facility under the Grants CCRT2024-valag and CCRT2025-valag
awarded by the Fundamental Research Division (DRF) of CEA.

\bibliography{ref}

\end{document}